\documentclass[iop,twocolappendix,numberedappendix,appendixfloats, tighten]{emulateapj}

\usepackage{graphicx}
\usepackage[space]{grffile}
\usepackage{latexsym}
\usepackage{amsfonts,amsmath,amssymb}
\usepackage{url}
\usepackage[utf8]{inputenc}
\usepackage{fancyref}
\usepackage{textcomp}
\usepackage{longtable}
\usepackage{multirow,booktabs}
\usepackage{subfigure}
\usepackage{natbib}
\usepackage{color}

\usepackage[backref,breaklinks,colorlinks,citecolor=blue]{hyperref}

\newcommand{\msol}{M$_\odot$}
\newcommand{\mass}{M$_*$}
\newcommand{\logmass}{$\mathrm{log_{10}(M_*/M_\odot)}$}
\newcommand{\zsol}{Z$_\odot$}
\newcommand{\hubble}{{\it Hubble}}

\newcommand{\Halpha}{H$\alpha$}
\newcommand{\Hbeta}{H$\beta$}
\newcommand{\SII}{[\hbox{{\rm S}\kern 0.1em{\sc ii}}]}
\newcommand{\AlIII}{\hbox{{\rm Al}\kern 0.1em{\sc iii}}}
\newcommand{\NII}{[\hbox{{\rm N}\kern 0.1em{\sc ii}}]}
\newcommand{\OII}{[\hbox{{\rm O}\kern 0.1em{\sc ii}}]}
\newcommand{\OIII}{[\hbox{{\rm O}\kern 0.1em{\sc iii}}]}
\newcommand{\MgII}{\hbox{{\rm Mg}\kern 0.1em{\sc ii}}}
\newcommand{\MgI}{\hbox{{\rm Mg}\kern 0.1em{\sc i}}}
\newcommand{\FeII}{\hbox{{\rm Fe}\kern 0.1em{\sc ii}}}
\newcommand{\CIII}{\hbox{{\rm C}\kern 0.1em{\sc iii}}}
\newcommand{\CIV}{\hbox{{\rm C}\kern 0.1em{\sc iv}}}

\newcommand{\CII}{\hbox{{\rm C}\kern 0.1em{\sc ii}}}
\newcommand{\OI}{\hbox{{\rm O}\kern 0.1em{\sc i}}}
\newcommand{\NeIII}{[\hbox{{\rm Ne}\kern 0.1em{\sc iii}}] }
\newcommand{\NeII}{[\hbox{{\rm Ne}\kern 0.1em{\sc ii}}] }
\newcommand{\NaI}{[\hbox{{\rm Na}\kern 0.1em{\sc i}}] }

\newcommand{\around}{$\sim$}
\newcommand{\AvSED}{$\mathrm{A_{v}(SED)}$}
\newcommand{\AvZFIRE}{$\mathrm{A_{v}(ZF)}$}
\newcommand{\Av}{$\mathrm{A_v}$}
\newcommand{\gr}{$\mathrm{(g-r)_{0.1}}$}
\newcommand{\NMAD}{$\sigma_{\mathrm{NMAD}}$}
\newcommand{\boxfil}{$\mathrm{[340]-[550]}$}
\newcommand{\dustfil}{$\mathrm{[150]-[260]}$}
\newcommand{\salpeter}{$\Gamma=-1.35$}
\newcommand{\vini}{$v_{ini}$}
\newcommand{\vcrit}{$v_{crit}$}

\newcommand{\sample}{ZFIRE-SP sample}
\newcommand{\nlimits}{56}
\newcommand{\ndetections}{46}

\slugcomment{To appear in The Monthly Notices of the Royal Astronomical Society (MNRAS)}

\shorttitle{The in situ IMF at $z\sim2$}
\shortauthors{Themiya Nanayakkara} 

\begin{document}



\title{ZFIRE: Using \Halpha\ Equivalent Widths to Investigate the
  In Situ Initial Mass Function at $z\sim2$}


\author{Themiya Nanayakkara\altaffilmark{1,*} }
\author{Karl Glazebrook\altaffilmark{1}}
\author{Glenn G. Kacprzak\altaffilmark{1}}
\author{Tiantian Yuan\altaffilmark{2}}
\author{David B. Fisher\altaffilmark{1}}
\author{Kim-Vy Tran\altaffilmark{3}}
\author{Lisa Kewley\altaffilmark{2}}
\author{Lee Spitler\altaffilmark{5,6}}
\author{Leo Alcorn\altaffilmark{3}} 
\author{Michael Cowley\altaffilmark{5,6}}
\author{Ivo Labbe\altaffilmark{4}}
\author{Caroline Straatman\altaffilmark{4,7}}
\author{Adam Tomczak\altaffilmark{8}}

\altaffiltext{1}{Centre for Astrophysics and Supercomputing, Swinburne University of Technology, Hawthorn, Victoria 3122, Australia.}
\altaffiltext{*}{tnanayak@astro.swin.edu.au}
\altaffiltext{2}{Research School of Astronomy and Astrophysics, The Australian National University, Cotter Road, Weston Creek, ACT 2611, Australia.}
\altaffiltext{3}{George P. and Cynthia W. Mitchell Institute for Fundamental Physics and Astronomy, Department of Physics and Astronomy, Texas A \& M University, College Station, TX 77843.}
\altaffiltext{4}{Leiden Observatory, Leiden University, PO Box 9513, 2300
RA Leiden, Netherlands.}
\altaffiltext{5}{Department of Physics \& Astronomy, Macquarie University,
Sydney, NSW 2109, Australia.}
\altaffiltext{6}{Australian Astronomical Observatory, PO Box 915, North
Ryde, NSW 1670, Australia.}
\altaffiltext{7}{Max-Planck Institute for Astronomy, Königstuh 17, 
D-69117, Heidelberg, Germany.}
\altaffiltext{8}{University of California Davis, Department of Physics, One Shields Avenue, Davis, California 95616, USA.}

\begin{abstract}

We use the ZFIRE\footnote{\url{http://zfire.swinburne.edu.au}} survey to investigate the high mass slope of the initial mass function (IMF) for a mass-complete ($\mathrm{log_{10}(M_*/M_\odot)\sim9.3}$) sample of 102 star-forming galaxies at $z\sim2$ using their \Halpha\ equivalent widths (\Halpha-EW) and rest frame optical colours. 
We compare dust-corrected \Halpha-EW distributions with predictions of star-formation histories (SFH) from PEGASE.2 and Starburst99 synthetic stellar population models. We find an excess of high \Halpha-EW galaxies that are up to 0.3--0.5 dex above the model-predicted Salpeter IMF locus and the \Halpha-EW distribution is much broader (10-500\AA) than can easily be explained by a simple monotonic SFH with a standard Salpeter-slope IMF. Though this discrepancy is somewhat alleviated when it is assumed that there is no relative attenuation difference between stars and nebular lines, the result is robust against observational biases, and no single IMF (i.e. non-Salpeter slope) can reproduce the data. 
We show using both spectral stacking and Monte Carlo simulations that starbursts cannot explain the EW distribution.
We investigate other physical mechanisms including models with variations in stellar rotation, binary star evolution, metallicity, and the IMF upper-mass cutoff.  
IMF variations and/or highly rotating extreme metal poor stars (Z$\sim0.1$\zsol) with binary interactions are the most plausible explanations for our data.  If the IMF varies, then the highest \Halpha-EWs would require very shallow slopes ($\Gamma>-1.0$) with no one slope able to reproduce the data. Thus, the IMF would have to vary stochastically. We conclude that the  stellar populations at $z\gtrsim2$ show distinct differences from local populations and there is no simple physical model to explain the large variation in \Halpha-EWs at $z\sim2$.

\end{abstract}



\section{Introduction}
\label{sec:introduction}

The initial mass function (IMF) or the mass distribution of stars formed in a volume of space at a given time is one of the most fundamental empirically derived relations in astrophysics \citep{Salpeter1955,Miller1979,Kennicutt1983,Scalo1986b,Kroupa2001,Chabrier2003,Baldry2003}.  
Since the mass of a star is the primary factor that governs its evolutionary path, the collective evolution of a galaxy is driven strongly by its distribution of stellar masses \citep{Bastian2010}.
Therefore, understanding and quantifying the IMF is of paramount importance since it affects  galactic star formation rates (SFR), galactic chemical evolution, formation and evolution of stellar clusters, stellar remnant populations,  galactic supernova rates, the energetics and phase balance of the interstellar medium (ISM), mass-to-light ratios, galactic dark matter content, and how we model galaxy formation and evolution \citep{Kennicutt1998b,Hoversten2007}.

The IMF of stellar populations can be investigated either via direct \citep[studies that count individual  stars to infer stellar ages to compute an IMF, eg.,][]{Bruzzese2015} or indirect methods \citep[studies that model the integrated light from stellar populations to infer an IMF, eg.,][]{Baldry2003}. 
Due to current observational constraints, the number of extragalactic IMF measurements that utilizes direct measures of the IMF is limited \citep{Leitherer1998}.  
Therefore, most studies employ indirect measures of the IMF, which are affected by numerous systematic uncertainties and limitations.

Indirect IMF measures can be  insensitive to low mass stellar populations since bright O, B, and red supergiant stars may outshine low mass stars. 
In contrast at the highest mass end, there can be an insufficient number massive stars to make a significant contribution to the detected light \citep{Leitherer1998,Hoversten2008}. 
In addition, degeneracies in stellar population models play a significant role in the uncertainties of the derived IMFs, especially since stellar age, stellar metallicity, galactic dust, galactic SFH, and stellar IMF cannot be easily disentangled  \citep{Conroy2013}.
Furthermore, indirect IMF results may depend strongly on more sophisticated features of stellar population models [mainly stellar rotation, binary evolution of O and B stars and the treatment of Wolf–Rayet (W-R) stars  \citep{Wolf2000}] and dark matter profiles of galaxies. 
\citet{Smith2014} showed that galaxy by galaxy comparisons of inferred IMF mass factors via dynamical and spectroscopic fitting techniques can lead to inconsistent results due to our limited understanding of element abundance ratios, dark matter contributions, and/or more sophisticated shape of the IMFs \citep{McDermid2014, Smith2014}.

The concept of an IMF was first introduced by \citet{Salpeter1955} as a logarithmic slope $\Gamma$ defined by,
\begin{equation}
\label{eq:imf_def_salp}
\Phi(\log\ m)=dN/d(\log\ m) \propto m^{\Gamma} 
\end{equation}
where $m$ is the mass of a star, $N$ the number of stars within a logarithmic mass bin, and $\Gamma=-1.35$ is the slope of the IMF.
Historic studies of the IMF slope at the high mass end (M $\gtrsim$ 1 \msol) showed no statistically significant differences from the value derived by Salpeter giving rise to the concept of IMF universality \citep{Scalo1986,Gilmore2001}. 
Theoretical studies attempt to explain the concept of universal IMF by invoking mechanisms such as fragmentation of molecular clouds \citep{Larson1973} or feedback from the ISM \citep{Klishin2016}. 
However, there is no definitive theoretical model that can predict a given universal IMF from first principles, which limits our theoretical understanding of the fundamental physics that govern the IMF.


\subsection{Should the IMF vary?}

We expect the IMF to vary since a galaxy’s metallicity, SFRs, and environment can change dramatically with time \citep[eg.,][]{Schwarzschild1953,Larson1998,Larson2005, Weidner2013a, Chattopadhyay2015,Ferreras2015,Lacey2016}.  
Lower metallicities, higher SFRs, and high cloud surface densities prominent at high-redshift can favour the formation of high-mass stars \citep{Chattopadhyay2015} while interactions between gas clumps in dense environments can suppress the formation of low mass stars \citep{Krumholz2010}.  
Furthermore, physically motivated models of early-type galaxies (ETGs) suggest scenarios in which star formation occurs in different periods giving rise to variability in the mass of the stars formed \citep{Vazdekis1996,Weidner2013a,Ferreras2015}.

Following from theoretical predictions, recent observational studies have started showing increasing evidence for a non-universal IMF \citep{Hoversten2008,vanDokkum2010,Meurer2011,Gunawardhana2011,Cappellari2012,Cappellari2013,LaBarbera2013,Ferreras2013,Conroy2013b,Navarro2015a,Navarro2015d,Navarro2015c,Navarro2015b}. These studies investigate both early and late-type galaxies in different physical and environmental conditions and use different techniques to probe the IMF at the lower and upper mass end.

IMF studies of ETGs in the local universe infers/has-shown a high abundance of low mass stars \citep{vanDokkum2010,Ferreras2013,LaBarbera2013} with strong evidence for IMF variations as a function of galaxy velocity dispersion \citep{Cappellari2012,vanDokkum2012,Cappellari2013,Conroy2013b}, metallicity \citep{Navarro2015c} and radial distance within a galaxy \citep{Navarro2015a}. These results suggest that the IMF of ETGs are most likely to depend on the physical conditions of the galaxy when it formed bulk of its stars.
Local star-forming galaxies show evidence for IMF variation as a function of galaxy luminosity \citep{Schombert1990,Lee2004,Hoversten2008,Meurer2009,Meurer2011}, metallicity \citep{Rigby2004}, and SFR \citep{Gunawardhana2011}.
Comparisons between HI selected low and high surface brightness galaxies have shown the need for a systematic variation of the upper mass cutoff and/or the slope of the IMF to model the far-ultraviolet (far-UV) and \Halpha\ luminosities \citep{Meurer2009,Meurer2011}.  
\Halpha\ EW and optical colour analysis of the SDSS \citep{York2000} data showed that low-luminous galaxies may be deficient in high mass stars \citep{Hoversten2008}, while a similar analysis on the GAMA survey \citep{Driver2009} showed an excess of high mass stars in high star-forming galaxies \citep{Gunawardhana2011}, both compared to expectations from a Salpeter IMF.

In spite of IMF being fundamental to galaxy evolution, our understanding of it at higher redshifts ($z\gtrsim2$) is extremely limited. IMF studies of strong gravitational lenses at $z>1$ have shown no deviation from Salpeter IMF \citep{Pettini2000,Steidel2004,Quider2009}, but quiescent galaxies at $z<1.5$ have shown systematic trends for the IMF with stellar mass \citep{Navarro2015b}. Using local analogues to $z\sim2$ galaxies, \citet{Navarro2015d} finds evidence for an abundance of low mass stars in the early universe.

Understanding the \emph{relic} IMF at high redshift requires populations of quiescent galaxies which are relatively rare at high redshift, extremely long integration times to obtain absorption line/kinematic features, and complicated modelling of stellar absorption line features. 
Since IMF defines the mass distribution of formed stars at a given time, in the context of understanding the role of IMF in galaxy evolution, it should be investigated \emph{in situ} at an era when most galaxies are in their star forming phase and evolving rapidly to produce large elliptical galaxies found locally. 
Furthermore, high mass stars are absent in ETGs and therefore star-forming galaxies are imperative to study the high mass end of the IMF. 
Simulations have shown that $z\sim2$ universe is ideal for such studies \citep{Hopkins2006}.
rest frame optical spectra of high redshift galaxies are dominated by strong emission lines produced by nebulae associated with high mass stars ($>$15\msol) and therefore provide a direct tracer of the high mass end of the IMF \citep{Bastian2010}. 
Due to the recent development of sensitive near infra-red (NIR) imagers and multiplexed spectrographs that take advantage of the Y, J, H, and K atmospheric windows, the $z\sim2$ universe is ideal to study rest frame optical features of galaxies.


\subsection{Investigating the IMF with \Halpha\ Emission}
\label{sec:method}

The total flux of a galaxy at \Halpha\ emission wavelength is the sum of the \Halpha\ emission flux, the continuum level at the same wavelength minus the \Halpha\ absorption. \Halpha\ absorption for galaxies at $z\sim2$ is $\leq$ 3\% of its flux level \citep{Reddy2015} and therefore can be ignored. 
In case B recombination \citep{Brocklehurst1971}, following the Zanstra principle \citep{Zanstra1927} the \Halpha\ flux of a galaxy is directly related to the number of Lyman continuum photons emitted by massive young O and B stars with masses $>$10\msol. The continuum flux at the same wavelength is dominated by red giant stars with masses between $0.7-3.0$ \msol. Therefore, \Halpha\ EW, that is the ratio of the strength of the emission line to the continuum level can be considered as the ratio of massive O and B stars to $\sim1$ \msol\ stars present in a galaxy.

The rest frame optical colours of a galaxy is tightly correlated with its \Halpha\ emission. The \Halpha\ flux probes the specific SFR (sSFR) of the shorter lived massive stars, while the optical colours probe the sSFR of the longer lived less massive stars. Therefore, in a smooth exponentially declining SFH, the optical colour of a galaxy will transit from bluer to redder colours with time due to the increased abundance of older less massive red stars. Similarly, with declining SFR the \Halpha\ flux will decrease and the continuum contribution of the older redder stars will increase, which will act to decrease the \Halpha\ EW in a similar SFH.  
The \Halpha\ EW and optical colours parameter space is degenerated in such a way that the slope of the function is equivalent to lowering the highest mass stars that are formed and/or increasing the fraction of intermediate-mass stars.

Multiple studies have investigated possibilities for IMF variation in galaxies using Balmer line flux in the context of probing SFHs \citep{Meurer2009,Weisz2012,Zeimann2014,Guo2016,Smit2016}. 
Modelling effects of IMF variation using \Halpha\ or \Hbeta\ to UV flux ratios have strong dependence on the assumed SFH and dust extinction of the galaxies and is only sensitive to the upper end of the high mass IMF. Apart from IMF variation \citep{Boselli2009,Meurer2009,Pflamm-Altenburg2009}, stochasticity in SFH \citep{Boselli2009,Fumagalli2011,Guo2016}, non-constant SFHs \citep{Weisz2012}, and Lyman leakage \citep{Relano2012} can provide viable explanations to describe offsets between expected Balmer line to UV flux ratios and observed values. 

\citet{Kauffmann2014} used SFRs derived via multiple nebular emission line analysis with the 4000\AA\ break and H$\delta_A$ absorption to probe the recent SFHs of SDSS galaxies with $\mathrm{log_{10}(M_*/M_\odot)<10}$ and infer possibilities for IMF variation. 
They did not find conclusive evidence for IMF variation, with contradictions in the 4000\AA\ features with \citet{Bruzual2003} stellar templates being attributed to errors in the spectro-photometric calibration. 
However, using absorption line analysis to probe possible IMF variations in actively star-forming galaxies suffers from strong Balmer line emissions that dominate and fill the absorption features. Furthermore, absorption lines probe older stellar populations, and linking them with current star-formation requires further assumptions about the SFH. 

\citet{Smit2016} used SED fitting techniques to probe discrepancies between \Halpha\ to UV SFRs ratios of $z\sim4-5$ galaxies and local galaxies.
They inferred an excess of ionizing photons in the $z\sim4-5$ galaxies but the origin couldn't be distinguished between a shallow high-mass IMF scenario or a metallicity dependent ionizing spectrum. 
Using broad-band imaging and SED fitting techniques to infer \Halpha\ flux has underlying uncertainties from the contamination of other emission lines that fall within the same observed filter (eg., \NII, \SII) and assumptions of IMF, SFH, metallicity, and dust law of the SED templates.

In this paper we use the MOSFIRE NIR spectra of galaxies obtained as a part of the ZFIRE survey \citep{Nanayakkara2016} along with multi-band photometric data from the ZFOURGE survey \citep{Straatman2016} to study the high-mass IMF of a mass complete sample of star-forming galaxies at $z\sim2$.  
MOSFIRE spectra has higher resolution and is able to probe into redder wavelengths (up to K band) compared to \emph{Hubble Space Telescope (HST)} grism spectra \citep[eg.,][]{Zeimann2014}, thus allows us to detect clear \Halpha\ nebular emissions up to $z\sim2.7$. 
Galaxies follow nearly the same locus in \Halpha\ EW, optical colour as long as their SFHs are smoothly increasing or decreasing \citep{Kennicutt1983} and we show this in later figures. The change in galaxy evolutionary tracks due to IMF is largely orthogonal to the changes in tracks due to effects of dust extinction.  
Therefore, our method allows stronger constraints to be made on the high-mass IMF compared to the studies that probe Balmer line to UV flux ratios and is an improvement of the recipe that was first implemented by \citet{Kennicutt1983} and subsequently used by \citet{Hoversten2008} and \citet{Gunawardhana2011} to study the IMF at $z\sim0$.

The paper is organized in the following way. 
Section \ref{sec:EW_calc} describes the sample selection, the continuum fitting procedure, \Halpha\ EW calculation, optical colour derivation, and the completeness of our selected sample.  Section \ref{sec:PEGASE_models} shows how the synthetic stellar population models (SSP) were computed. 
In Section \ref{sec:results} we show the first results of our IMF study and identify stellar population effects that could describe the distribution of our sample. We discuss effects of dust to our analysis in Section \ref{sec:dust}, observational bias in Section \ref{sec:observational_bias}, and starbursts in Section \ref{sec:star_bursts}. In Section \ref{sec:other_exotica} we discuss the effects of other properties such as stellar rotation, binary stars, metallicity, and the high mass cutoff of the stellar models to our analysis. 
Section \ref{sec:other_observables} investigates dependencies of our parameters with other observables. 
Section \ref{sec:discussion} gives a through discussion of our results investigating the change of IMF and other possible scenarios. We conclude the paper in Section \ref{sec:summary} by describing further improvements needed in the field to determine the IMF of the galaxies in the high redshift universe. Throughout the paper we refer to the IMF slope of $\Gamma=-1.35$ computed by \citet{Salpeter1955} as the Salpeter IMF. 
We assume various IMFs and a cosmology with \hubble $= 70$ km/s/Mpc, $\Omega_\Lambda=0.7$ and $\Omega_m= 0.3$. All magnitudes are expressed using the AB system.


\section{Observations \& Data}
\label{sec:EW_calc}

\subsection{Galaxy Sample Selection}
\label{sec:sample_selection}

The sample used in this study was selected from the ZFIRE \citep{Nanayakkara2016} spectroscopic survey, which also consists of photometric data from the ZFOURGE survey \citep{Straatman2016}.
In this section, we describe the sample selection process from the ZFIRE survey for our analysis.

ZFIRE is a spectroscopic redshift survey of star-forming galaxies at $1.5<z<2.5$, which utilized the MOSFIRE instrument \citep{McLean2012} on Keck-I to primarily study galaxy properties in rich environments. ZFIRE has observed $\sim300$ galaxy redshifts with typical absolute accuracy of $\mathrm{\sim15\ kms^{-1}}$ and derived basic galaxy properties using multiple emission line diagnostics. \citet{Nanayakkara2016} give full details on the ZFIRE survey. 
In this study we use the subset of ZFIRE galaxies observed in the COSMOS field \citep{Scoville2007} based on a stellar mass limited sample reaching up to 5$\sigma$ emission line flux limits of $\mathrm{\sim3\times10^{-18}erg/s/cm^2}$ selected from deep NIR data $\mathrm{K_{AB}<25}$ obtained by the ZFOURGE survey.

ZFOURGE\footnote{\url{http://zfourge.tamu.edu}} (PI I. Labb\'e) is a Ks selected deep 45 night photometric legacy survey carried out using the purpose built FourStar imager \citep{Persson2013} in the 6.5 meter Magellan Telescopes located at Las Campanas observatory in Chile. The survey covers 121 arcmin$^2$ in each of the COSMOS, UDS \citep{Beckwith2006}, and CDFS \citep{Giacconi2001} legacy fields. Deep FourStar medium band imaging (5$\sigma$ depth of Ks$\leq$25.3 AB ) and the wealth of public multi-wavelength photometric data (UV to Far infra-red) available in these fields were used to derive photometric redshifts with accuracies $\lesssim1.5\%$ using EAZY \citep{Brammer2008}. Galaxy masses, ages, SFRs, and dust properties were derived using FAST \citep{Kriek2009} with a \citet{Chabrier2003} IMF, exponentially declining SFHs, and \citet{Calzetti2000} dust law. At $z\sim2$ the public ZFOURGE catalogues are 80\% mass complete to $\sim10^9$\msol \citep{Nanayakkara2016}. Refer to \citet{Straatman2016} for further details on the ZFOURGE survey.

ZFIRE and ZFOURGE are ideal surveys to use in this study since both provide mass complete samples. The total ZFIRE sample in the COSMOS field contains 142 \Halpha\ detected [$>5\sigma$, redshift quality flag (Q$_z$)=3 ] star-forming galaxies that is mass complete down to \logmass$>9.30$  (at 80\% for $Ks=24.11$). Thus, our \Halpha\ selected sample  contains no significant systematic biases towards SFH, stellar mass, and magnitude. Furthermore, ZFIRE contains a large cluster at $z=2$ containing 51 members with $5\sigma$ \Halpha\ detections \citep{Yuan2014} and therefore we are able to examine if the IMF is affected by the local environment of galaxies. 

For this study, we apply the following additional selection criteria to the 142 \Halpha\ detected galaxies.\\  

$\bullet$ We remove active galactic nucleus (AGN) using photometric \citep{Cowley2016} and emission line ($\mathrm{log_{10}}$(f(\NII)/f(\Halpha))$>-0.5$; \citet{Coil2015}) criteria resulting in identifying 26 AGN with our revised sample containing $N=116$ galaxies.
We note that all galaxies selected as AGN from ZFOURGE photometry by \citet{Cowley2016} are flagged as AGN by the \citet{Coil2015} selection. We further discuss contamination to \Halpha\ from sub-dominant AGN in Appendix \ref{sec:AGN}.\\

$\bullet$ Galaxies must have a matching ZFOURGE counterpart such that we can obtain galaxy properties, resulting in N=109 galaxies.\\

$\bullet$ We compute the total spectroscopic flux for these galaxies and remove 4 galaxies with negative fluxes resulting in N=105 galaxies. 
We perform stringent \Halpha\ emission quality cuts to the spectra for these 105 galaxies and remove 2 galaxies due to strong sky line subtraction issues. We further remove 1 galaxy due to an overlap of the galaxy spectra with a secondary object that falls within the same slit.\\

\begin{figure}
\includegraphics[trim= 10 10 10 10 , clip, scale=0.9]{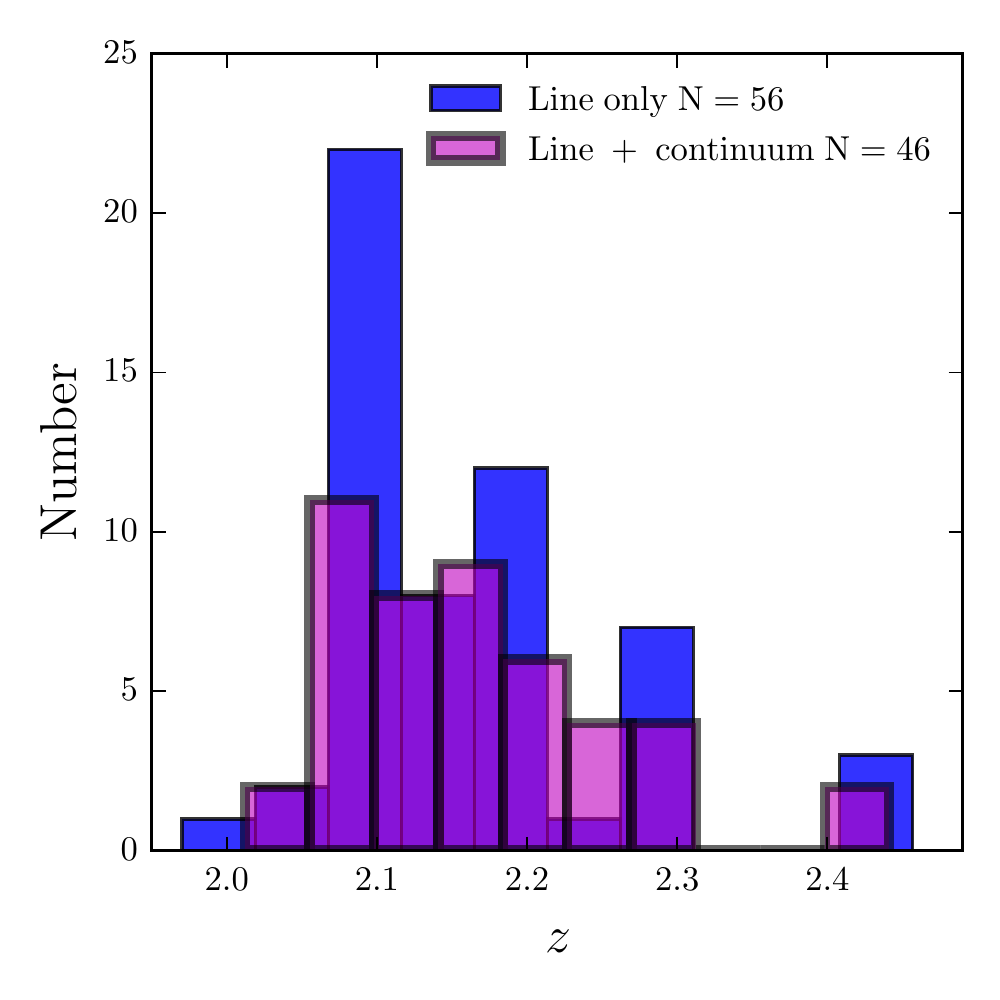}
\caption{The redshift distribution of the \sample. Galaxies with line+continuum detection are shown by magenta and galaxies only with \Halpha\ line detection are shown by blue.}
\label{fig:specz_distribution}
\end{figure}

Our final sample of galaxies used for the IMF analysis in this paper comprise of 102 galaxies. 
Henceforth we refer to this sample of galaxies as the ZFIRE stellar population (SP) sample. \\
The redshift distribution for the \sample\ is shown by Figure \ref{fig:specz_distribution}. The \sample\ is divided into continuum detected and non-detected galaxies as described in section \ref{sec:cont_fit}. Galaxies in our sample lie within redshifts of $1.97<z<2.46$ corresponding to a $\Delta t \sim 650$ Myr.

\subsection{Completeness}
\label{sec:completeness}

In order to determine any significant detection biases in our \sample, we evaluate the completeness of the galaxies selected in this analysis. 
We define a redshift window for analysis between $1.90<z<2.66$ ($\sim8.6$ Gpc), which corresponds to the redshifts that \Halpha\ emission will fall within the MOSFIRE K band.  Note that here we discuss galaxies with \Halpha\ detections and Q$_z>1$, while in Section \ref{sec:sample_selection} we discussed the Q$_z=3$ \Halpha\ detected sample.

In the ZFOURGE catalogues used for the ZFIRE sample selection (see \citet{Nanayakkara2016} for details), there were 1159 galaxies (including star-forming and quiescent galaxies) in the COSMOS field with photometric redshifts ($z_{photo}$) within $1.90<z_{\mathrm{photo}}<2.66$.  
160 of these galaxies with $1.90<z_{\mathrm{photo}}<2.66$ were targeted in K band out of which 
128\footnote{Note that this is different from the 142 galaxies mentioned in Section \ref{sec:sample_selection} because the sample of 142 galaxies has a Q$_z=3$, includes galaxies with no ZFOURGE counterparts (see \citet{Nanayakkara2016} for further details) and galaxies with non-optimal ZFOURGE photometry (see \citet{Straatman2016} for further details).} 
were detected with at least one emission line with SNR $>5$. 
 None of the \Halpha\ detected galaxies had spectroscopic redshifts outside the considered redshift interval. 
However, 3 additional galaxies (1 object with Q$_z=2$, 2 objects with Q$_z=3$) fell within $1.90<z<2.66$ due to inaccurate photometric redshifts. 
There were 8 galaxies targeted in K band that did not have \Halpha\ detections but do have other emission line detections (i.e. No \Halpha\ but have \NII, \OIII, \Hbeta\ etc). Furthermore, there were no galaxies that were targeted in K band expecting \Halpha\ but resulted in other emission line detections.

There were 151 objects within $1.90<z<2.66$ with \Halpha\ detections (Q$_z>1$) and 26 of them were flagged as AGN following selection criteria from \citet{Coil2015} and \citet{Cowley2016}. In the remaining 125 galaxies, 8 galaxies did not have matching ZFOURGE counterparts and 8 galaxies had low confidence for redshift detection (Q$_z=2$) from \citet{Nanayakkara2016}. We removed those 16 galaxies from the sample. 
Out of the 109 remaining galaxies, seven are removed due to the following reasons: four galaxies due to negative spectroscopic flux, one galaxy due to multiple objects overlapping in the spectra, and two galaxies due to extreme sky line interference.

Our sample constitutes of the remaining 102 galaxies out of which, 46 have continuum detections (see Section \ref{sec:cont_fit}). Furthermore, 38 (out of which, 16 are continuum detected) galaxies are confirmed cluster members \citep{Yuan2014} and the remaining 64 (out of which, 30 are continuum detected) galaxies comprise of field galaxies. 
32 galaxies targeted with photometric redshifts between $1.90<z<2.66$ show no \Halpha\ emission detection. 
We divide our sample into 3 mass bins with masses between $\log_{10}(\mathrm{M}_\odot)<9.5$, $\mathrm{9.5\leq log_{10}(M_\odot)\leq 10.0}$, $\mathrm{10.0<log_{10}(M_\odot)}$ and show the corresponding data as described above in Table \ref{tab:sample_details}.

We define observing completeness as the percentage of detected galaxies (Q$_z>1$) with photometric redshifts between $1.90<z<2.66$ and calculate it to be  $\sim80\%$.
However, it is possible that the 32 null detections with photometric redshifts  within $1.90<z<2.66$ to have been detected if the ZFIRE survey was more sensitive. We stack the the photometric redshift likelihood functions ($P(z)$) of the ZFIRE targeted galaxies within this redshift range, to compute the expectation of detections based of photometric redshift accuracies (See \citet{Nanayakkara2016} Section 3.2 to further details on how $P(z)$ stacking is performed) . The calculated expectation for \Halpha\ to be detected within K band is $\sim80\%$, which is extremely similar to the observed completeness. Therefore, non-detections rate is consistent with uncertainties in the photometric redshifts. 
To further account for any detection bias, we employ a stacking technique of the non-detected spectra in order to calculate a lower-limit to the stacked EW values. This is further discussed in Section \ref{sec:EW_stacking}.

\begin{deluxetable*}{ccccccccccc}
\tabletypesize{\scriptsize}
\tablecaption{ Galaxies selected for the IMF study.
\label{tab:sample_details}}
\tablecolumns{10}
\tablewidth{0pt} 
\tablehead{
\colhead{\mass\tablenotemark{a}} &
\colhead{$\mathrm{N_{ZFOURGE}}$} &
\colhead{$\mathrm{N_{ZFIRE}}$} &
\colhead{$\mathrm{N_{detections}}$} &
\colhead{$\mathrm{N_{outliers}}$} &
\colhead{$\mathrm{N\tablenotemark{b}_{AGN}}$} &
\colhead{$\mathrm{N_{sky}}$} &
\colhead{$\mathrm{N_{selected}}$} &
\colhead{$\mathrm{N_{line\_only}}$} &
\colhead{$\mathrm{N_{null\_detection}}$} &
}
\startdata
$<9.5$     & 568 & 59 &  44 & 0 &  2 & 1 & 41 & 34 &  9 & \\ 
$9.5-10.0$ & 318 & 47 &  39 & 0 &  1 & 0 & 35 & 16 &  6 & \\ 
$10.0<$    & 273 & 54 &  45 & 0 & 20 & 1 & 26 &  6 &  5 & \\ 
Total      &1159 &160 & 128 & 0 & 23 & 2 &102 & 56 & 20 & \\
\tablecomments{The columns keys are as follows:\\
\mass: The Mass bin of the galaxies in log$\mathrm{_{10}}$(\msol).\\ 
$\mathrm{N_{ZFOURGE}}$: Number of ZFOURGE galaxies with photometric redshifts within $1.90<z<2.66$ .\\
$\mathrm{N_{ZFIRE}}$: Number of ZFIRE targeted galaxies in K band with photometric redshifts within $1.90<z<2.66$ .\\
$\mathrm{N_{detections}}$: Number of ZFIRE detected galaxies in K band (Q$_z>1$) with  spectroscopic redshifts within $1.90<z<2.66$ .\\
$\mathrm{N_{outliers}}$: Number of ZFIRE detected galaxies with spectroscopic redshifts outside $1.90<z<2.66$ .\\
$\mathrm{N_{AGN}}$: Number of ZFIRE detected galaxies identified as AGN with  spectroscopic redshifts within $1.90<z<2.66$ .\\
$\mathrm{N_{sky}}$: Number of ZFIRE detected galaxies with spectroscopic redshifts within $1.90<z<2.66$  removed from \sample\ due to sky line interference.\\
$\mathrm{N_{selected}}$: Number of ZFIRE detected galaxies selected for the IMF study with  spectroscopic redshifts within $1.90<z<2.66$.\\
$\mathrm{N_{line\_only}}$: Number of galaxies selected for the IMF study which shows no continuum detection with  spectroscopic redshifts within $1.90<z<2.66$ .\\
$\mathrm{N_{null\_detection}}$: Number of ZFIRE K band targeted galaxies with photometric redshifts within $1.90<z<2.66$  and no \Halpha\ detection.}
\tablenotetext{a}{Where applicable spectroscopic redshifts have been used to calculate the stellar masses from FAST.}
\tablenotetext{b}{One galaxy flagged as an AGN doesn't have a matching ZFOURGE counterpart.}
\end{deluxetable*}

\subsection{Continuum fitting and \Halpha\ EW calculation}
\label{sec:cont_fit}

In this section, we describe our continuum fitting method for our 102 \Halpha\ detected galaxies selected from the ZFIRE survey. 
Fitting a robust continuum level to a spectrum requires nebular emission lines and sky line residuals to be masked. Furthermore, the wavelength interval used for the continuum fit should be sufficient enough to perform an effective fit but should be smaller enough to not to be influenced by the intrinsic SED shape. After extensive testing of various measures used to fit a continuum level, we find the method outlined below to be the most effective to fit a continuum level for our sample.

By visual inspection and spectroscopic redshift of the galaxies in our sample, we mask out the \Halpha\ and \NII\ emission line regions in the spectra. 
We further mask all known sky lines by selecting a region $\times 2$ the spectral resolution ($\pm5.5$\AA) of MOSFIRE K band. 
We then use the \texttt{astropy} \citep{Astropy2013} sigma-clipping algorithm to mask out remaining strong features in the spectra.  
These spectra are then used to fit an inverse variance weighted constant level fit, which we consider as the continuum level of the galaxy. 
Three objects fail to give a positive continuum level using this method and for these we perform a $3\sigma$ clip with two iterations without masking nebular emission lines and sky lines.  
Using this method we are able to fit positive continuum levels to all galaxies in our sample.  
We further investigate the robustness of our measures continuum levels in Appendix \ref{sec:cont_Halpha_test} using ZFOURGE photometric data and conclude that our measured continuum level is consistent (or in agreement) with the photometry.

We use two approaches to calculate the \Halpha\ line flux: 1) direct flux measurement and 2) Gaussian fit to sky-line blended and kinematically unresolved emission lines. Our two methods provide consistent results for emission-lines that are not blended with sky lines (see Appendix \ref{sec:cont_Halpha_test}).
By visual inspection, we selected kinematically resolved (due to galaxy rotation etc.) \Halpha\ emission lines that were not blended with sky-lines and computed the EW by integrating the line flux.
Within the defined emission line region, we calculated the  \Halpha\ flux by subtracting the flux at each pixel ($F_i$) by the corresponding continuum level of the pixel ($F_{cont_i}$). 
For the remaining sample, which comprises of galaxies with no strong velocity structure and galaxies with \Halpha\ emission with little velocity structure and/or \Halpha\ contaminated by sky lines, we perform Gaussian fits to the emission lines, to calculate the \Halpha\ flux values. We then subtract the continuum level from the computed \Halpha\ line flux.

Next, we use the calculated \Halpha\ flux along with the fitted continuum level to calculate the observed \Halpha\ EW ($H\alpha_{EW_{obs}}$) as follows:
\begin{subequations}
\begin{equation}
\label{eq:Ha_EW_obs}
H\alpha\ EW_{obs} =  \sum_{i} (1-\frac{F_i - F_{cont_i}}{F_{cont_i}}) \times \Delta \lambda_i
\end{equation} 
where $\Delta \lambda_i$ is the increment of wavelength per pixel.
Finally, using the spectroscopic redshift ($z$) we calculate the rest frame \Halpha\ EW ($H\alpha_{EW_{rest}}$), which we use throughout the paper:
\begin{equation}
\label{eq:Ha_EW_rest}
H\alpha\ EW = \frac{H\alpha\ EW_{obs}}{1+z}
\end{equation} 
\end{subequations}

We calculate EW errors by bootstrap re-sampling the flux of each spectra randomly within limits specified by its error spectrum. We re-calculate the EW iteratively 1000 times and use the 16$\mathrm{^{th}}$ and 84$\mathrm{^{th}}$ percentile of the distribution of the EW measurements as the lower and upper limits for the EW error, respectively. 
Since the main uncertainty arises from the continuum fitting, we do not consider the error of the \Halpha\ flux calculation in our bootstrap process.

The robustness of an EW measurement relies on the clear identification of the nebular emission line and the underlying continuum level. The latter becomes increasingly hard to quantify at high redshift for faint star forming sources due to the continuum not being detected. 
Therefore we derive continuum detection limits to identify robustly measured continua from non-detections.

In order to establish the limit to which our method can reliably measure the continuum, we select 14 2D slits with no continuum or nebular emission line detections to extract 1D spectra. We define extraction apertures using a representative profile of the flux monitor star and perform multiple extractions per slit depending on their spatial size. 
A total of 93 1D sky spectra are extracted and their continuum level is measured by masking out sky lines and performing a sigma-clipping algorithm. 
The error of the sky continuum fit is calculated by bootstrap re-sampling of the sky fluxes 1000 times. We consider the $1\sigma$ scatter of the bootstrapped continuum values to be the error of the sky continuum fit and $1\sigma$ scatter of the flux values used for the continuum fit as the RMS of the flux.

\begin{figure}
\includegraphics[trim = 10 10 10 0, clip, scale=0.9]{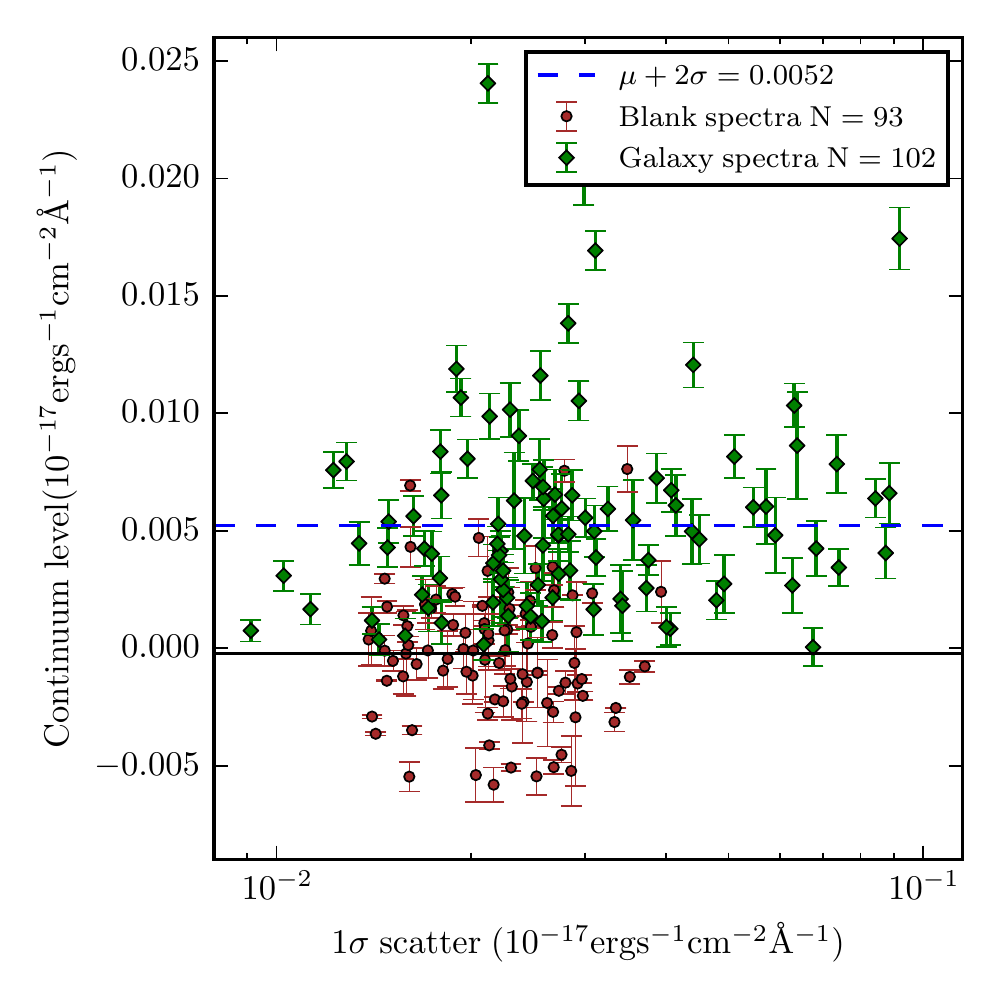}
\caption{ The figure illustrates the continuum detection levels for the \sample. 
The measured continuum level is plot against the $1\sigma$ scatter of the flux values used to fit the continuum level. 
The brown circles represent the continuum levels calculated for the blank slits and the green diamonds represent the continuum level calculated for the IMF sample. 
The blue horizontal line is the $2\sigma$ scatter above the median ($\sim$ 0) for the blank sky regions. Any continua detected above this level of $5.2\mathrm{\times 10^{-20} ergs/s/cm^2/\AA}$ are considered as detected continuum levels. 
\label{fig:noise_levels}
}
\end{figure}

The comparison between the flux continuum level for the sky spectra with the \sample\ spectra are shown in Figure \ref{fig:noise_levels}. 
The median and the 2$\sigma$ standard deviation for the continuum levels of the sky spectra are $\mathrm{-2.3\times 10^{-21} ergs/s/cm^2/\AA}$ and $\mathrm{5.4 \times 10^{-20} ergs/s/cm^2/\AA}$ respectively. 
We consider the horizontal blue dashed line in the Figure \ref{fig:noise_levels}, which is $2\sigma$ above the median sky level, to be our lower limit for the continuum detections in our sample. The \ndetections\ galaxies in our \sample\ with continuum levels above this flux level detections are considered to have a robust continuum detection. For the remaining \nlimits\ galaxies we consider the continuum  measurement as a limit and use it to calculate a lower limit to the \Halpha\ EW values. The redshift distribution of these galaxies is shown by Figure \ref{fig:specz_distribution}.

\subsection{Calculating optical colours}
\label{sec:col_calculation}

Rest frame optical colours for the \sample\ are computed using an updated version of EAZY\footnote{Development version: \url{https://github.com/gbrammer/eazy-photoz/}} \citep{Brammer2008}, which derives the best-fitting SEDs for galaxies  using high quality ZFOURGE photometry to compute the colours. We investigate the robustness of the rest frame colour calculation of EAZY in Appendix \ref{sec:EAZY colour comparision}.    
The main analysis of our sample is carried out using optical colours derived using two idealized, synthetic box car filters, which probes the bluer and redder regions of the rest frame SEDs. We select these filters to avoid regions of strong nebular emission lines as explained in Section \ref{sec:PEGASE_models} and Appendix \ref{sec:box car filters}. 

In order to allow direct comparison between ZFIRE $z\sim2$ galaxies with $z=0.1$ SDSS galaxies from HG08, we further calculate optical colours for the \sample\ at $z=0.1$ using blue-shifted SDSS g and r filters.  
Blueshifting the filters simplifies the (g$-$r) colour calculation at $z=0.1$ (\gr) by avoiding additional uncertainties, which may arise due to K corrections if we are to redshift the galaxy spectra to $z=0.1$ from $z=0$.


\section{Galaxy Spectral Models}
\label{sec:PEGASE_models}

In this section, we describe the theoretical galaxy stellar spectral models employed to investigate the effect of IMF, SFHs, and other fundamental galaxy properties in \Halpha\ EW vs optical colour parameter space. We use PEGASE.2 detailed in \citet{Fioc1997} as our primary spectral synthesis code to perform our analysis and further employ Starburst99 \citep{Leitherer1999} and BPASSv2 \citep{Eldridge2016} models to investigate the effects of other exotic stellar features.

PEGASE is a publicly available spectral synthesis code developed by the Institut d'Astrophysique de Paris. 
Once the input parameters are provided, PEGASE produces galaxy spectra for varying time steps, which can be used to evaluate the evolution of fundamental galaxy properties over cosmic time.

\subsection{Model Parameters}

In this paper, we primarily focus on the effect of varying the IMF, SFH, and metallicity on \Halpha\ EW and  optical colour of galaxies. A thorough description of the behaviour of PEGASE models in this parameter space can be found in \cite{Hoversten2008}. 
The parameters we vary are as follows. 
\begin{itemize}
	\item {\bf The IMF :} we follow HG08 and use an IMF with a single power law  as shown by Equation \ref{eq:imf_def_salp}. Models were calculated with varying IMF slopes ($\Gamma$) ranging between -0.5 to -2.0 in logarithmic space. The lower and upper mass cutoffs were set to 0.5\msol\ and 120\msol, respectively. 
	The IMF indication method used in this analysis is dependant on the ability of a star with a specific mass to influence the \Halpha\ emission and the optical continuum level. Stars below 1\msol\ cannot strongly influence the optical continuum and hence this method is not sensitive to probe the IMF below 1\msol. 
	Furthermore stars below $\sim0.5$\msol\ gives no significant variation to the parameters investigated by this method \citep{Hoversten2008}. Therefore we leave the lower mass cutoff at 0.5\msol.  
	The higher mass cutoff of 120\msol\ used is the maximum mass allowed by PEGASE. 
	Though the high mass end has a stronger influence on the IMF identified using this method we justify the 120\msol\ cutoff due to the ambiguity of the stellar evolution models above this mass. 
	Varying the upper mass cutoff has a strong effect on \Halpha\ EW and optical colours. As HG08 showed this is strongly degenerated with changing $\Gamma$. In this work we focus on $\Gamma$ parametrization, noting that changing the cutoff could produce similar effects. 
	We further discuss the degeneracy between the high mass cutoff and the \Halpha\ EW vs optical colours slope in Section \ref{sec:mass_cutoff}. 
 
	\item {\bf The SFH :} exponentially increasing/declining SFHs, constant SFHs, and starbursts are used. Exponentially declining SFHs are in the form of $\mathrm{SFR(t) = p_2\ exp(-t/p_1) / p_1 }$, with p$_1$ varying from 500 to 1500 Myr. 
	Star bursts are used on top of constant SFHs with varying burst strength and time-scales. 
	Further details are provided in Section \ref{sec:simulations}. 

	\item {\bf Metallicity :} models with consistent metallicity evolution and models with fixed metallicity of 0.02 are used.  
\end{itemize}

The other parameters we use for the PEGASE models are as follows.
We use Supernova ejecta model B from \cite{Woosley1995} with the stellar wind option turned on.
The fraction of close binary systems are left at 0.05 and the initial metallicity of the ISM is set at 0.
We turn off the galactic in-fall function and the mass fractions of the substellar objects with star formation are kept at 0. Galactic winds are turned off, nebular emissions are turned on and we do not consider extinction as we extinction correct our data.

\begin{figure}
\begin{center}
 \includegraphics[trim=0 0 0 0, clip, scale=0.85]{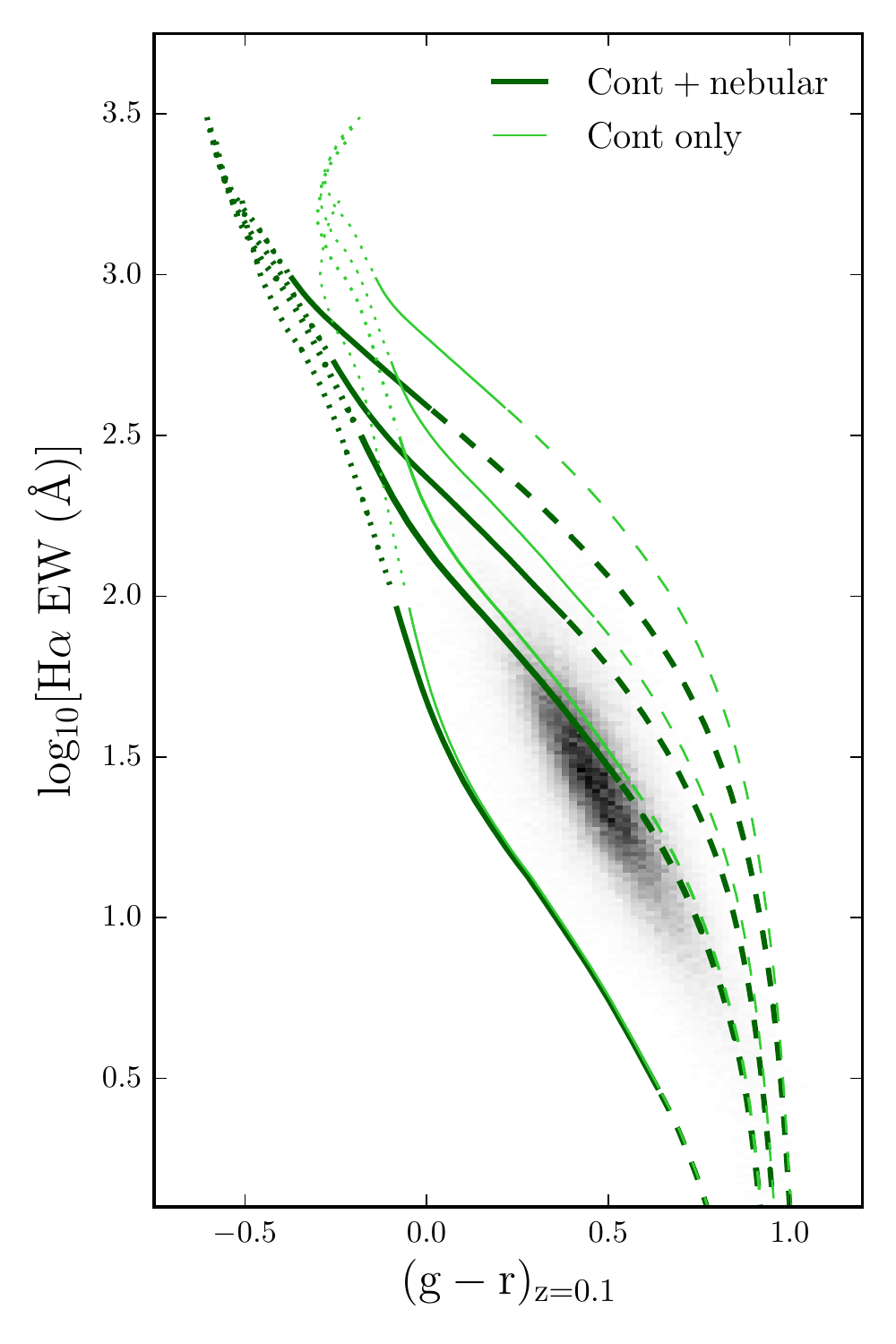} 
 \caption{ The evolution of PEGASE SSP galaxies in the \Halpha\ EW vs \gr\ colour space. We show four galaxy models with  exponentially declining SFHs computed using identical parameters but varying IMFs. The thick dark green tracks show from top to bottom  galaxies with $\Gamma$ values of -0.5,-1.0,-1.35, and -2.0, respectively. The thin light green tracks follow the same evolution as the thick ones, but the nebular line contribution is not considered for the \gr\ colour calculation. All tracks commence at the top left of the figure and are divided into three time bins. The dotted section of the track corresponds to the first 100 Myr of evolution of the galaxy. The solid section of the tracks show the evolution between 100 Myr - 3100 Myr ($z\sim2$) and the final dashed section shows evolution of the galaxy up to 13100 Myr ($z\sim0$). 
 The distribution of the galaxies from the SDSS HG08 sample are shown by 2D histogram.
 }
\label{fig:PEG_neb_comp_gr}
\end{center}
\end{figure}

As a comparison with HG08, in Figure \ref{fig:PEG_neb_comp_gr} we show the evolution of 4 model galaxies from PEGASE in the \Halpha\ EW vs \gr\ colour space. The models computed with exponentially declining SFHs with p$_{1}=1000$ Myr, varying IMFs, and nebular emission lines agrees well with the SDSS data. However, the evolution of the \gr\ colour shows strong dependence on the nebular emission contribution, specially for shallower IMFs. HG08 never considered the effect of emission lines in \gr\ colours and the significant effect at younger ages/bluer colours are likely to be important for $z\sim2$ galaxies.

\begin{figure}
\begin{center}
 \includegraphics[trim=10 0 0 0, clip, scale=0.9]{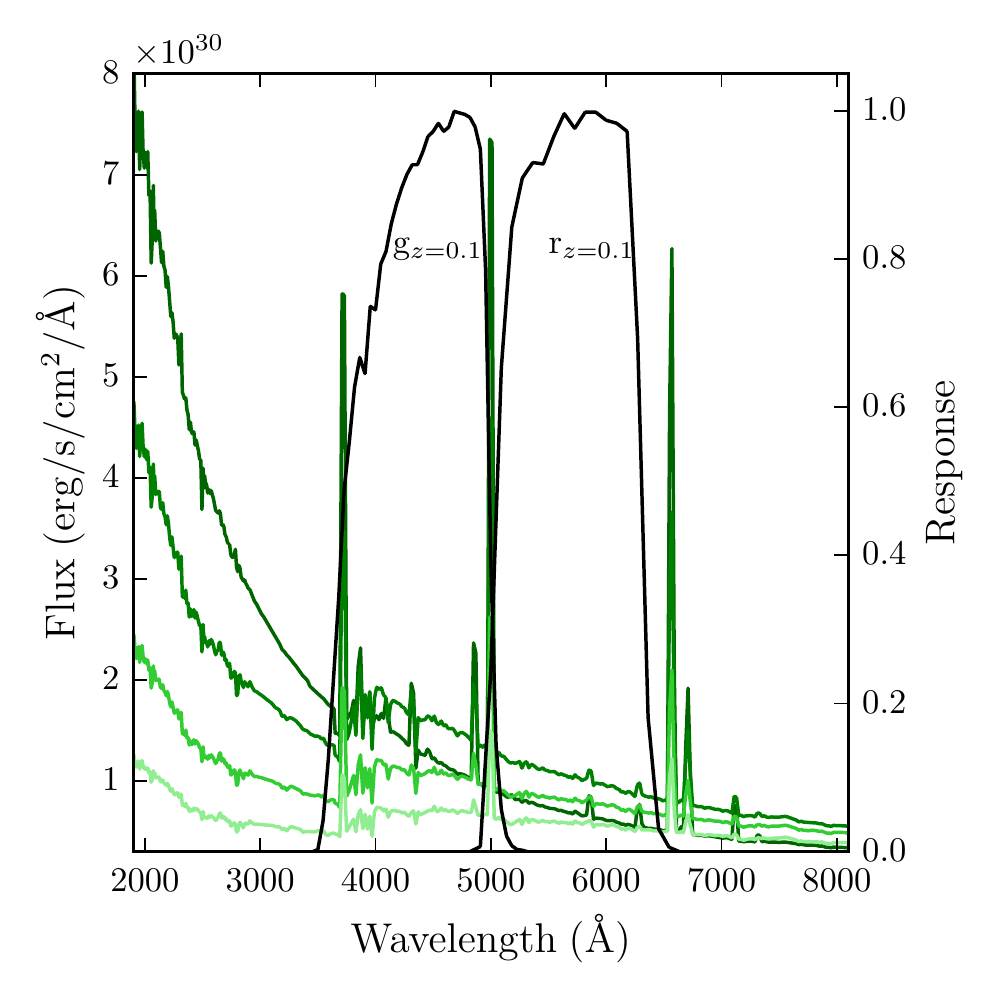} 
 \caption{An example of a model galaxy spectrum generated by PEGASE. Here we show the evolution of the optical wavelength of a galaxy spectra with an exponentially declining SFH and a \salpeter\ with no metallicity evolution. The time steps of the models from top to bottom are: 100 Myr (dark green), 1100 Myr (green), 2100 Myr (limegreen), and 3100 Myr (lightgreen).
 The $g_{z=0.1}$ and $r_{z=0.1}$ filter response functions are overlaid on the figure.  
 }
\label{fig:PEG_spectra}
\end{center}
\end{figure}

Figure \ref{fig:PEG_spectra} shows an example of a synthetic galaxy spectra generated by PEGASE. The galaxy is modelled to have an exponentially declining SFH with $p_1=1000$Myr and a \salpeter\  IMF. Due to the declining nature of the SFR, the stellar and nebular contribution of the galaxy spectra decreases with cosmic time. We overlay the filter response functions of the $g_{z=0.1}$ and $r_{z=0.1}$ filters used in the analysis by HG08. As evident from the spectra, this spectral region covered by the $g_{z=0.1}$ and $r_{z=0.1}$ filters includes strong emission lines such as \OIII\ and \Hbeta. Therefore, the computed \gr\ colours will have a strong dependence on photo-ionization properties of the galaxies.

To mitigate uncertainties in photo-ionization models in our analysis, we employ synthetic filters specifically designed to avoid regions with strong nebular emission lines. We design two box-car filters centred at 3400\AA\ and 5500\AA\ with a width of 450\AA. The rest frame wavelength coverage of these filters corresponds to a similar region covered by the FourStar $\mathrm{J_1}$ and $\mathrm{H_{long}}$ filters in the observed frame for galaxies at $z=2.1$ and therefore requires negligible K corrections. 
Further details on this filter choice is provided in Appendix \ref{sec:filter choice 340}. 
Henceforth, we refer to the blue filter as [340], the redder filter as [550], and the colour of blue filter - red filter as \boxfil. The \boxfil\ colour evolution of a galaxy is independent of the nebular emission lines.

We also compare results using Starburst99 (S99) \citep{Leitherer1999} models in Appendix \ref{sec:SSP comparision}. We find that PEGASE and S99 models show similar evolution and find that our choice of SSP model (PEGASE or S99) to interpret the IMF of the \sample\ at $z\sim2$ to be largely independent to our conclusions. 
However, stellar libraries that introduce rotational and/or binary stars used in these models do have an influence of the \Halpha\ EWs and \boxfil\ colours, which we discuss in detail in Section \ref{sec:ssp_issues}.


\subsection{Comparison to \Halpha\ EW \& optical colours at $z\sim2$}
\label{sec:results}

\begin{figure*}
\includegraphics[trim=5 10 10 0, clip, scale=0.9]{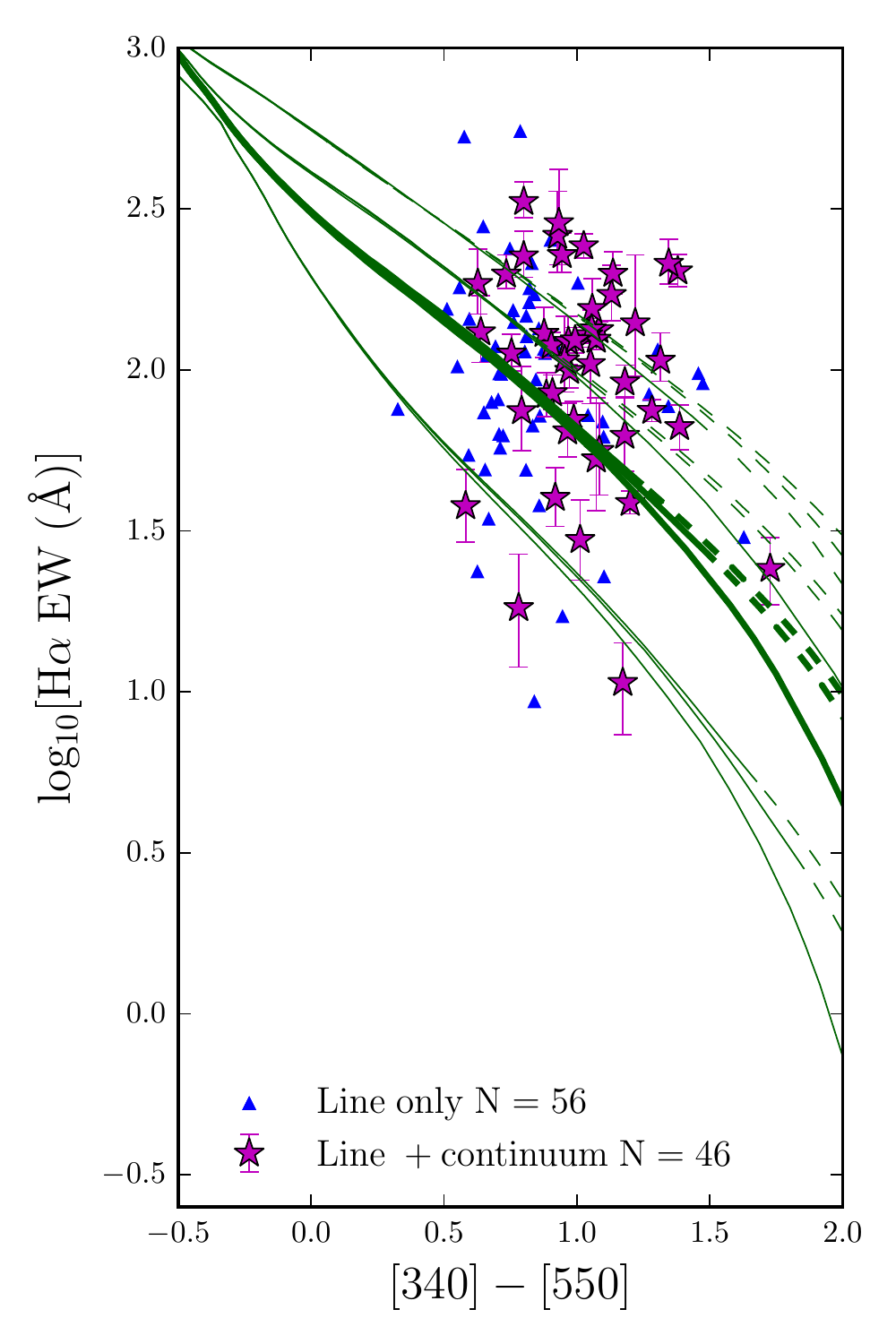}
\includegraphics[trim=5 10 10 0, clip, scale=0.9]{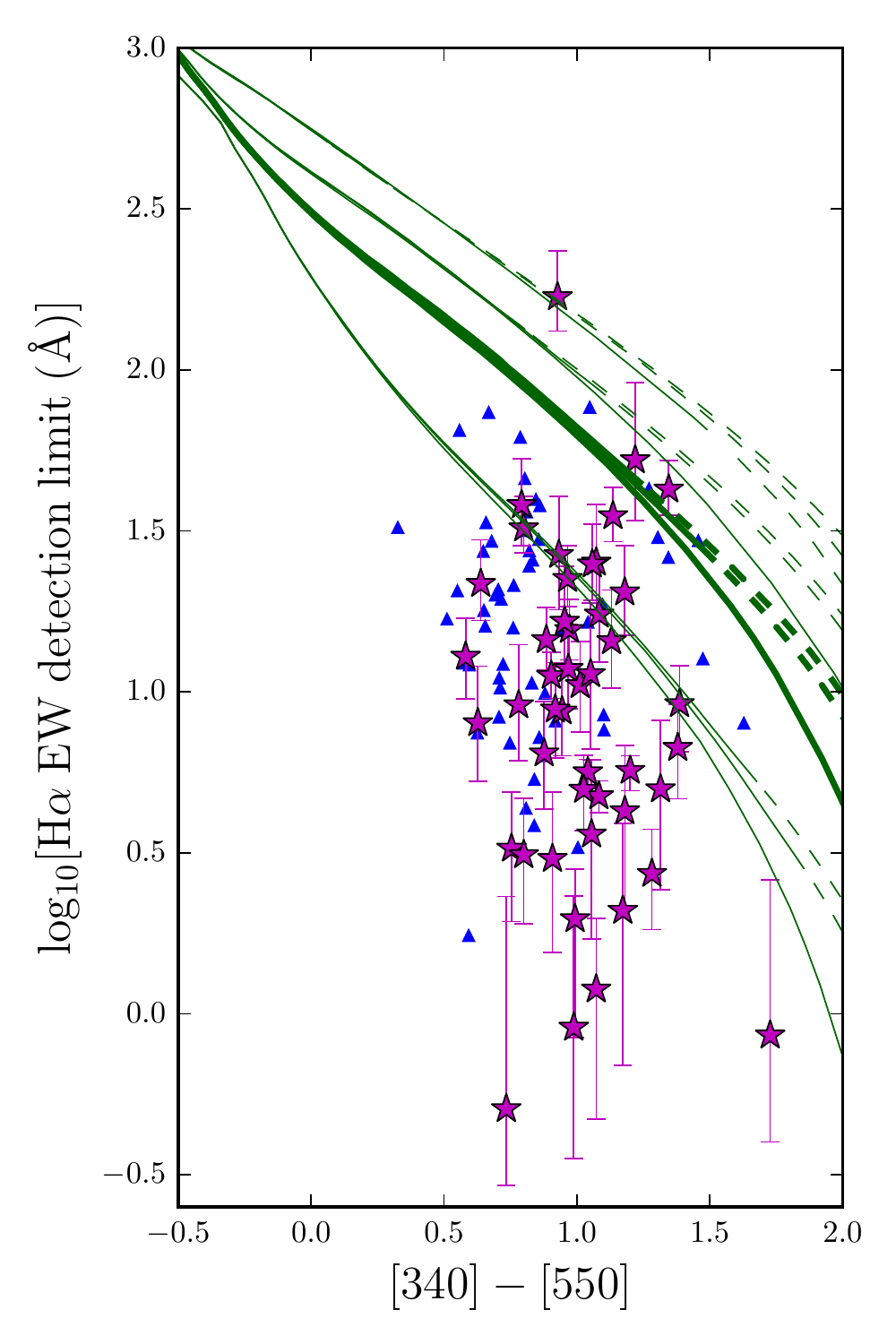}
\caption{The \Halpha\ EW vs \boxfil\ colour distribution of the \sample. No dust corrections have been applied to the observed data. Galaxies with \Halpha\ and continuum detections are shown by magenta stars while galaxies only with \Halpha\ detections (and continuum from $1\sigma$ upper limits) are shown as $1\sigma$ lower limits on EW by blue triangles. The errors for the continuum detected galaxies are from bootstrap re-sampling.
The solid (t $<3200$ Myr) and dashed lines (t $>3200$ Myr) are SSP models computed from PEGASE. Similar to Figure \ref{fig:PEG_neb_comp_gr}, we compute models for 4 IMFs with $\Gamma$ values of $-0.5,-1.0,-1.35$ (this is the thick set of tracks which is similar to the IMF slope inferred by Salpeter), and $-2.0$. Each set of tracks from top to bottom represent these IMF in order. 
For each IMF we compute three models with exponentially declining SFHs with varying p$_1$ values. From top to bottom for each IMF these tracks represent p$_1$ values of 1500 Myr, 1000 Myr, and 500 Myr.
{\bf Left:} The \Halpha\ EW vs \boxfil\ colours of the \sample.  
{\bf Right:} Similar to the left figure but the \Halpha\ EW has been calculated using $1\sigma$ detection limits of the \Halpha\ flux values to demonstrate the sensitivity limits of our EW measurements. }
\label{fig:EW_no_dust_corrections}
\end{figure*}

We explore the IMF of $z\sim2$ star-forming galaxies using \Halpha\ EW values from ZFIRE spectra and rest frame optical colours from ZFOURGE photometry. Our observed sample used in our analysis is shown in Figure \ref{fig:EW_no_dust_corrections}.
The left panel shows the distribution of \Halpha\ EW and \boxfil\ colours of the \sample\ before dust corrections are applied. 
We overlay model galaxy tracks generated by PEGASE for various IMFs. All models are computed using an exponentially declining SFH, but with varying time constants (p$_1$) as shown in the figure caption. For a given IMF, smoothly varying monotonic SFHs have very similar loci in this parameter space. 
The thick set of models (third from top) shows a slope with $\Gamma=-1.35$, which is similar to the Salpeter slope. 
Galaxies above these tracks are expected to contain a higher fraction of higher mass stars in comparison to the mass distribution expected following a Salpeter  IMF. Similarly galaxies below these tracks are expected to contain a lower fraction of high mass stars. 
Galaxies have a large spread in this parameter space but we expect this scatter to decrease when dust corrections are applied to the data as outlined in Section \ref{sec:dust_corrections}.

We note the large scatter of the \Halpha\ EW values with respect to the Salpeter IMF, especially the large number of high EW objects ($\gtrsim0.5$ dex above the Salpeter locus). Could this simply be due to the \sample\ only detecting \Halpha\ emissions in bright objects?\ i.e. a sample bias. First, we note our high completeness of $\sim80\%$ for \Halpha\ detections (Section \ref{sec:completeness}). Second, our \Halpha\ flux limits are actually quite faint. To show this explicitly, we define \Halpha\ flux detection limits for our sample using $1\sigma$ detection thresholds for each galaxy parametrised by the integration of the error spectrum within the same width as the emission line. 
Figure \ref{fig:EW_no_dust_corrections} (right panel) shows the \Halpha\ EW calculated using \Halpha\ flux detection limits, which illustrates the distribution of the \sample\ if the \Halpha\ flux was barely detected.
The \Halpha\ EW of the continuum detected galaxies decrease by $\sim1$ dex which suggest that our EW detection threshold is not biased towards higher \Halpha\ EW values. \\

Similar to IMF, there are a number of effects that may account for the clear disagreement between the observed data and models. In subsequent sections we explore effects from\\
$\bullet$ dust (Section \ref{sec:dust}),\\
$\bullet$ observational bias (Section \ref{sec:observational_bias}),\\
$\bullet$ star bursts (Section \ref{sec:star_bursts}),\\
$\bullet$ stellar rotation (Section \ref{sec:stellar_rotation}),\\
$\bullet$ binary stellar systems (Section \ref{sec:binaries}),\\ 
$\bullet$ metallicity (Section \ref{sec:model_Z}), and\\
$\bullet$ high mass cutoff (Section \ref{sec:mass_cutoff})\\
in SSP models to explain the distribution of \Halpha\ EW vs optical colours of the \sample\ without invoking IMF change.

\section{Is Dust the Reason?}
\label{sec:dust}

As summarized by \citet{Kennicutt1983}, the dust vector is nearly orthogonal to IMF change vector, and therefore, we expect the tracks in the \Halpha\ EW vs optical colour parameter space to be independent of galaxy dust properties. 
In this section, we describe galaxy dust properties.We explain how dust corrections were applied to the data and their IMF dependence and explore the difference in reddening between stellar and nebular emission line regions as quantified by \citet{Calzetti2000} for $z\sim0$ star-forming galaxies.

We use FAST \citep{Kriek2009} with  ZFIRE spectroscopic redshifts from \citet{Nanayakkara2016} and multi-wavelength photometric data from ZFOURGE \citep{Straatman2016} to generate estimates for  stellar attenuation (Av) and stellar mass for our galaxies. 
FAST uses SSP models from \citet{Bruzual2003} and a $\chi^2$ fitting algorithm to derive ages, star-formation time-scales, and dust content of the galaxies. All FAST SED templates  have been calculated assuming solar metallicity, \citet{Chabrier2003} IMF, and \citet{Calzetti2000} dust law.  We refer the reader to \citet{Straatman2016} for further information on the use of FAST to derive stellar population properties in the ZFOURGE survey.


\subsection{Applying SED derived dust corrections to data}
\label{sec:dust_corrections}

We use stellar attenuation values calculated by FAST to perform dust corrections to our data. 
First, we consider the dust corrections for rest frame \Halpha\ EWs and then we correct the \boxfil\ colours. 

By using \cite{Cardelli1989} and \citet{Calzetti2000} attenuation laws to correct nebular and continuum emission lines, respectively, we derive the following equation to obtain dust corrected \Halpha\ EW ($EW_i$) values:
\begin{equation}
\label{eg:EW_dust_corrected}
log_{10}(EW_i) = log_{10}(EW_{obs}) + 0.4A_c(V)(0.62f-0.82)
\end{equation}
where $EW_{obs}$ is the observed EW, $A_c$ is the SED derived continuum attenuation, and $f$ is the difference in reddening between continuum and nebular emission lines. 

\cite{Calzetti2000} found a $f\sim2$ for $z\sim0$ star-forming galaxies, which we use for our analysis under the assumption that the actively star forming galaxies at $z\sim0$ are analogues to star-forming galaxies at $z\sim2$. 
Henceforth, for convenience we refer to $f=1/0.44$ \cite{Calzetti2000} value as $f=2$.
We further show key plots in this analysis using a dust correction of $f=1$ to consider equal dust extinction between stellar and ionized gas regions. This is driven by the assumption that A and G stars that contribute to the continuum of $z\sim2$ star-forming galaxies are still associated within their original birthplaces similar to O and B stars due to insufficient time for the stars to move away from the parent birth clouds within the $<3$ Gyr time-scale.

Similarly, using \citet{Calzetti2000} attenuation law we obtain dust corrected fluxes for the [340] and [550] filters as follows:
\begin{subequations}
\begin{equation}
\label{eq:BC340 dust corrected}
f([340]) = f([340]_{obs}) \times 10^{0.4 \times 1.56 A_c(V)}
\end{equation}
\begin{equation}
\label{eq:BC550 dust corrected}
f([550]) = f([550]_{obs}) \times 10^{0.4 \times 1.00 A_c(V)}
\end{equation}
\end{subequations}
A complete derivation of the dust corrections presented here are shown in Appendix \ref{sec:dust_derivation}.

Figure \ref{fig:EW_with_dust_corrections} shows the distribution of our sample before and after dust corrections are applied.  
In the left panels we show our sample before any dust corrections are applied, with arrows in cyan denoting dust vectors for varying $f$ values. 
It is evident from the figure that the galaxies in this parameter space are very dependent on the $f$ value used. For $f$ values of 1 and 2, the effect of dust is orthogonal to IMF change, while values above 2 may influence the interpretation of the IMF. 
We note that $f>2$ makes the problem of high \Halpha\ EW objects worse, so we do not consider such values further. 

Figure \ref{fig:EW_with_dust_corrections} right panels show the dust corrections applied to both \Halpha\ EW and the \boxfil\ colours for the \sample. Without the effect of dust, we expect the young star forming galaxies to show similar bluer colours and therefore, the narrower \boxfil\ colour space occupied by our dust corrected sample is expected. 
With a dust correction of $f=1$, majority of the galaxies lie below the $\Gamma=-1.35$ IMF track with only $\sim1/5$th of galaxies showing higher \Halpha\ EWs. 
However, with $f\sim2$ dust correction, there is a significant presence of galaxies with extremely high \Halpha\ EW values for a given \boxfil\ colour inferred from a $\Gamma=-1.35$ IMF and \around60\% of the galaxies lie above this IMF track. 

Even $\sim\times2$ larger errors for the individual \Halpha\ EW measurements cannot account for the galaxies with the largest deviations from the Salpeter tracks. 
The change of $f$ from $2\Rightarrow1$ decreases the median \Halpha\ EW value by $\sim0.2$dex. However, galaxies still show a large scatter in \Halpha\ EW vs \boxfil\ colour parameter space with points lying well above the Salpeter IMF track.


\begin{figure*}
\includegraphics[scale=0.9]{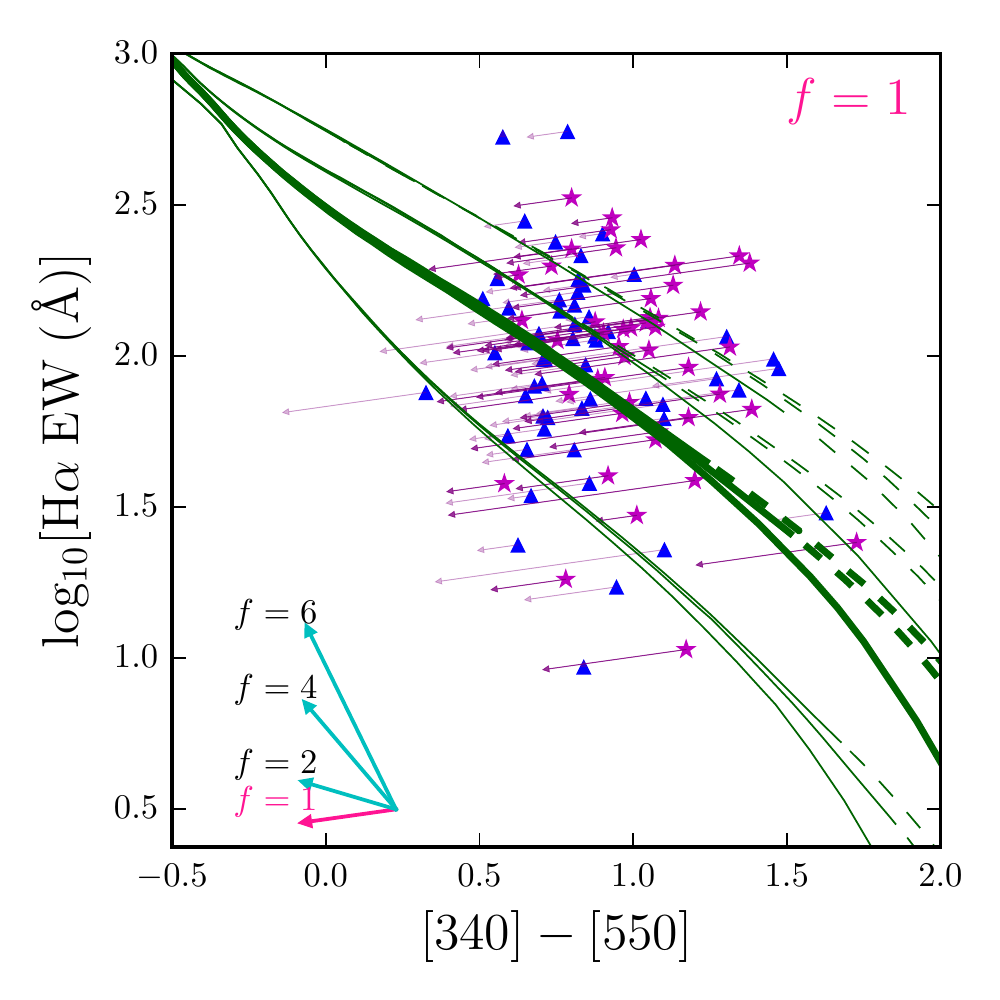}
\includegraphics[scale=0.9]{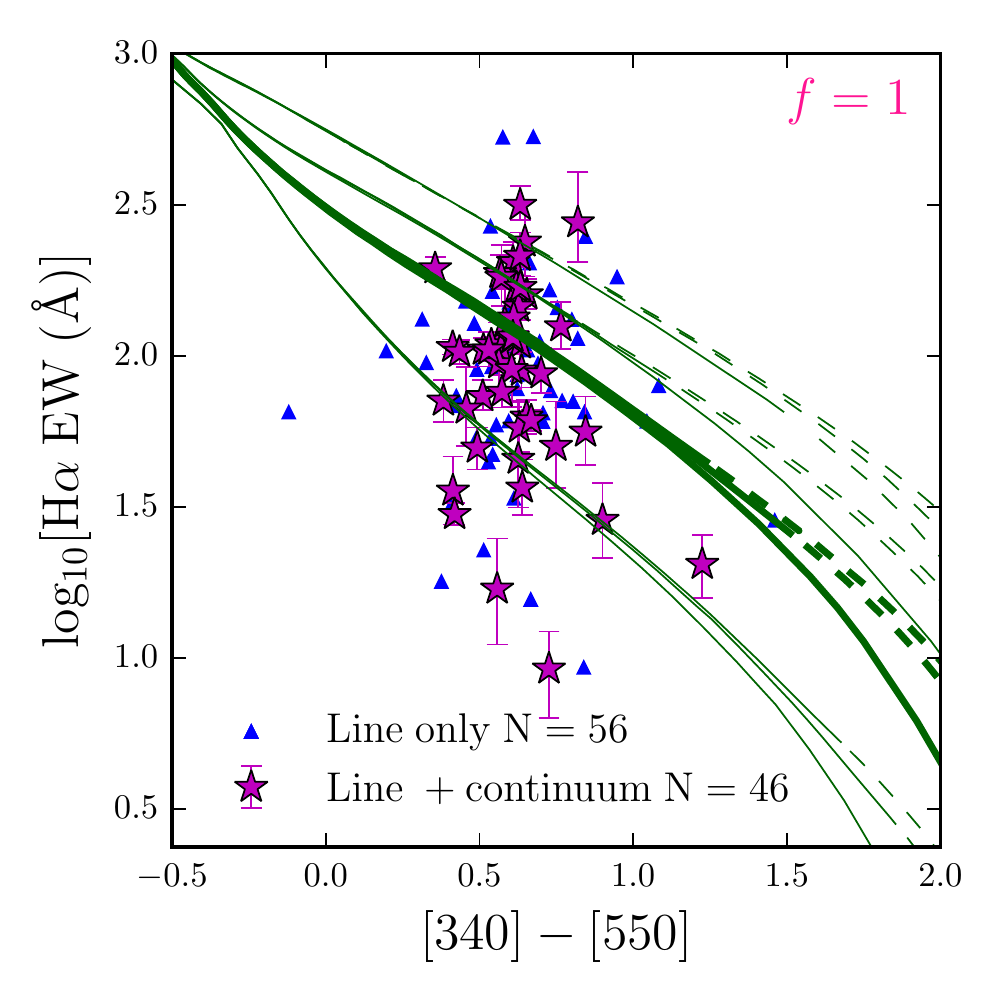}
\includegraphics[scale=0.9]{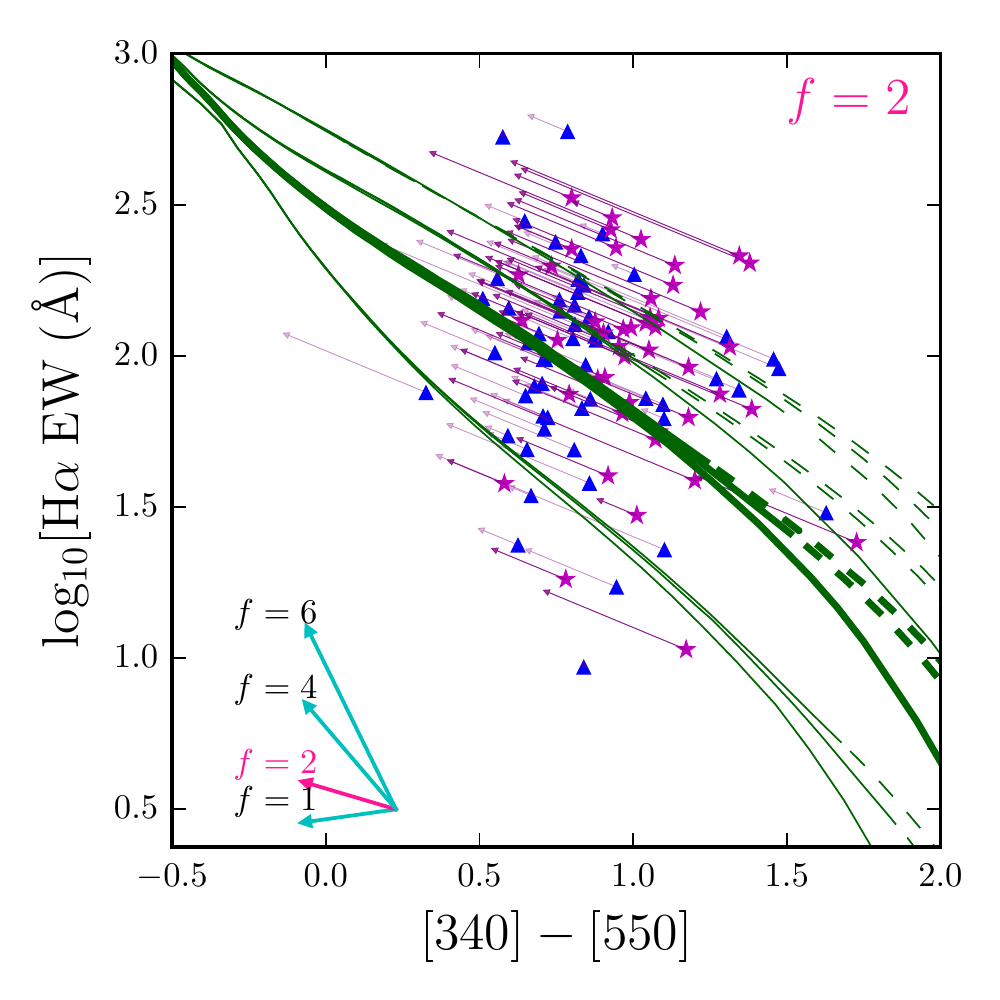}
\includegraphics[scale=0.9]{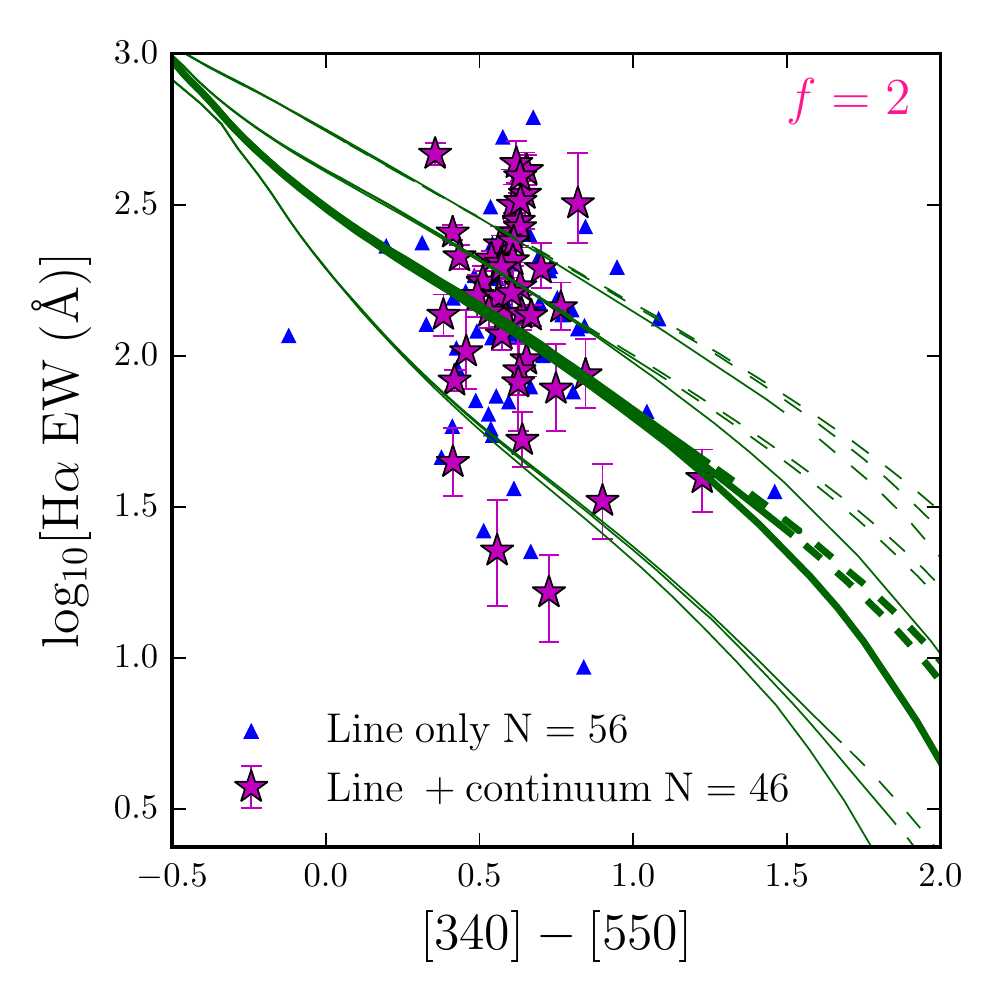}
\caption{The dust correction process of the \sample. This figure is similar to Figure \ref{fig:EW_no_dust_corrections} but shows the intermediate and final step of the dust correction process. 
{\bf Top Left:} Here we show the dust correction vector for each galaxy in our sample, computed following the prescriptions explained in Section \ref{sec:dust_corrections}. In summary we use \cite{Calzetti2000} attenuation law to correct the continuum levels and the optical \boxfil\ colours. We use \cite{Cardelli1989} attenuation law to dust correct the nebular emission lines. We use attenuation values calculated by FAST and apply equal amount of extinction to continuum and nebular emission line regions. The purple arrows denote the dust vector for the individual galaxies. Galaxies with no arrows have 0 extinction. 
The arrows in the bottom left corner show the dust vector for a galaxy with Av=0.5 but with varying \cite{Calzetti1994} factors, which is shown as \emph{f} next to each arrow. 
{\bf Top Right:} The final \Halpha\ EW vs \boxfil\ colour distribution of the dust corrected \sample\ with $f=1$. Most galaxies lie at $([340]-[550])\sim0.6$, which corresponds to $\sim850$ Myr of age following the Salpeter IMF tracks. 
{\bf Bottom Left:} Similar to top left panel, but with a higher amount of extinction to nebular emission line regions compared ($\sim\times2$) to the continuum levels.
{\bf Bottom Right:} The final \Halpha\ EW vs \boxfil\ colour distribution of the dust corrected \sample\ with $f=2.27$. 
}
\label{fig:EW_with_dust_corrections}
\end{figure*}

The form of the attenuation law of galaxies at $z>2$ show conflicting results between studies.  Observations from the Atacama Large Millimeter Array have indicated the presence of galaxies with low infra-red (IR) luminosities suggesting galaxies with attenuation similar to the Small Magellanic Cloud \citep[SMC.,][]{Capak2015,Bouwens2016b}. 
\citet{Reddy2015} showed an SMC like attenuation curve for $z\sim2$ galaxies at $\lambda\gtrsim2500$\AA\ and a \citet{Calzetti2000} like attenuation curve for the shorter wavelengths.
However, \emph{HST} grism and SED fitting analysis of galaxies at $z\sim2-6$ has shown no deviation in the attenuation law derived by \citet{Calzetti2000} for local star-forming galaxies. 
Such conflicts are also apparent in simulation studies, where \citet{Mancini2016} showed evidence for SMC like attenuation with clumpy dust regions while \citet{Cullen2017} has shown that galaxies contain similar dust properties as inferred by \citet{Calzetti2000}. 

\begin{figure}
\centering
\includegraphics[trim=0 0 0 0 , clip, scale=0.7]{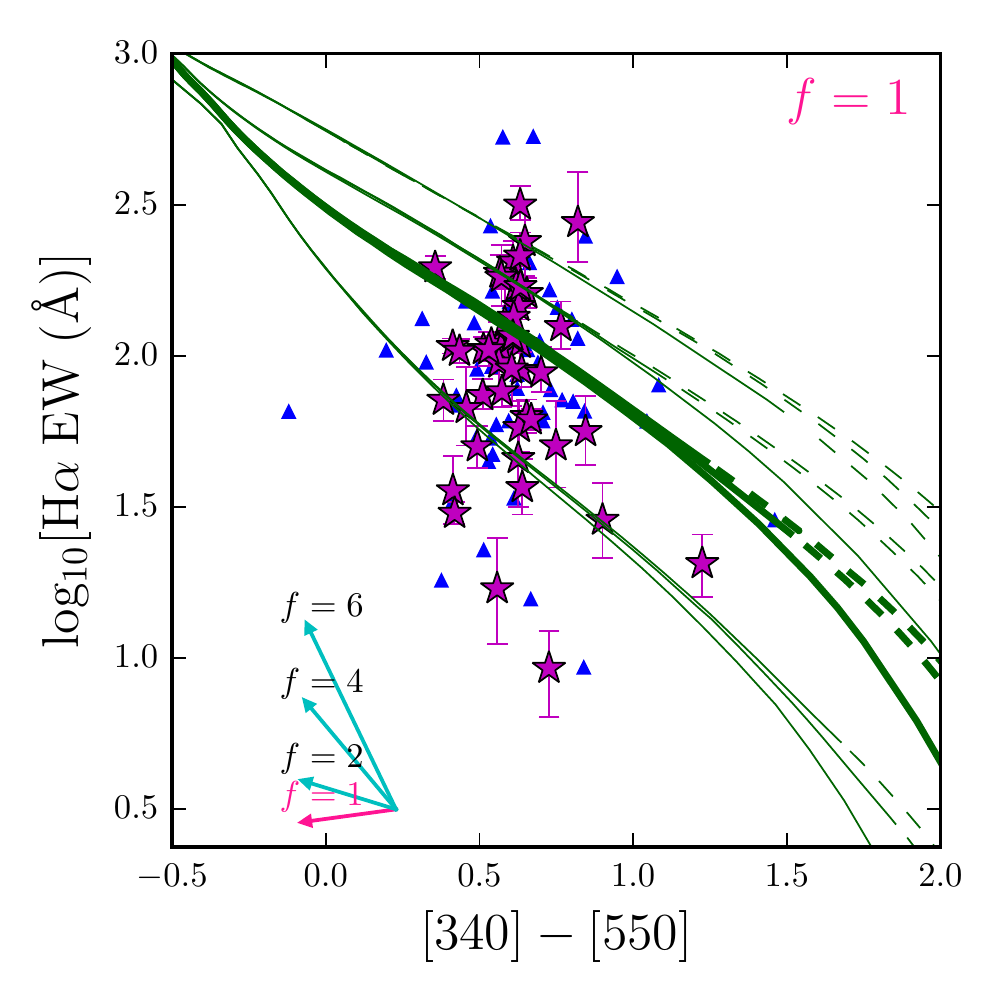}
\includegraphics[trim=0 0 0 0 , clip, scale=0.7]{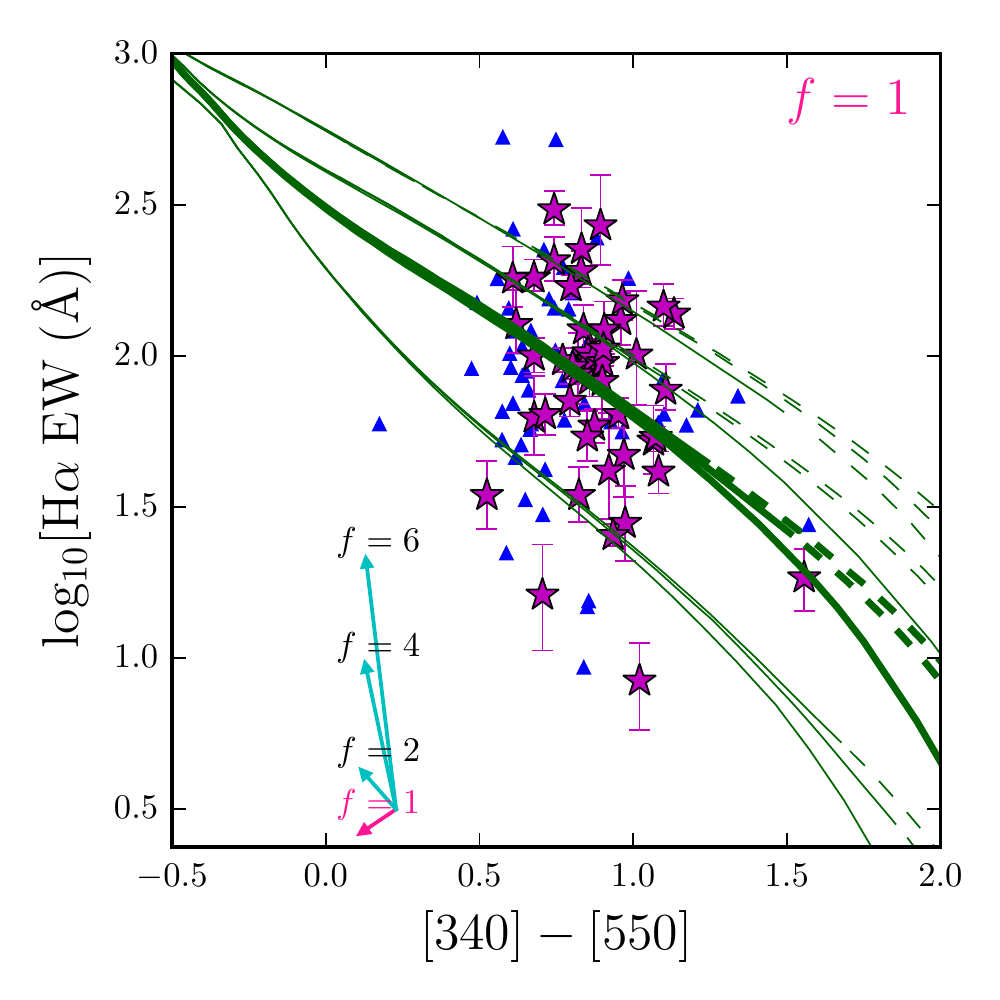}
\includegraphics[trim=0 0 0 0 , clip, scale=0.7]{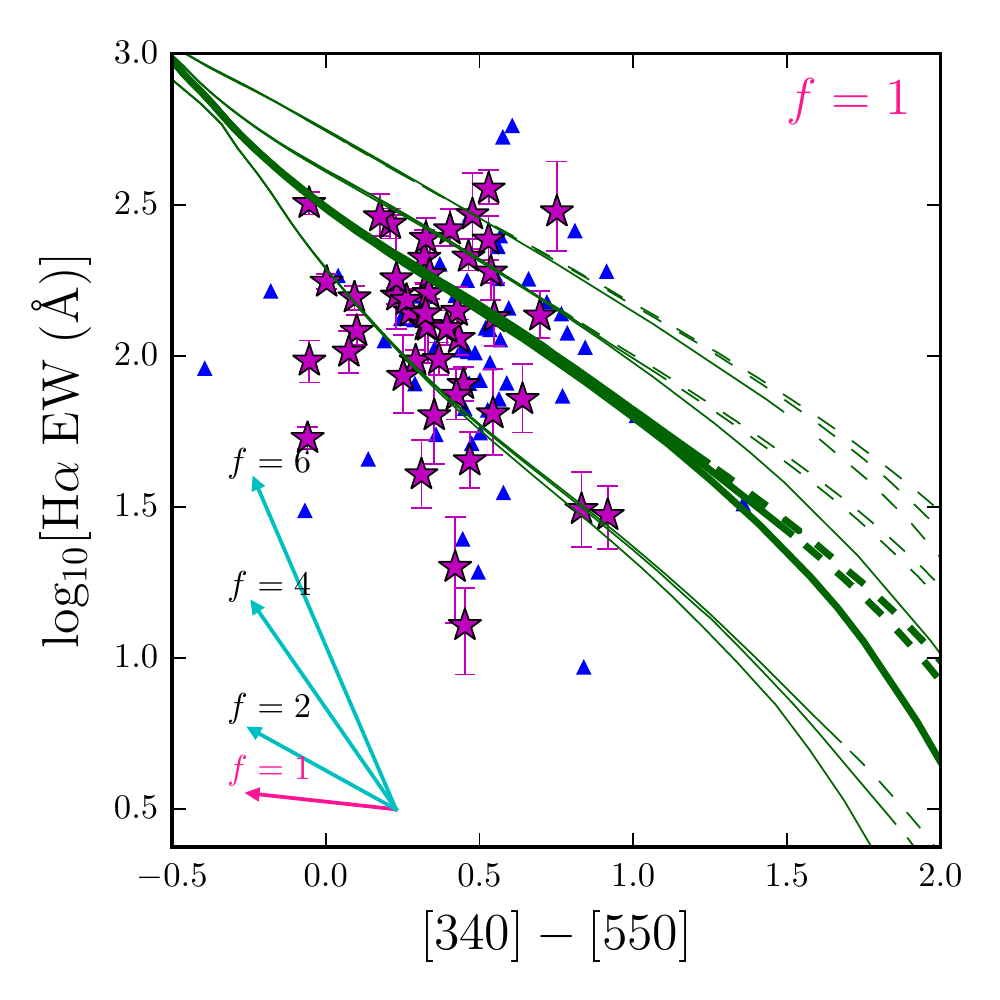}
\caption{Here we show the distribution of the \sample\ in the \Halpha\ EW vs \boxfil\ colour parameter space with the \Halpha\ continuum and optical colours dust corrections applied following 
{\bf Top:} \cite{Calzetti2000} attenuation law, 
{\bf Centre:} \cite{Pei1992} SMC attenuation law, and 
{\bf Bottom:} \cite{Reddy2015} attenuation law. 
In all panels \cite{Cardelli1989} attenuation law has been used to dust correct the nebular emission lines with equal amount of extinction applied to continuum and nebular emission line regions $(f=1)$.
The arrows in the bottom left corner show the dust vector for a galaxy with Av=0.5 but with varying \cite{Calzetti1994} factors, which is shown as \emph{f} next to each arrow. 
}
\label{fig:EW_with_dust_corrections_with_various_dust_laws}
\end{figure}

In order to understand the role of  dust laws in the \Halpha\ EW vs \boxfil\ colour parameter space we compare the results using other dust laws such as \citet{Pei1992} SMC dust law and \citet{Reddy2015} $z\sim2$ dust law to correct the stellar contributions (\Halpha\ continuum and optical colours).
A comparison between the distribution of galaxies obtained with different dust laws for a given $f$ is shown by Figure \ref{fig:EW_with_dust_corrections_with_various_dust_laws}. 
The fraction of galaxies with $\Delta$EW $>2\sigma$ from the $\Gamma=-1.35$ IMF track with $f=1 (f=2)$ dust corrections are $\sim20\% (\sim45\%),\sim35\% (\sim75\%),$ and $\sim15\% (\sim55\%)$ for \citet{Calzetti2000}, \citet{Pei1992} SMC, and \citet{Reddy2015} dust laws, respectively.  
However, we refrain from interpreting the differences in the distributions of the sample between the considered dust laws because the attenuation values used in the ZFIRE/ZFOURGE surveys have been derived from SED fitting by FAST using a \citet{Calzetti2000} dust law. 
Compared to the adopted dust law, the change in the value of $f$ has a stronger influence on the galaxies in our parameter space and can significantly affect the EW values, which is discussed further in Section \ref{sec:calzetti_factor}.

To investigate differences between our $z\sim2$ sample with HG08 $z\sim0$ sample, we derive dust corrections to the \gr\ colours. 
Using the following equations to apply dust corrections to g$_{0.1}$ and r$_{0.1}$ fluxes we recalculate the \gr\ colours for the \sample.   
\begin{subequations}
\begin{equation}
\label{eq:g dust corrected}
f(g_i)_{0.1} = f(g_{obs})_{0.1} \times 10^{0.4 \times 1.25 A_c(V)}
\end{equation}
\begin{equation}
\label{eq:r dust corrected}
f(r_i)_{0.1} = f(r_{obs})_{0.1} \times 10^{0.4 \times 0.96 A_c(V)}
\end{equation}
\end{subequations}

\begin{figure}
\includegraphics[scale=0.9]{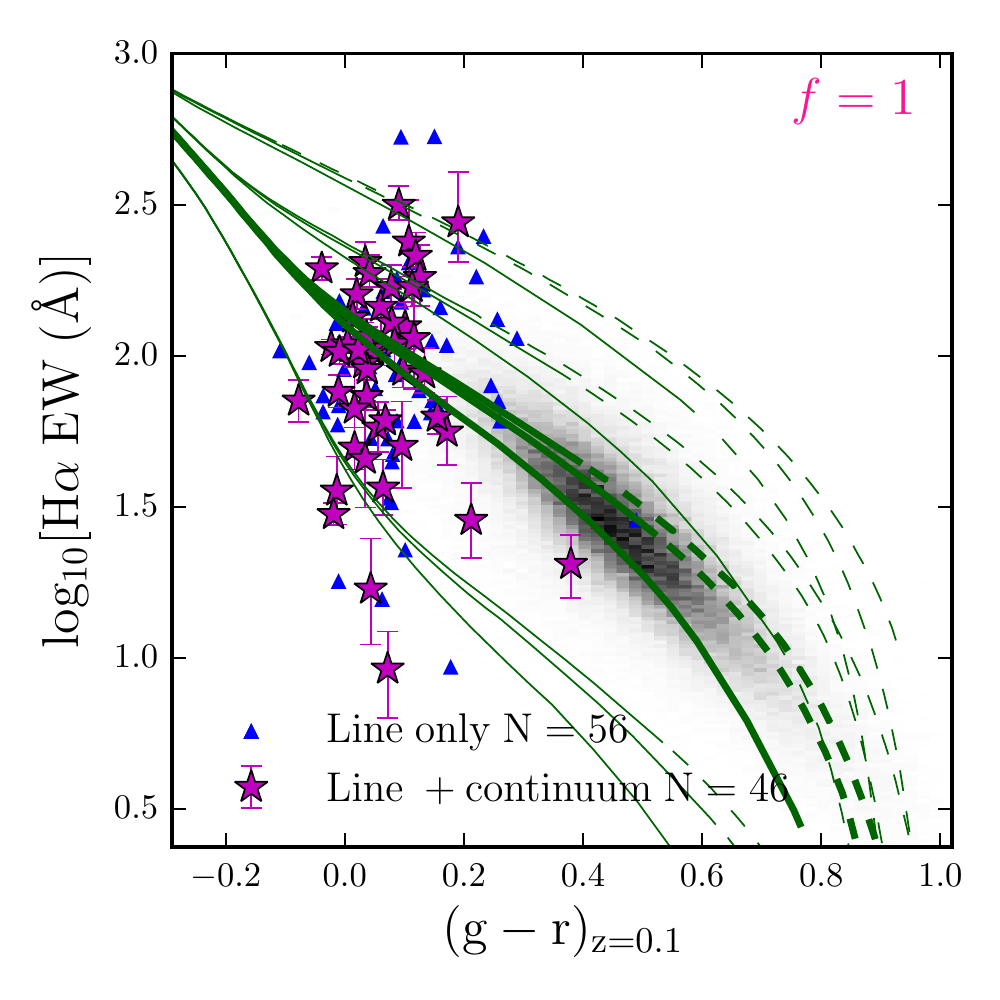}
\includegraphics[scale=0.9]{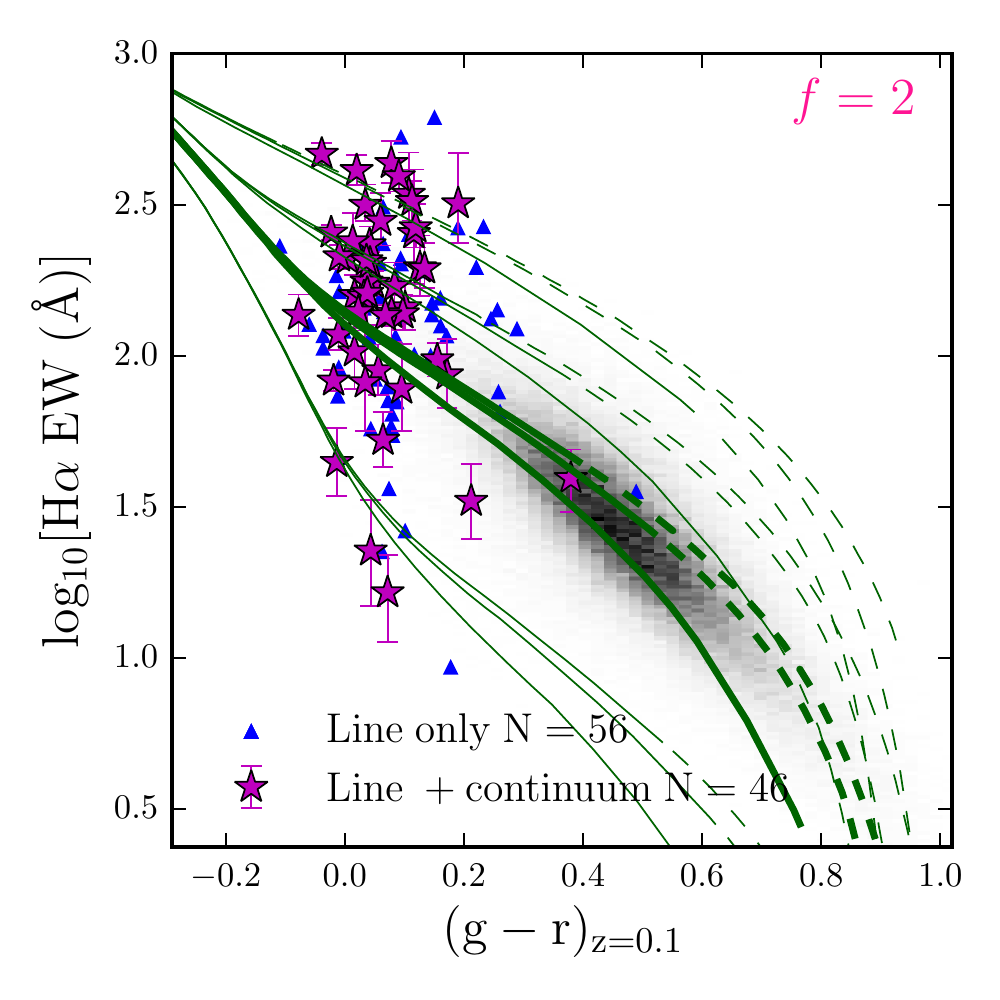}
\caption{Comparison of the \Halpha\ EW and \gr\ colours of the $z\sim2$ \sample\ with the HG08 $z\sim0$ sample. The HG08 sample is shown by the 2D grey histogram. The PEGASE models shown correspond to varying IMFs: from top to bottom $\Gamma=-0.5,-1.0,-1.35,\ \mathrm{and}\ -2.0$. Similar to Figure \ref{fig:EW_no_dust_corrections}, for each IMF we show three models with exponentially declining SFHs with varying p$_1$ values (from top to bottom p$_1$=1500 Myr, 1000 Myr, and 500 Myr). Model tracks at t$>3200$ Myr are shown by the dashed lines. 
{\bf Top:}  \sample\ with dust corrections applied with a $f=1$.
{\bf Bottom:} \sample\ with dust corrections applied with a $f=2$.
Note that HG08 uses a $f=2$ in their dust corrections. 
}
\label{fig:EW_HG08_comp}
\end{figure}

We show the \Halpha\ EW vs \gr\ colour comparison between ZFIRE and SDSS samples in Figure \ref{fig:EW_HG08_comp}. The dust corrections for the \sample\ has been performed using a $f=1$ and $f=2$. 
Similar to the \boxfil\ colour relationship, there is a significant presence of galaxies with extremely high \Halpha\ EW values and \around60\%of the galaxies lie above the Salpeter IMF track when dust corrections are applied with a $f=2$.
Furthermore, the $z\sim2$ sample shows much bluer colours compared to HG08 sample, which we attribute to the younger ages ($\sim850$ Myr inferred from tracks with a Salpeter IMF) and the higher SFRs of galaxies at $z\sim2$.

\begin{figure}
\includegraphics[trim=10 10 10 0, clip, scale=0.9]{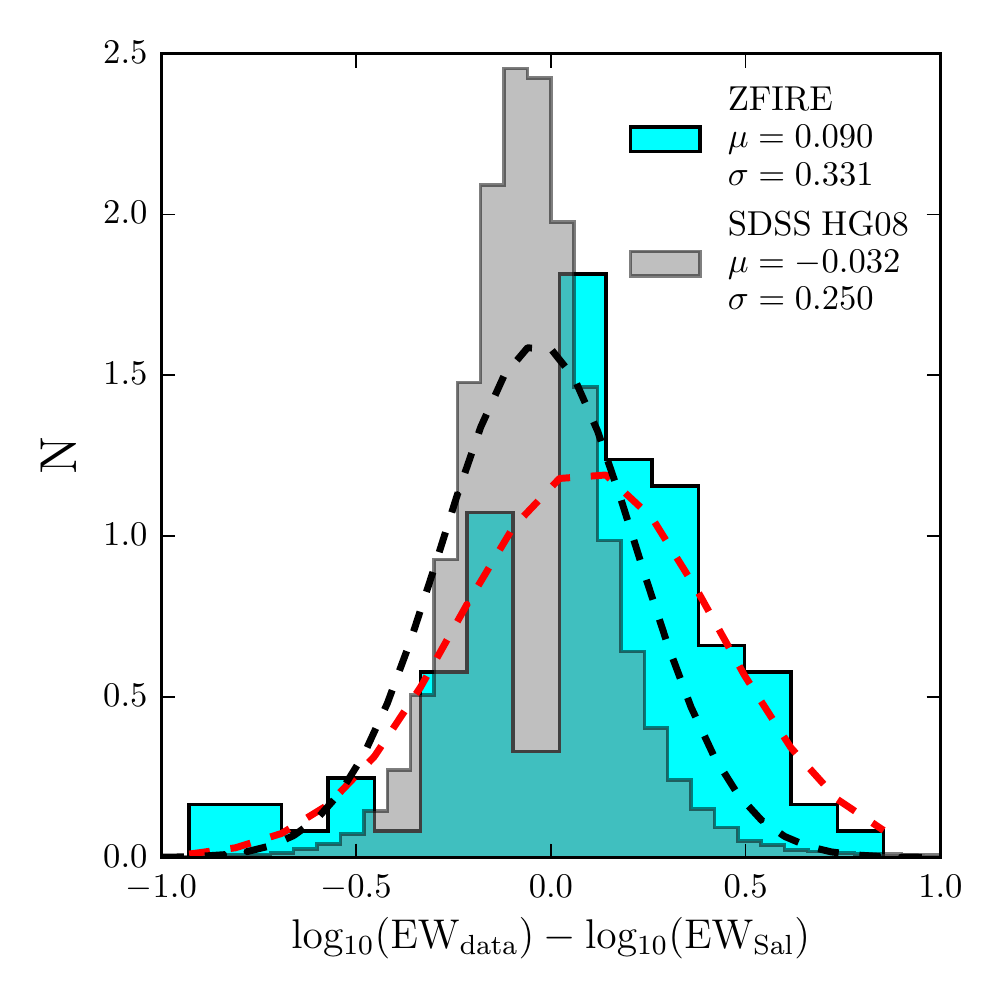}
\caption{  Deviations of the observed \Halpha\ from the canonical Salpeter like IMF tracks in the \Halpha\ EW vs \gr\ colour space. We show the $z\sim0$ SDSS HG08 sample (grey--black) and the $z\sim2$ \sample (cyan--red). Both histograms are normalized to an integral sum of 1 and best-fitting Gaussian functions are overlaid. The parameters of the Gaussian functions are shown in the legend.
}
\label{fig:EW_deviations_from_salp}
\end{figure}

In Figure \ref{fig:EW_deviations_from_salp} we use the $\Gamma=-1.35$ IMF tracks to compute the deviation of observed \Halpha\ EW values from a canonical Salpeter like IMF. 
For each \gr\ galaxy colour we calculate the expected \Halpha\ EW using the standard PEGASE model computed using an exponential decaying SFH with a p$_1=1000$ Myr.  
We then calculate the deviation between the observed values to the expected values. Only the $f=2$ scenario is considered here to be consistent with the dust corrections applied by HG08.
Our results suggest that the ZFIRE sample exhibits a log-normal distribution with a mean and a standard deviation of 0.090 and 0.321 units, respectively.  Similarly for the HG08 sample, the values are distributed with a mean and a standard deviation of -0.032 and 0.250 units. 
Compared to HG08, the \sample\ shows a larger scatter and favours higher \Halpha\ EW values for a given Salpeter like IMF. 
A simple two sample K-S test for the \sample\ and HG08 gives a Ks statistic of 0.37 and a P value of $1.32\times10^{-12}$, which suggests that the two samples are distinctively different from each other. 
In subsequent sections, we further explore whether the differences between the $z\sim0$ and $z\sim2$ populations are driven by IMF change or other stellar population parameters.

\subsection{IMF dependence of extinction values}
\label{sec:further_dust}

Dust corrections applied to the \sample, as explained in Section \ref{sec:dust_corrections}, are derived from FAST \citep{Kriek2009} using best-fitting model SEDs to ZFOURGE photometric data. 
FAST uses a grid of SED template models to fit galaxy photometric data to derive the best fit redshift, metallicity, SFR, age, and Av  values for the galaxies via a $\chi^2$  fitting technique. Even though these derived properties may show degeneracy with each other (see \citet{Conroy2013} for a review), in general FAST successfully describes observed galaxy properties of deep photometric redshift surveys \citep{Whitaker2011,Skelton2014,Straatman2016}. FAST has a limited variety of stellar templates, and therefore, we cannot explore the effect of varying IMFs on the FAST derived extinction values.

In order to examine the role of IMF on derived extinction values, we compare the distribution of ZFIRE rest frame UV and optical colours with PEGASE model galaxies. 
Following the same procedure used to derive the [340] and [550] filters, we design two boxcar filters centred at 1500\AA\ ([150]) and 2600\AA\ ([260]) with a length of 675\AA. The wavelength regime covered by these two filters approximately correspond to the B and I filters in the observed frame for galaxies at $z\sim2$ (further information is provided in Appendix \ref{sec:filter choice 150}). Therefore, K corrections are small and the computed values are robust.

\begin{figure*}
\includegraphics[trim=10 0 0 0 , clip, scale=0.91]{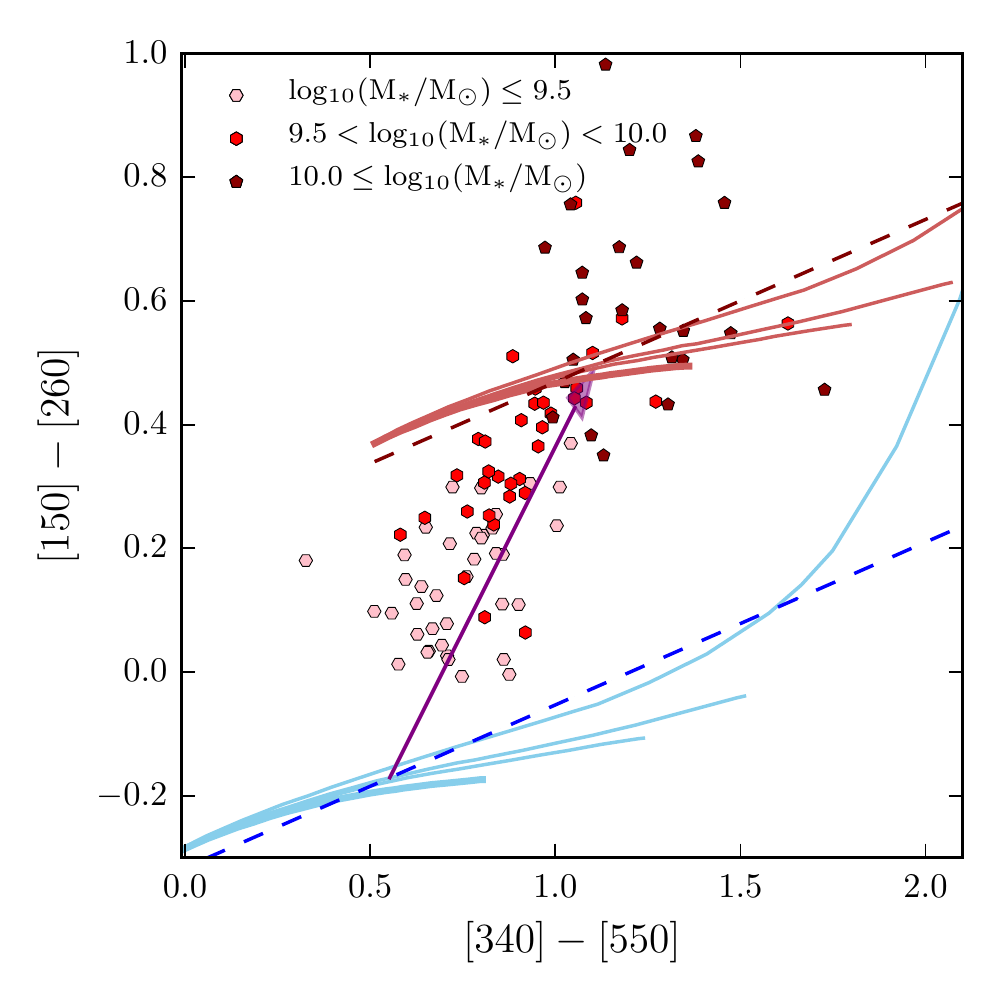}
\includegraphics[trim=10 0 0 0 , clip, scale=0.90]{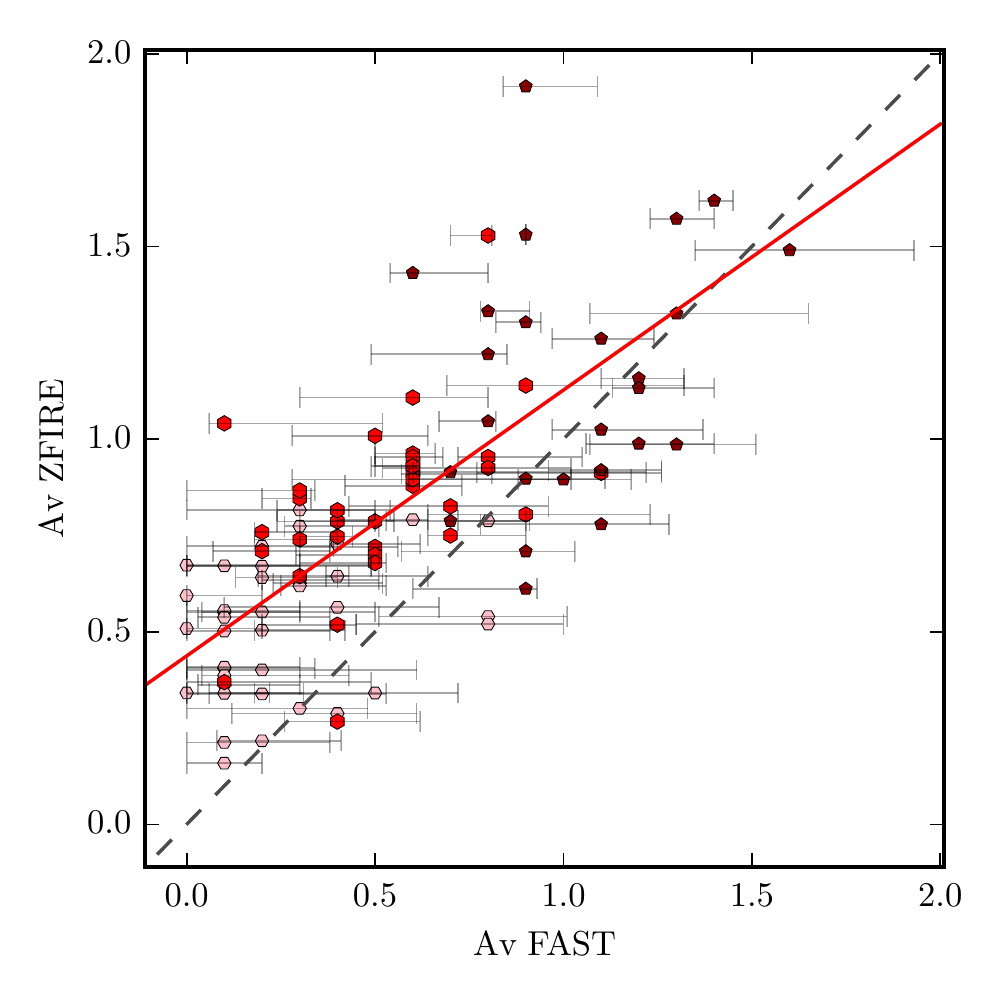}
\caption{ Dust parametrisations to investigate IMF dependencies of dust extinction values. Galaxies are divided to 3 mass bins. 
{\bf Left:} The dust content of the galaxies are parametrized using UV/optical colours. 
The cyan solid lines are PEGASE models for $\Gamma=-1.35$ IMF with varying SFHs. The dashed blue line is the best-fitting line for these models. 
The pink lines are similar to the cyan lines, but with an extinction of 1 mag. 
The magenta dashed line is the best-fitting line for these models with \Av=1. 
The purple arrow denotes the direction of the dust vector and connects the two best-fit lines at time $t$.
{\bf Right:} Comparison between the extinction derived by Equation \ref{eq:Av_ZFIRE} with the extinction values derived by FAST. The error bars are the upper and lower 68th percentile of the \Av\ values compiled by FAST and the diagonal red line is the error weighted least squares fit to the data.
The diagonal dashed line is the \AvSED=\AvZFIRE\ line.
}
\label{fig:Av_ZFIRE}
\end{figure*}

By binning galaxies in stellar mass, we find massive galaxies to be dustier than their less less massive counterparts.
We show the distribution of our sample in the rest frame UV vs rest frame optical parameter space in Figure \ref{fig:Av_ZFIRE} (left panel).   PEGASE model galaxies with $\Gamma=-1.35$ and varying SFHs are shown by the solid model tracks.
When we apply a \Av=1 extinction, the models show a strong diagonal shift due to reddening of the colours in both axis.  For each set of tracks, we perform a best-fitting line to the varying SFH models.
The dust vector (shown by the arrow) joins the two best fit lines drawn to the models with \Av=0 and \Av=1 at time $t$. 
We define \AvZFIRE\ to be the correction needed for each individual galaxy to be brought down parallel to the dust vector to the best fit  line with \Av=0,  and is parametrized by the following equation:

\begin{equation}
\label{eq:Av_ZFIRE}
\begin{split}
A_{v}(\mathrm{ZF}) = -0.503\times ([340]-[550])\\
+ 1.914\times ([150]-[260])+ 0.607
\end{split}
\end{equation}

Our simple method of dust parametrization is similar to the technique used by FAST, which fits SED templates to the UV continuum to derive the extinction values.  
The \dustfil\ colour probes the UV continuum slope, which is ultra sensitive to dust, while the 
\boxfil\ probes the optical continuum slope which is less sensitive to dust. 
\Halpha\ emission does not fall within these filters, and hence, is not strongly sensitive to the SFR of the galaxies.

In the right panel of Figure \ref{fig:Av_ZFIRE}, we compare the derived extinction values from our method (\AvZFIRE) with the extinction values derived by FAST (\AvSED). Since \citet{Chabrier2003} IMF at $\mathrm{M_*>1M_\odot}$ is similar to the slope of Salpeter IMF ($\Gamma=-1.35$), the comparison is largely independent of the IMF. 
The median and \NMAD\ scatter of the Av values derived via FAST and our method is $\sim-0.3$ and $\sim-0.3$ respectively. Therefore, the values agree within $1\sigma$.
There is a systematic bias for \AvZFIRE\ to overestimate the extinction at lower \AvSED\ values and underestimate at higher \AvSED\ values. 
We attribute this residual pattern to age metallicity degeneracy, which is not considered in the derivation of \AvZFIRE.

The choice of IMF will affect dust corrections derived from UV photometry (using FAST or our empirical method) as there is a modest dependence of the rest frame UV continuum slope on IMF for star-forming populations. We are primarily interested in IMF slopes shallower than Salpeter slope ($\Gamma> -1.35$) to explain our population of high \Halpha\ EW galaxies. 
For $\Gamma=-0.5$ we find the best fit model line in the left panel of Figure \ref{fig:Av_ZFIRE} shifts down by \around0.1 mag. 
This increases the magnitude of dust corrections 
and extends the arrows in Figure \ref{fig:EW_with_dust_corrections} to bluer colours and higher EWs and does not explain the presence of high \Halpha\ EW objects. For the purpose of comparing with our default hypothesis (Universal IMF with $\Gamma=-1.35$) we adopt the FAST derived dust corrections.

\subsection{Balmer decrements}
\label{sec:balmer_decrements}

Stellar attenuation values computed by fitting a slope to galaxy SEDs in UV, estimates the extinction of old stellar populations that primarily contributes to the galaxy continuum. 
Nebular emission lines originates from hot ionized gas around young and short-lived O and B stars. Given their short lifetime ($\sim10-20$ Myr), O and B stars are not expected to move far from their birthplace (dusty clouds), thus,  the nebular emission-lines are expected to have high levels of extinction.
Next, we investigate the dust properties of the stars in different star-forming environments using the luminosity ratios of nebular hydrogen lines and observed UV colours.

Luminosity ratios of nebular hydrogen lines are insensitive to the underlying stellar population and IMF parameters for a fixed electron temperature \citep{Osterbrock1989}. 
These line ratios are governed by quantum mechanics, and therefore, can be used to probe the reddening of nebular emission lines and dust geometry under the assumption that ionized gas attenuation resembles that of the underlying stellar population.

With the recent development of sensitive NIR imagers and multi-object spectrographs, studies have now started to investigate the properties of dust at $z\sim2$ \citep{Shivaei2015, Reddy2015, deBarros2015}. These studies show conflicting results on the fraction of stellar to nebular attenuation of galaxies at $z\sim2$. 
Here we show  Balmer decrement results for a sub-sample of our \sample\ which shows SNR $>5$ detections for both \Halpha\ and \Hbeta. The data presented herein are a combination of data released by the ZFIRE data release \citet{Nanayakkara2016} and additional MOSFIRE observations carried out during January 2016. 
Our sample comprises of 42 galaxies with both \Halpha\ and \Hbeta\ emissions line detections with a SNR $>5$ and 35 galaxies are part of the \sample. Further details on \Hbeta\ detection properties are explained in Appendix \ref{sec:Balmer decrement extended}.

\begin{figure*}
\includegraphics[trim = 10 0 10 0, clip,scale=0.95]{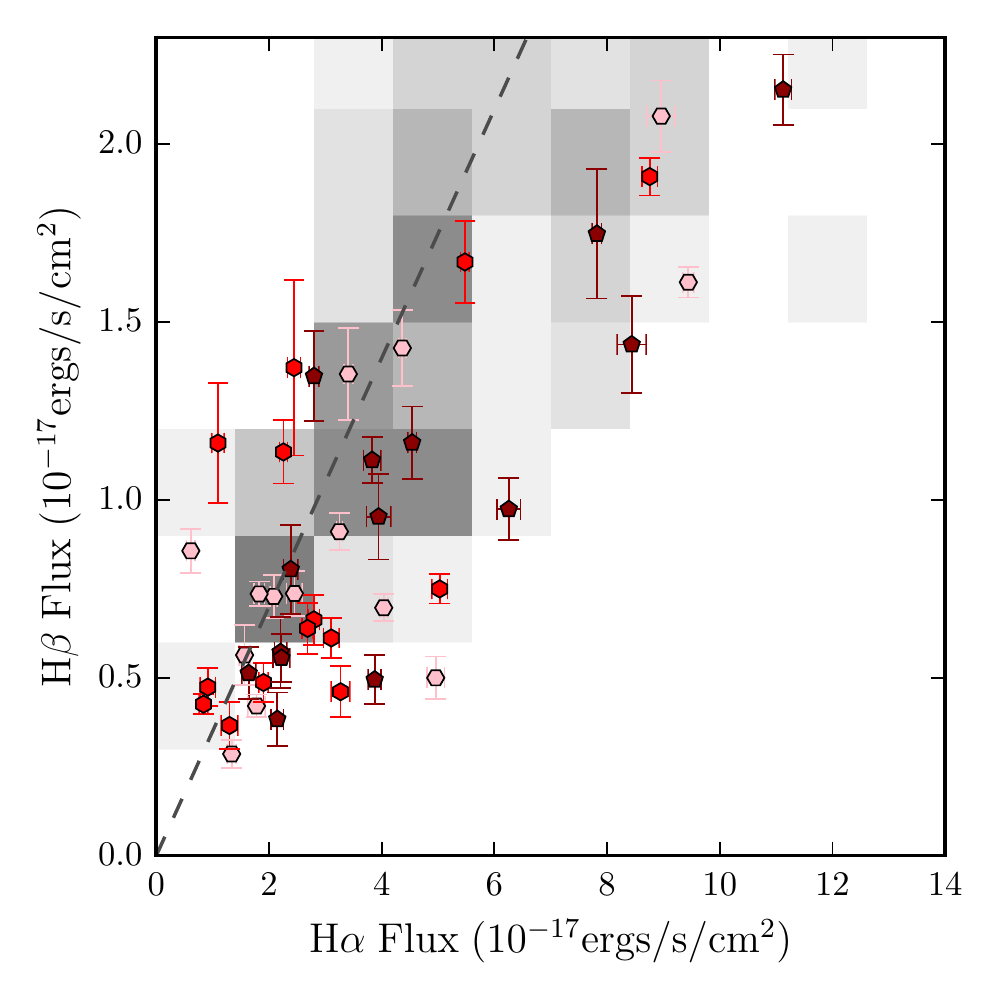}
\includegraphics[trim = 10 0 10 0, clip,scale=0.95]{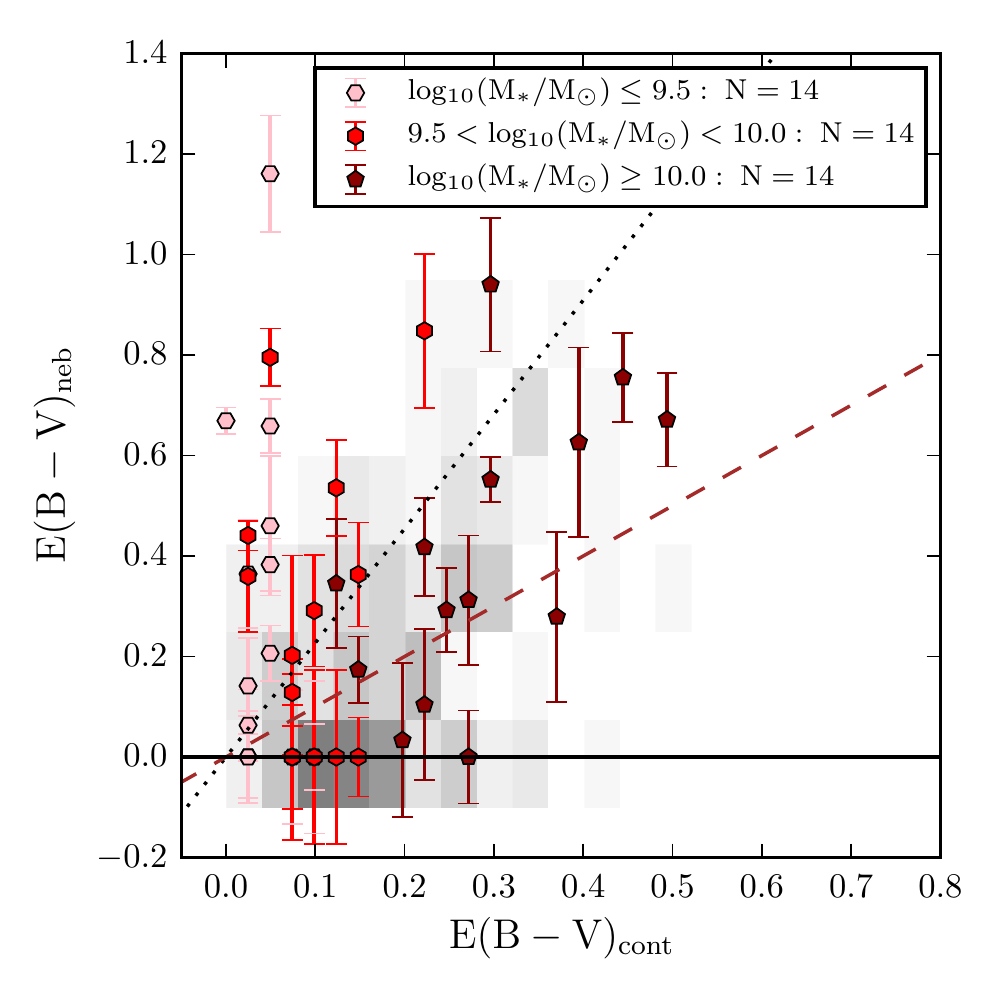}
\caption{Balmer decrement properties of the ZFIRE sample. Here we show the subset of \Halpha\ and \Hbeta\ detected galaxies in our \sample. The 2D density histogram shows the distribution of values from \citet{Reddy2015}
{\bf Left:} \Hbeta\ flux vs \Halpha\ flux measurements for the \sample.  
The individual galaxies are divided into three mass bins. 
The diagonal dashed line is the $f(H\alpha)=2.86\times f(H\beta)$ line which denotes the Balmer decrement for case B recombination.  
{\bf Right:} Comparison between SED derived extinction values with extinction computed from the Balmer decrement. The black dotted line is the $E(B-V)_{neb}=E(B-V)_{cont} /0.44$ relationship expected from \citet{Calzetti2000} to compute the extra extinction for nebular emission line regions. The brown dashed line is the $E(B-V)_{neb}=E(B-V)_{cont}$ line and the horizontal solid black line is the $E(B-V)_{neb}=0 $ line. Galaxies with $E(B-V)_{neb}$ $<0$ have been assigned a value of 0.
}
\label{fig:balmer_decrement}
\end{figure*}

We show the \Hbeta\ flux vs \Halpha\ flux for our total ZFIRE galaxies Figure \ref{fig:balmer_decrement} (left panel). 
The diagonal dashed line of the left panel shows the Balmer decrement for Case B recombination models with \Halpha/\Hbeta = 2.86 \citep{Osterbrock1989}.
Galaxies that fall below this criteria are expected to have realistic dust models. 
In Figure \ref{fig:balmer_decrement} (right panel) we show the comparison between extinction computed for stars by FAST with the extinction computed for ionized gas regions using the Balmer decrement. The colour excess is computed from the Balmer decrement using:
\begin{equation}
\label{eq:balmer_dec}
E(B-V)_{neb} = \frac{2.5}{1.163} \times \mathrm{log}_{10} \left \{ \frac{f(H\alpha)}{2.86 \times f(H\beta)} \right \}
\end{equation}
The distribution of our sample in these panels is similar to \citet{Reddy2015} results as shown by the 2D density histogram and therefore, highlights the complicated dust properties of $z\sim2$ star-forming galaxy populations.

\subsection{The difference in extinction between stellar and nebular regions}
\label{sec:calzetti_factor}

In this section, we investigate how the differences in dust properties between stellar and ionized gas regions can effect the distribution of our galaxies in \Halpha\ EW vs \boxfil\ colour space. 
\cite{Calzetti2000} showed that, for $z\sim0$ star-forming galaxies, the nebular lines are $\sim\times2$ more attenuated than the continuum regions, but at $z\sim2$ studies show conflicting results \citep{Shivaei2015, Reddy2015, deBarros2015}. 
Using ZFIRE data in Figure \ref{fig:balmer_decrement}, we show that galaxies occupy a large range of $f$ values in our \Hbeta\ detected sample. We attribute the scatter in extinction to the properties of sight-lines of the nebular line regions.



Since the universe is only  $\sim3$ Gyr old at $z\sim2$, the dense molecular clouds collapsed to form stars would only have had limited time to evolve into homogeneous structures within galaxies. This can give rise to differences in the dust geometries within ionizing clouds resulting in non-uniform dust sight-lines for galaxies at $z\sim2$. 
By varying the value of $f$ as a free parameter, we calculate the $f$ values required for our galaxies to be consistent with a universal IMF with slope $\Gamma=-1.35$. 
For each dust corrected \boxfil\ colour, we compute the \Halpha\ EW of the PEGASE $\Gamma=-1.35$ IMF track with p$_1=1000$ Myr. We then use the observed and required \Halpha\ EW values to compute the $f$ as follows:
\begin{equation}
\begin{split}
f =  \left \{ \frac{\mathrm{log}_{10}(\mathrm{H\alpha\ EW})_{\Gamma=-1.35} - \mathrm{log}_{10}(\mathrm{H\alpha\ EW})_{obs}}{0.44 \times 0.62 \times A_c(v)} 
+ \frac{0.82}{0.62}  \right \}  \\
\end{split}
\label{eq:varying_f}
\end{equation}
where log$_{10}(\mathrm{H\alpha\ EW})_{\Gamma=-1.35}$ is the \Halpha\ EW of the PEGASE model galaxy for dust corrected \boxfil\ colours of our sample and log$_{10}(\mathrm{H\alpha\ EW})_{obs}$ is the observed \Halpha\ EW. 

\begin{figure}
\includegraphics[trim=5 10 10 0, clip, scale=0.9]{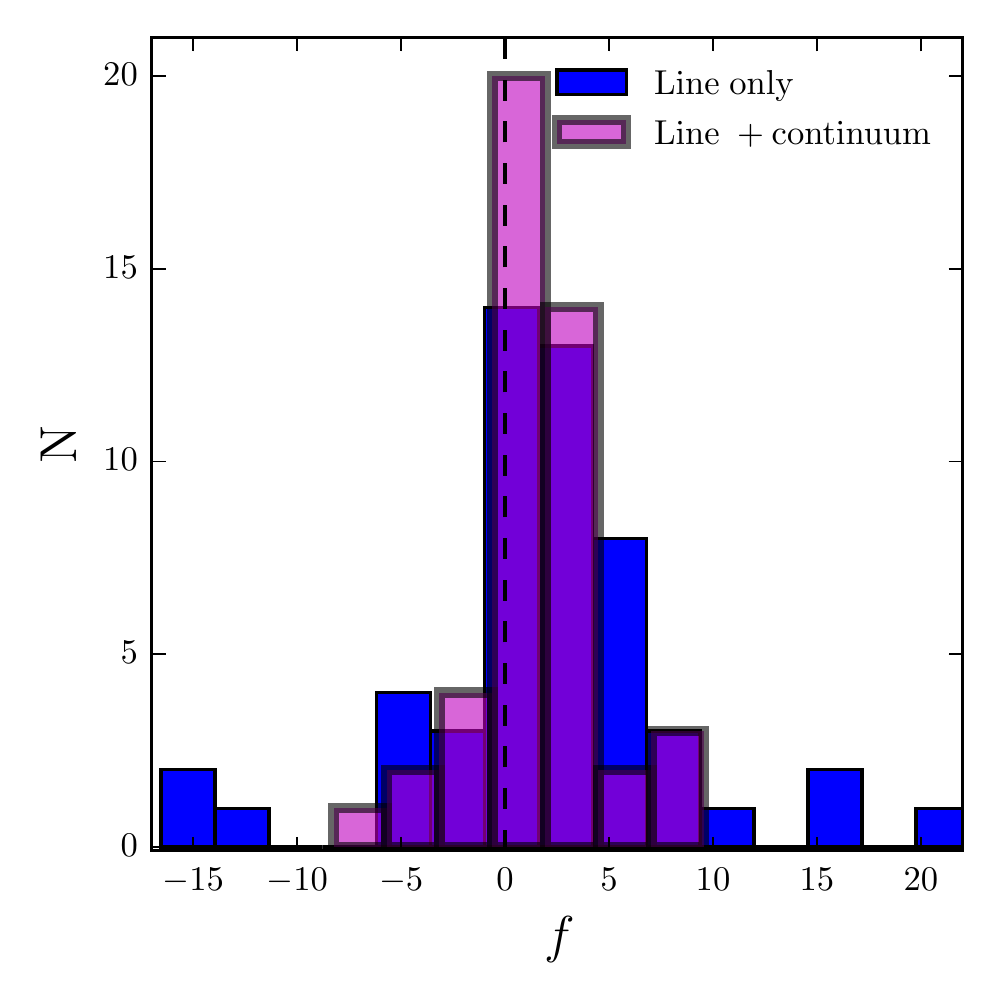}
\caption{ The distribution of $f$ values required for galaxies in the \sample\ to agree with a $\Gamma=-1.35$ Salpeter like IMF. For each dust corrected \boxfil\ colour of the \sample\ galaxies, we use the corresponding \Halpha\ EW of the $\Gamma=-1.35$ PEGASE track with an exponentially declining SFH with a p$_1=1000$ Myr to compute the $f$ value required for the observed \Halpha\ EW to agree with the $\Gamma=-1.35$ IMF. The vertical dashed line is the f=0 line. 
}
\label{fig:varying_f}
\end{figure}

In Figure \ref{fig:varying_f}, we show the $f$ values required for our galaxies to agree with a universal IMF with a slope of $\Gamma=-1.35$. 
For the 46 continuum detected galaxies, $\sim30\%$  show $f<1$. It is extremely unlikely that galaxies at $z\sim2$ would have $f<1$, which suggests that ionizing dust clouds where the nebular emission lines originate from are less dustier than regions with old stellar populations.
Furthermore, galaxies that lie above the Salpeter track (log$_{10}$(\Halpha\ EW) $>2.2$) requires $f<1$ and therefore, even a varying $f$ hypothesis cannot account for the high EW galaxies. 
$\sim17\%$ of continuum detections have $f<0$ which is not physically feasible since it requires dust to have the opposite effect to attenuation. 
Therefore, we reject the hypothesis that varying $f$ values could explain the high \Halpha\ EWs of our galaxies.

\subsection{Observational bias}
\label{sec:observational_bias}

The \sample\ spans a large range of \Halpha\ EWs, suggesting a considerable variation in the sSFRs of the ZFIRE galaxies at $z\sim2$. High \Halpha\ EW can result due to two reasons:
\begin{enumerate}
\item High line flux: suggests a higher SFR in time-scales of \around10 Myr.
\item Lower continuum level: suggests lower stellar mass for galaxies.
\end{enumerate}
These two scenarios should be considered together: i.e., a higher line flux with lower continuum level would suggest the galaxy to be going through an extreme star-formation phase. 
We investigate any detection bias that could explain the our distribution of  \Halpha\ EWs.

\begin{figure*}
\includegraphics[trim=10 0 0 0 , clip, scale=0.90]{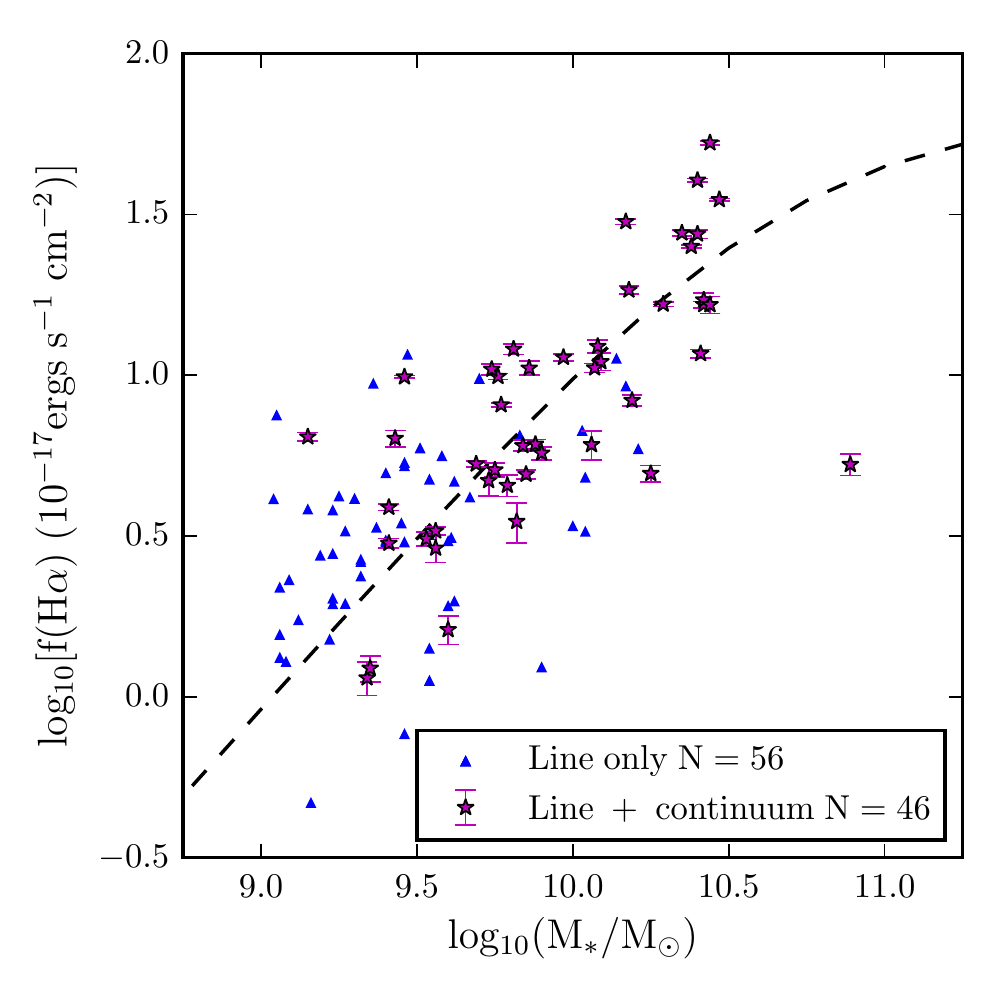}
\includegraphics[trim=10 0 0 0 , clip, scale=0.90]{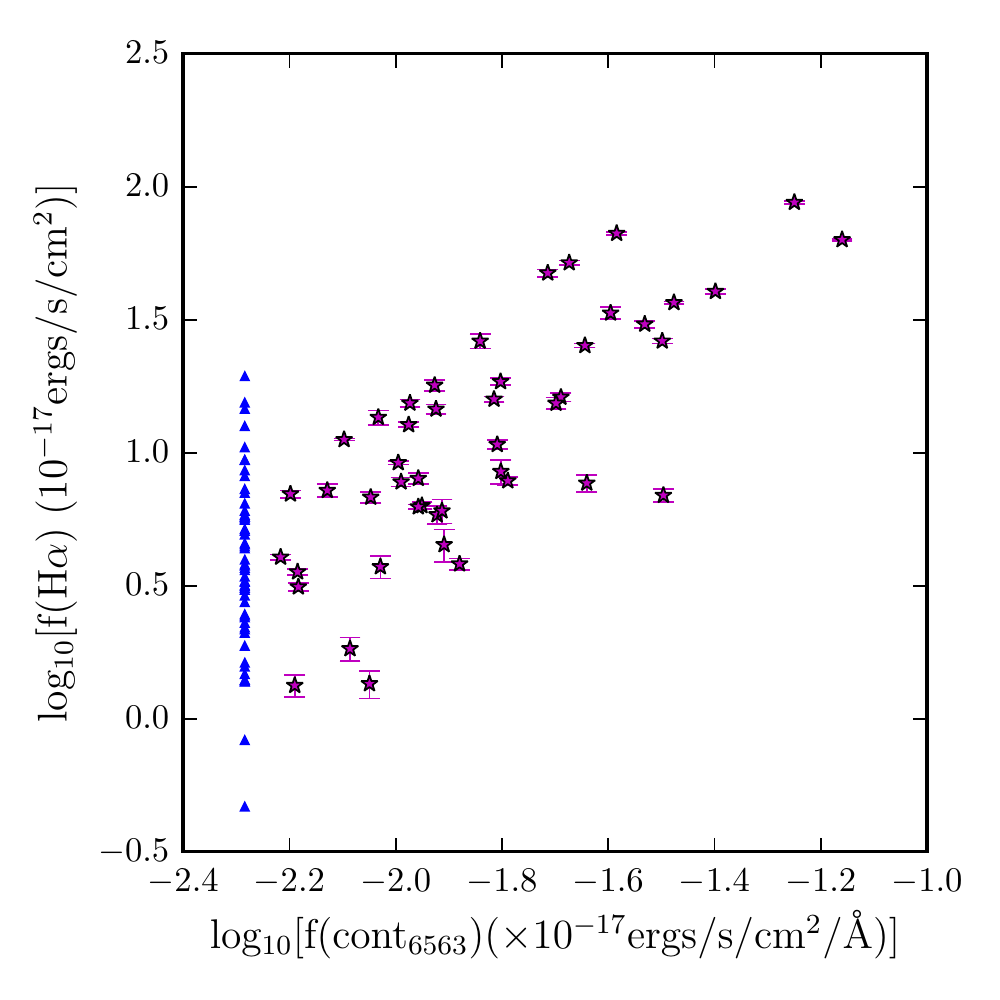}
\caption{ Here we investigate which observable parameter/s drive the high \Halpha\ EW values compared to $\Gamma=-1.35$ Salpeter like IMF expectations in \Halpha\ EW vs \boxfil\ colour space. 
{\bf Left:} The logarithmic \Halpha\ flux of the \sample\ as a function of stellar mass. The \Halpha\ flux has been dust corrected following Equation \ref{eq:Halpha dust corrected} with $f=1/0.44$. 
The black line is derived from the star-formation main sequence from \citet{Tomczak2014}, converted to \Halpha\ flux using \citet{Kennicutt1998} \Halpha\ star-formation law at $z=2.1$. 
{\bf Right:} The \Halpha\ flux vs the continuum level at 6563\AA. Both parameters (\Halpha\ flux as above, continuum level following Equation \ref{eq:cont dust corrected} ) have been corrected for dust extinction and are plot in logarithmic space. The \Halpha\ fluxes are $\sim2$ orders of magnitude brighter than the continuum levels. 
}
\label{fig:Ha_MS_and_cont}
\end{figure*}

In \citet{Nanayakkara2016}, we show that the ZFIRE COSMOS K band detections are mass complete to $\mathrm{\log_{10}(M_*/M_\odot)\sim9.3}$.
In Figure \ref{fig:Ha_MS_and_cont} we show the distribution of the \Halpha\ flux and continuum levels of our sample.  It is evident from Figure \ref{fig:Ha_MS_and_cont} (left panel) that our galaxies evenly sample the star-forming main-sequence described by \citet{Tomczak2014} without significant bias towards extreme \Halpha\ flux values.
Therefore, we conclude that the \Halpha\ fluxes we detect are typical of star-forming galaxies at $z=2.1$. 

In Figure \ref{fig:Ha_MS_and_cont} (right panel), we compare our \Halpha\ flux values with the derived continuum levels. 
Continuum detected galaxies show continuum levels that are in order of $\sim2$ mag smaller compared to the \Halpha\ fluxes, and therefore, the higher \Halpha\ EWs in our sample are primarily driven by the low continua. Note that our continuum detection level is $\sim-2.3$ log flux units. Therefore, for galaxies with only line detection, the difference between \Halpha\ flux and continuum level is much higher, which suggests much larger \Halpha\ EWs.

Several studies investigated the \Halpha\ EW of galaxies at higher redshifts ($z\geq1.5 $) \citep{Erb2006b,Shim2011,Fumagalli2012,Kashino2013,Stark2013,Masters2014,Sobral2014,Speagle2014,Marmol-Queralto2016,Rasappu2016} using SED fitting techniques and/or grism spectra. 
We find that our \Halpha\ EWs show good agreement with EWs expected at $z\sim2$ \citep{Marmol-Queralto2016} and conclude that our observed \Halpha\ EW values are typical of $z\sim2$ galaxies.

However, there are no studies that use high quality spectra to study the \Halpha\ EW at $z\sim2$. Even though our \Halpha\ fluxes and EWs are typical of $z\sim2$ galaxies, in the \Halpha\ EW vs \boxfil\ colour space, a large fraction of our galaxies show high EWs for a given \boxfil\ colour compared to the expectation by a Salpeter like IMF. 
Our high EWs are driven by lower continuum levels, for which we consider two possible explanations. 
\begin{enumerate}
\item Most galaxies have quenched their starburst phase in a time-scale of $\sim10$ Myr. Therefore, the old stellar populations are still being built up explaining the lack of continuum level from the older stars.
\item Stars are being formed continuously at $z\sim2$ with a higher fraction of high mass stars.
\end{enumerate}

In Section \ref{sec:star_bursts}, we investigate the effects of starbursts on our study to examine how probable it is for $\sim1/3$rd of our galaxies to have quenched their star-formation within a time-scale of  $\lesssim10$ Myr.

\section{Can Star bursts Explain the high \Halpha-EWs?}
\label{sec:star_bursts}

Galaxies at $z\sim2$ are at the peak of their star formation history \citep{Hopkins2006}. We expect these galaxies to be rapidly evolving with multiple stochastic star formation scenarios within their stellar populations. If our sample consists of a significant population of starburst galaxies, it may cause significant systemic biases to our IMF analysis.

In this section, we investigate the effects of bursts on the SFHs of the galaxies. We study how the distribution of galaxies in \Halpha\ EW vs \boxfil\ colour space may be affected by such bursts and how we can mitigate their effects.  We demonstrate that our final conclusions are not affected by starbursts.


\subsection{Effects of starbursts}

A starburst event would abruptly increase the \Halpha\ EW of a galaxy within a very short time-scale ($\lesssim5$ Myr). The increase in ionizing photons is driven by the extra presence of O and B stars during a starburst which increases the amount of Lyman continuum photons. Assuming that a constant factor of Lyman continuum photons get converted to \Halpha\ photons via multiple scattering events, we expect the number of \Halpha\ photons to increase as a proportion to the number of O and B stars. Furthermore, the increase of the O and B stars would drive the galaxy to be bluer causing the \boxfil\ colours to decrease.

The ability of a starburst to drive the points away from the monotonic Salpeter track is limited. The deviations are driven by the burst fraction, which we define as the burst strength divided by the length of the starburst. 
If the burst fraction is small, it has a small effect. However if it is very large it dominates both the \Halpha\ and the optical light, the older population is `masked', and it heads back towards the track albeit at a younger age, i.e. one is seeing the monotonic history of the burst component. The maximum deviation in our study occurs for burst mass fractions of 20--30\% occurring in time-scales of 100-200 Myr or fractions of thereof, which can cause excursions of up to $\sim 1$ dex. However as we will see this only occurs for a short time.

\begin{figure}
\includegraphics[trim = 5 10 10 10, clip, scale=0.90]{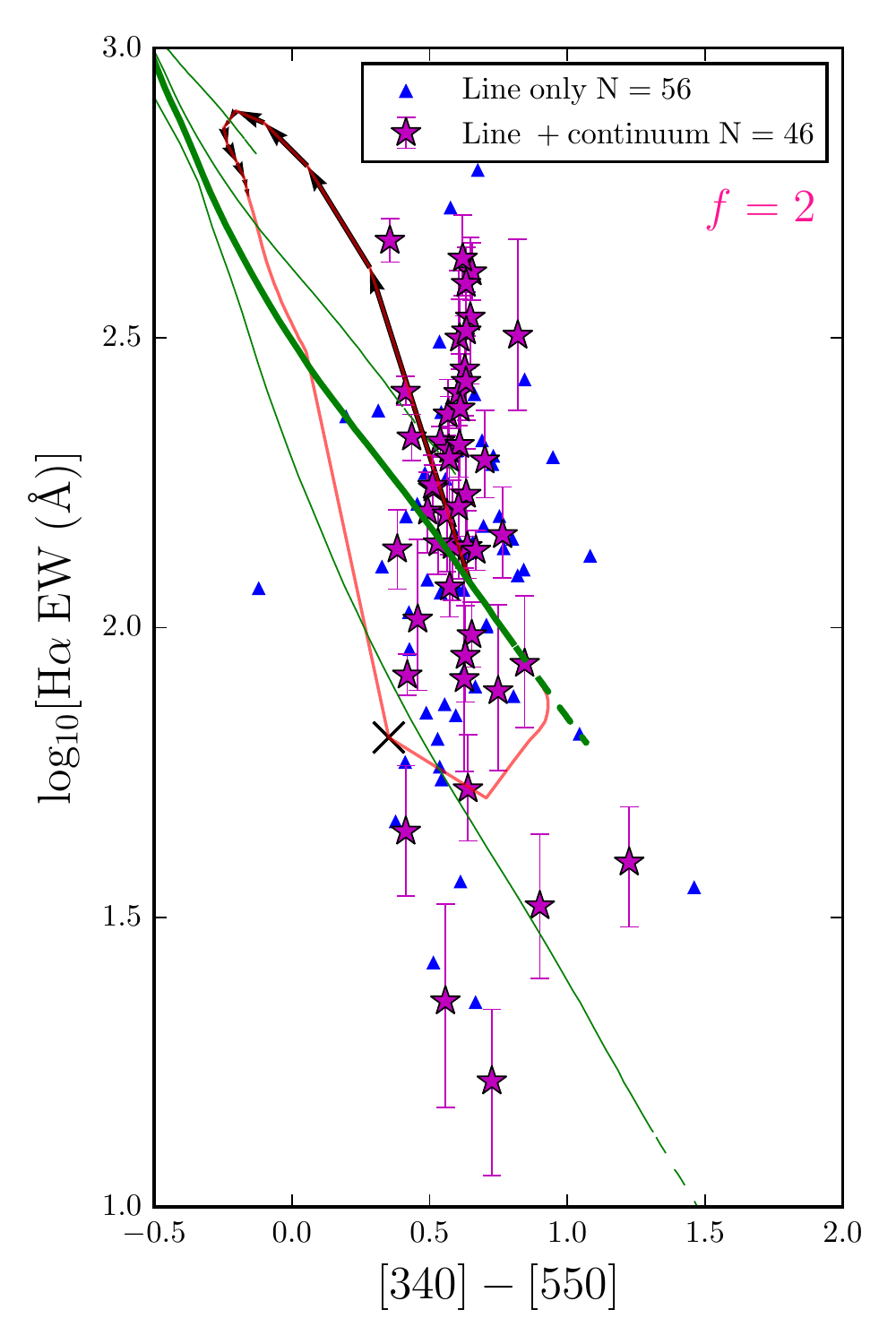}
\caption{ The effect of a star burst on a PEGASE model galaxy track.   
The green tracks are computed with constant SFHs but with different IMFs. From top to bottom they have $\Gamma$ values of respectively $-0.5, -1.0, -1.35$, and $-2.0$. All solid tracks end \around 3.1 Gyr and the continuation up to 13 Gyr is shown by the dashed lines.
The track in red follows the constant SFH model with a IMF slope $\Gamma=-1.35$ (thick green track) up to t=1500 Myr. A burst with a duration of $t_b=100$ Myr is superimposed on the track at t=1500 Myr. 
The burst generates 30\% of the galaxies' total mass at \around3000 Myr. The black arrows superimposed on the tracks show the direction of the burst in the first 10 Myr and are plot every 1 Myr to distinctively demonstrate the short time-scales in \Halpha\ EW evolution on the aftermath of a starburst. The cross denotes $t=1600$ Myr signalling the end of the burst. 
The \sample\ is dust corrected with a $f=2$ following prescriptions outlined in Section \ref{sec:dust_corrections}. The magenta stars show galaxies with continuum detections while solid blue triangles show galaxies only with \Halpha\ emission lines. 
}
\label{fig:burst_model_EW_BC}
\end{figure}

We show the effect of a starburst on a PEGASE model galaxy with a monotonic SFH in Figure \ref{fig:burst_model_EW_BC}.
A starburst with a time-scale of $\tau_b=100$ Myr and a strength $f_m=0.2$ (fraction of mass at $\sim3000$ Myr generated during the burst) is overlaid on the constant SFH model at time = 1500 Myr. The starburst drives the increase of \Halpha\ EW which occurs in a very short time-scale. In Figure \ref{fig:burst_model_EW_BC} the galaxy deviates from the constant SFH track as soon as the burst occurs and reaches a maximum \Halpha\ EW within 4 Myr. 
At this point, the extremely high-mass stars made during the burst will start to leave the main sequence. This will increase the number of red-giant stars resulting in higher continuum level around the \Halpha\ emission line.
Therefore, the \Halpha\ EW starts to decrease slowly after $\sim4$ Myr. Once the burst stops the \Halpha\ EW drops rapidly to values lower than pre-burst levels.
The galaxy track will eventually join the $\Gamma=-1.35$ smooth SFH track at a later time than what is expected by a smooth SFH model.

We further investigate the effect of starbursts with smaller time-scales ($t_b<20$ Myr) and find that the evolution of \Halpha\ EW in the aftermath of the burst to be more extreme for similar $f_m$ values. This is driven by more intense star-formation required to generate the same amount of mass within $\sim1/10$th of the time-scale. Since both \Halpha\ EW and \boxfil\ colours are a measure of sSFR we expect the evolution to strongly dependent on $f_m$ and $\tau_b$ of the burst and to be correlated with each other.

To consider effects of bursts, we adopt two complimentary approaches. First, we stack the data in stellar mass and \boxfil\ colour bins. Stellar populations are approximately additive and by stacking we smooth the stochastic SFHs in individual galaxies and also account the effect from galaxies with no \Halpha\ detections. Secondly, we use PEGASE to model starbursts to generate Monte Carlo simulations to predict the distribution of the galaxies in \Halpha\ EW vs \boxfil\ colour space. Using the simulations we investigate whether it is likely that the observed discrepancy is driven by starbursts and also double check whether the stacking of galaxies would generate smooth SFH models.

\subsection{stacking}
\label{sec:EW_stacking}

In order to remove effects from stochastic SFHs of individual galaxies in our sample, we employ a spectral stacking technique. We first divide the galaxies into three mass and dust corrected \boxfil\ colour bins as follows.
\begin{itemize}
\item Mass bins: log$\mathrm{_{10}(M_*/M_\odot)}$ $\leq9.5$, $9.5<$ log$\mathrm{_{10}(M_*/M_\odot)} <10$, log$\mathrm{_{10}(M_*/M_\odot)}\geq10$ 
\item \boxfil\ colour bins:  (\boxfil) $\leq0.56$, $0.56<$ (\boxfil) $<0.65$, (\boxfil) $\geq0.65$
\end{itemize} 
We select a wavelength interval of \around1500\AA\ centred around the \Halpha\ emission for each spectra and mask out the sky lines with approximately $2\times$ the spectral resolution. 
In order to avoid systematic biases arisen from narrowing down the sampled wavelength region in the rest frame, we instead redshift all spectra to a common $z=2.1$ around which most of the galaxies reside. 
We sum all the spectra at this redshift, in their respective bins.
The error spectra are stacked in quadrature following standard error propagation techniques. 

We mask out the nebular emission-line regions of the stacked spectra and use a sigma-clipping algorithm to fit a continuum (c1). 
The error in the continuum is assigned as the standard deviation of the continuum values of 1000 bootstrap iterations.

We visually inspect the stacked spectra to identify the \Halpha\ emission line profiles to calculate the integrated flux. Stacked \Halpha\ EW is calculated following equations \ref{eq:Ha_EW_obs} and \ref{eq:Ha_EW_rest}.

To estimate the error on the stack due to the stochastic variations between galaxies, we use a bootstrapping technique to calculate the error of the stacked \Halpha\ EW values.
We bootstrap galaxies with replacement in each bin to produce 1000 stacked spectra for each of which we calculate the \Halpha\ EW.
The standard deviation of the logarithmic EW values for each bin is considered as the error of the \Halpha\ EW of the stacked spectra. 
We expect the bootstrap errors to include stochastic variations in the SFHs between galaxies. If our sample comprises of galaxies undergoing extreme starbursts, the effects from such bursts should be quantified within these error limits. 

We stack the individual [340] and [550] fluxes of the galaxies in similar mass and \boxfil\ colour bins. The average extinction value of the galaxies in each bin is considered as the extinction  of the stacked spectra. 
We use this extinction value to dust correct the \Halpha\ EW and \boxfil\ colours of the stacked spectra, following recipes explained in Section \ref{sec:dust_corrections}.

\begin{figure*}
\includegraphics[scale=0.60]{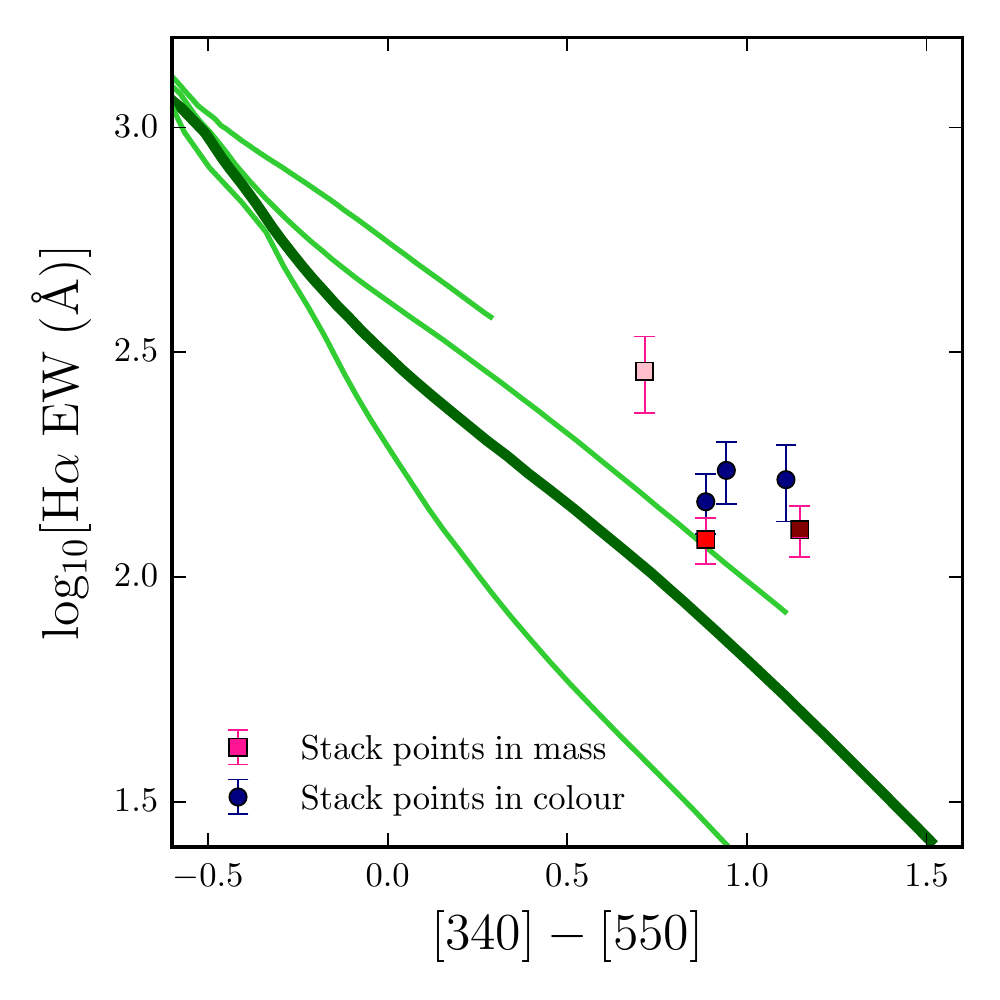}
\includegraphics[scale=0.60]{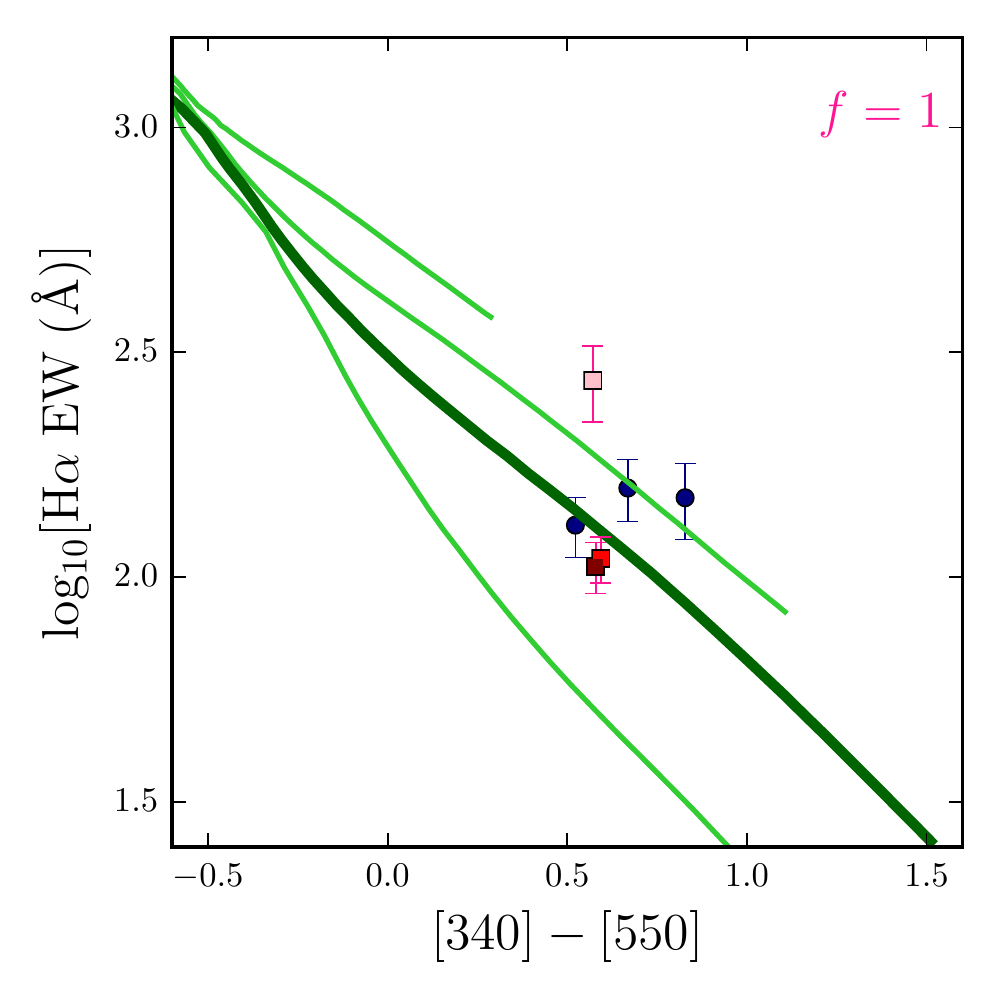}
\includegraphics[scale=0.60]{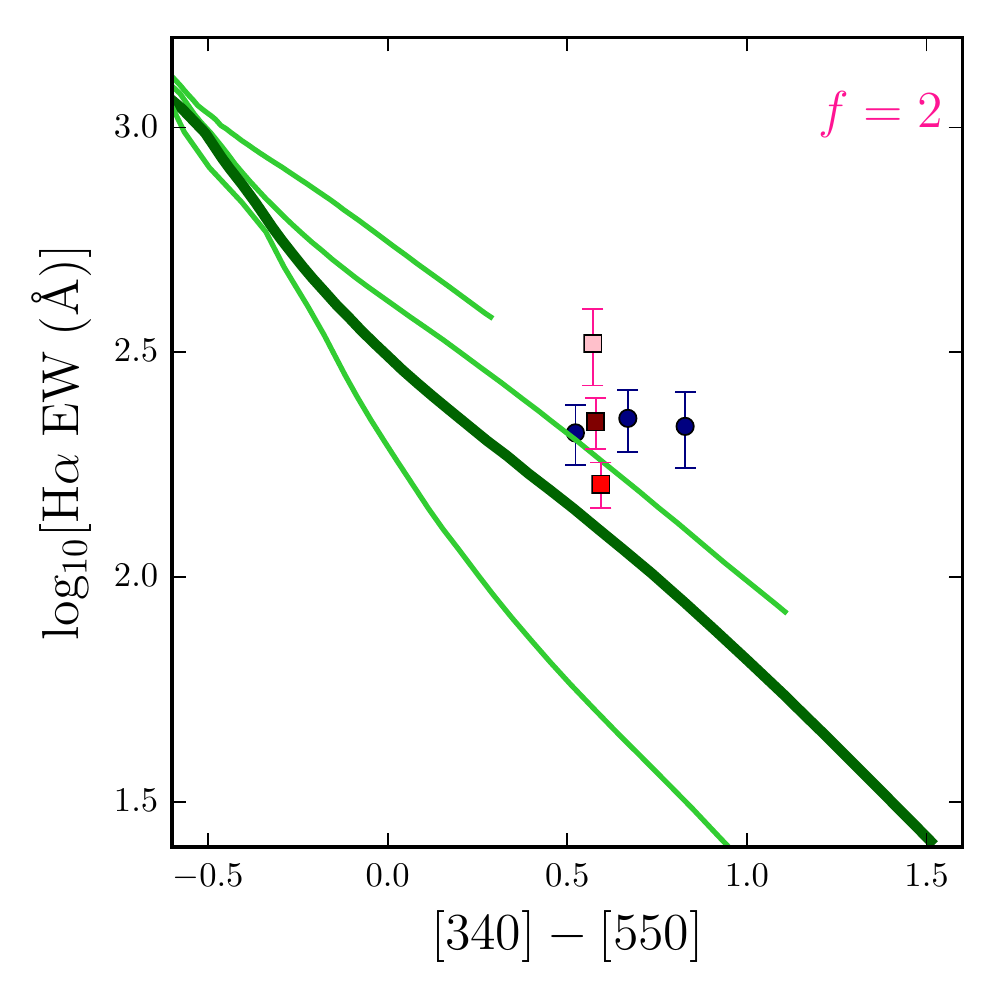}
\caption{ The \Halpha\ EW vs \boxfil\ colour distribution of the stacked \sample.  
The spectra are stacked in mass (squares) and \boxfil\ colour (circles) bins. The errors are from bootstrap re-sampling of the objects in each bin. The mass bins are colour coded where the higher masses have darker colours.
The tracks are SSP models computed from PEGASE with exponentially decaying SFHs of p$_1=1000$ Myr and varying IMFs. 
From top to bottom the tracks have $\Gamma$ values of respectively $-0.5, -1.0, -1.35$ (thick dark green line), and $-2.0$. 
All tracks end  \around3.1 Gyr ($z\sim2$).
{\bf Left:} Stacked galaxies before any dust corrections are applied. 
{\bf Centre:} Stacked galaxies after dust corrections are applied following recipes outlined in Section \ref{sec:dust_corrections} with a $f=1$.
{\bf Right:} Similar to the centre panel but with a $f=2$. 
}
\label{fig:EW_stacked}
\end{figure*}

Figure \ref{fig:EW_stacked} shows the distribution of the stacked spectra in \Halpha\ EW vs \boxfil\ colour space before and after dust corrections are applied. 
We consider dust corrections with $f=1$ and $f=2$.
For $f=1$ dust correction scenario the bluest colour bin and the medium and high mass stacked data points agree with the $\Gamma=-1.35$ track. However, the redder colours bins and lowest mass galaxies on average prefer shallower IMFs. 
With $f=2$ dust correction, the distribution of stacked data points even with bootstrap errors suggest values $\sim0.1-0.4$ dex above the Salpeter IMF, if interpreted as an IMF variation. 
In both scenarios, redder galaxies show larger deviation from the canonical Salpeter IMF. If starbursts drive the distribution of the galaxies, we expect the bluer galaxies on average to show larger deviation from the $\Gamma=-1.35$ tracks.

To further account for any detection bias arisen from \Halpha\ undetected galaxies, we use the ZFIRE COSMOS field K band targeted galaxies with no \Halpha\ emission detections to compute a continuum (c2) contribution to the stacked spectra. 
We use the photometric redshifts to select 37 galaxies within $1.90<z<2.66$, which is the redshift interval the \Halpha\ emission line falls within the MOSFIRE K band. 
Next we perform a cut to select galaxies with similar stellar masses ($9.04<\mathrm{log_{10}(M_\odot)}<10.90$) and \boxfil\ colours ($-0.12<$(\boxfil)$<1.46$) to the galaxies in the \sample. 
The final sample comprises of 21 galaxies which we use to stack the 1D spectra in mass and \boxfil\ colour bins.

\begin{figure}
\centering
\includegraphics[scale=0.70]{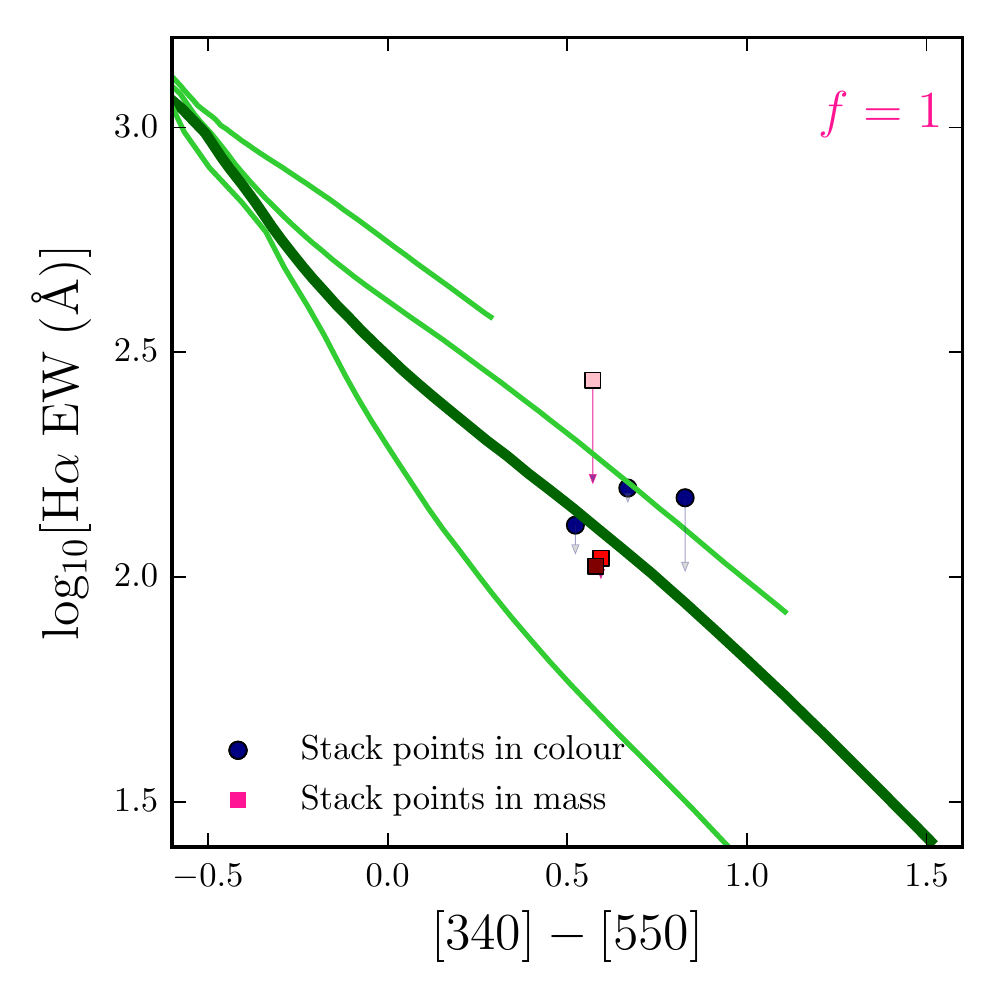}
\includegraphics[scale=0.70]{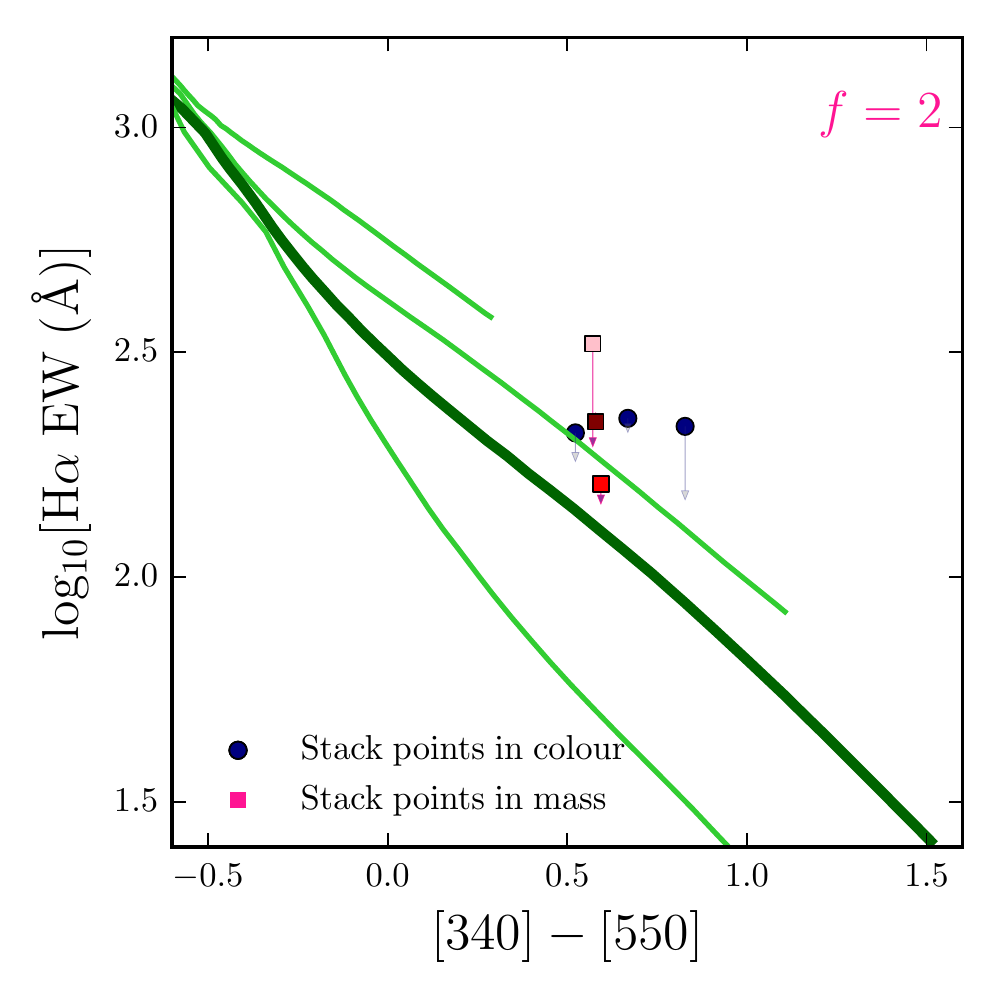}
\caption{ Here we show the effect of adding the continuum contribution (c2) of galaxies with no \Halpha\ detections to the \sample\ stacks. Galaxies are stacked in mass and colour bins and the arrows show the change in EW when the contribution from c2 is added to the data. All tracks shown are similar to Figure \ref{fig:EW_stacked}. 
{\bf Top:} Dust corrections applied with $f=1$. The errors for the data points are similar to Figure \ref{fig:EW_stacked} (centre panel).
{\bf Bottom:} Dust corrections applied with $f=2$. The errors for the data points are similar to Figure \ref{fig:EW_stacked} (right panel).
}
\label{fig:EW_c1c2_stacked}
\end{figure}

In order to stack the spectra, first we mask out the sky regions and assume that all galaxies are at a common $z=2.1$.
We mask out \Halpha\ and \NII\ emission line regions and fit a continuum similar to how c1 was derived. 
We add c1+c2 to re-calculate the \Halpha\ EW for each of the mass and colour bins. Since the continuum level is increased by the addition of c2, the \Halpha\ EW of the spectra reduces. We note that the highest mass bin contains no \Halpha\ undetected galaxies.
Figure \ref{fig:EW_c1c2_stacked} shows the change in stacked data points when the \Halpha\ non-detected continuum is considered with a dust correction of $f=1$ and $f=2$. 
The maximum deviation of the stacked \Halpha\ EW values is \around0.2 dex and the lowest mass and the reddest \boxfil\ colour bins show the largest deviation.  This is driven by the higher number of lower mass redder galaxies which have been targeted but not detected by the ZFIRE survey. 
The magnitude of the deviations are independent of the $f$ value used for the dust corrections and for both $f=1$ and $f=2$, the galaxies that show an excess of \Halpha\ EW compared to $\Gamma=1.35$ tracks still show an excess when the added c2 continuum contribution is considered.  
For $f=2$ dust corrections, even with considering the effect of non-detected continuum levels, majority of the stacked galaxies in our sample are significantly offset from the canonical Salpeter like IMF value.

\subsection{Simulations of starbursts}
\label{sec:simulations}

By employing spectral stacking and bootstrap techniques, we showed in Section \ref{sec:EW_stacking}, that our galaxies on average favour shallower IMFs than the universal Salpeter IMF.  
In this section, we use PEGASE SSP models to generate simulations with starbursts to calculate the likelihood for half of the \sample\ to be undergoing simultaneous starbursts at $z\sim2$. 
Furthermore, by randomly selecting galaxies from the simulation at random times, we stack the galaxies in mass and \boxfil\ colour bins to make comparisons with the stack properties of the \sample.

We empirically tune our burst parameters to produce the largest number of galaxies above the Salpeter track. 
A single starburst with time-scales of $t_b\sim100-300$ Myr and with $f_m\sim0.1-0.3$ are overlaid on constant SFH models with the starburst occurring at any time between $0-3250$ Myr of the galaxies' lifetime. Simulation properties and the evolution of \Halpha\ EW and \boxfil\ colours during starbursts are further discussed in Appendix \ref{sec:simulation_properties}.

Our final simulation grid contains 8337 possible time steps, which we use to randomly select galaxies within $2.0<z<2.5$ (similar to the time window where our observed sample lies) to perform a density distribution study and a stacking technique similar to the method described in Section \ref{sec:EW_stacking}.

\begin{figure*}
\includegraphics[scale=0.65]{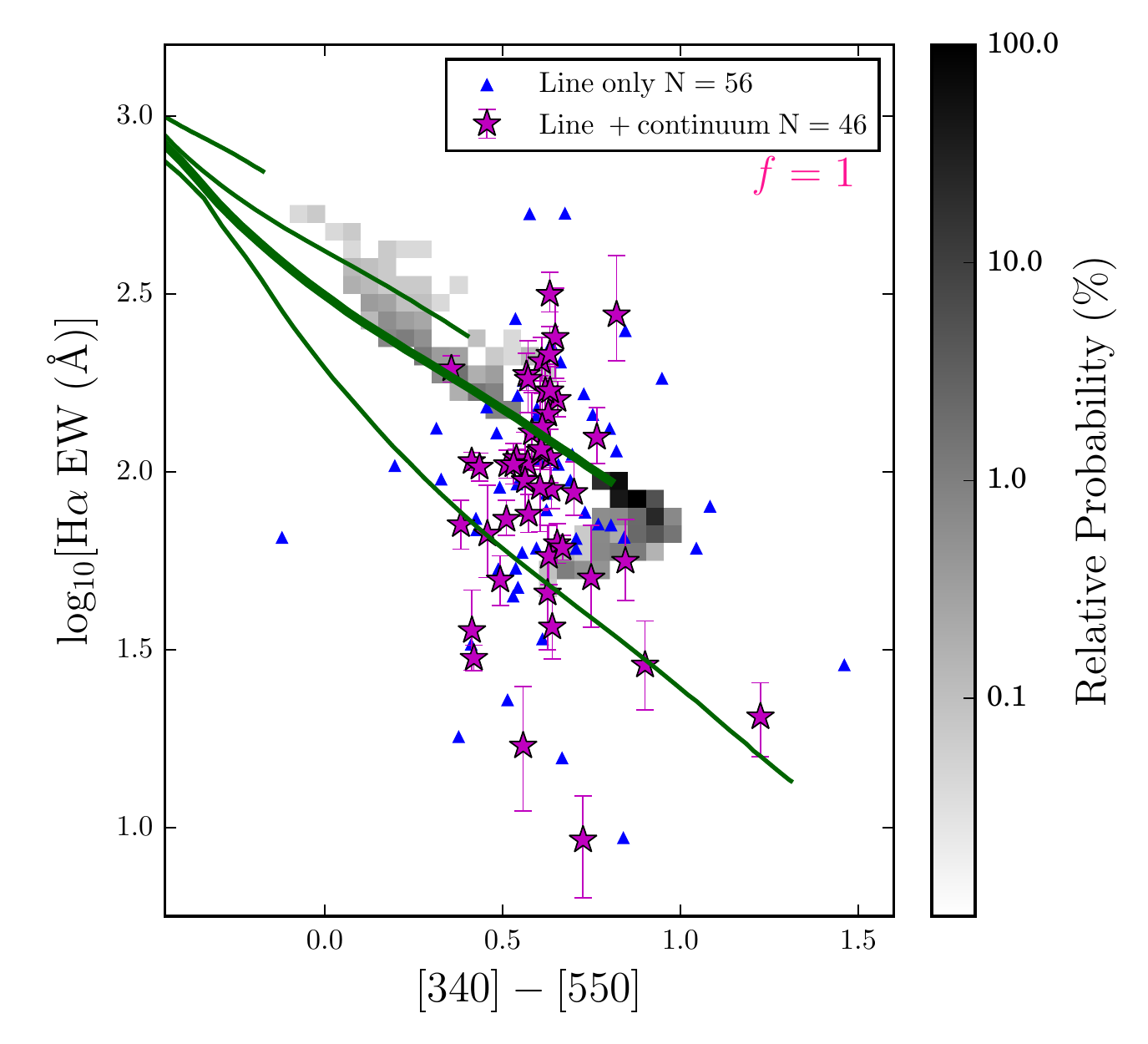}
\includegraphics[scale=0.65]{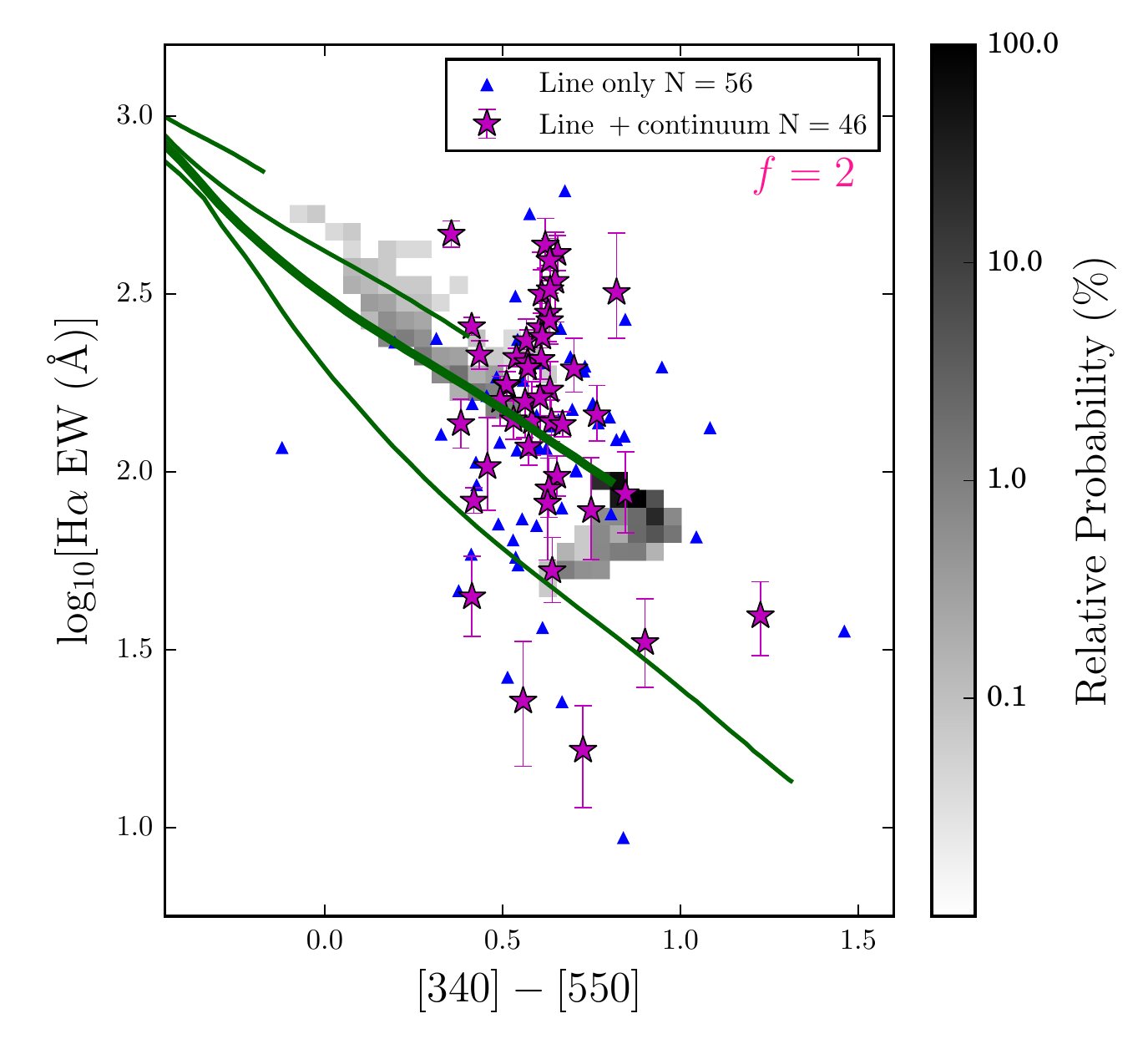}
\includegraphics[scale=0.65]{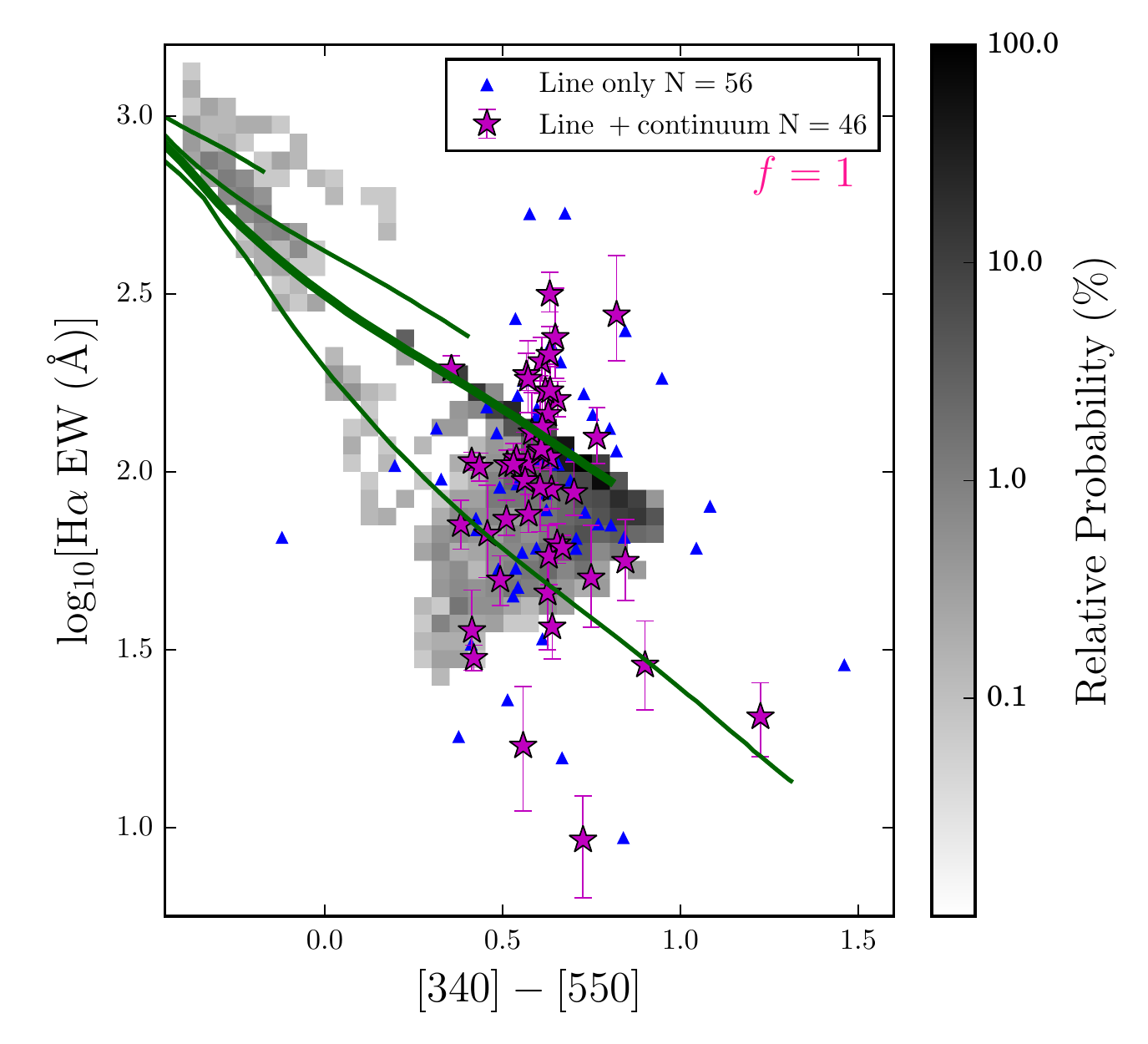}
\includegraphics[scale=0.65]{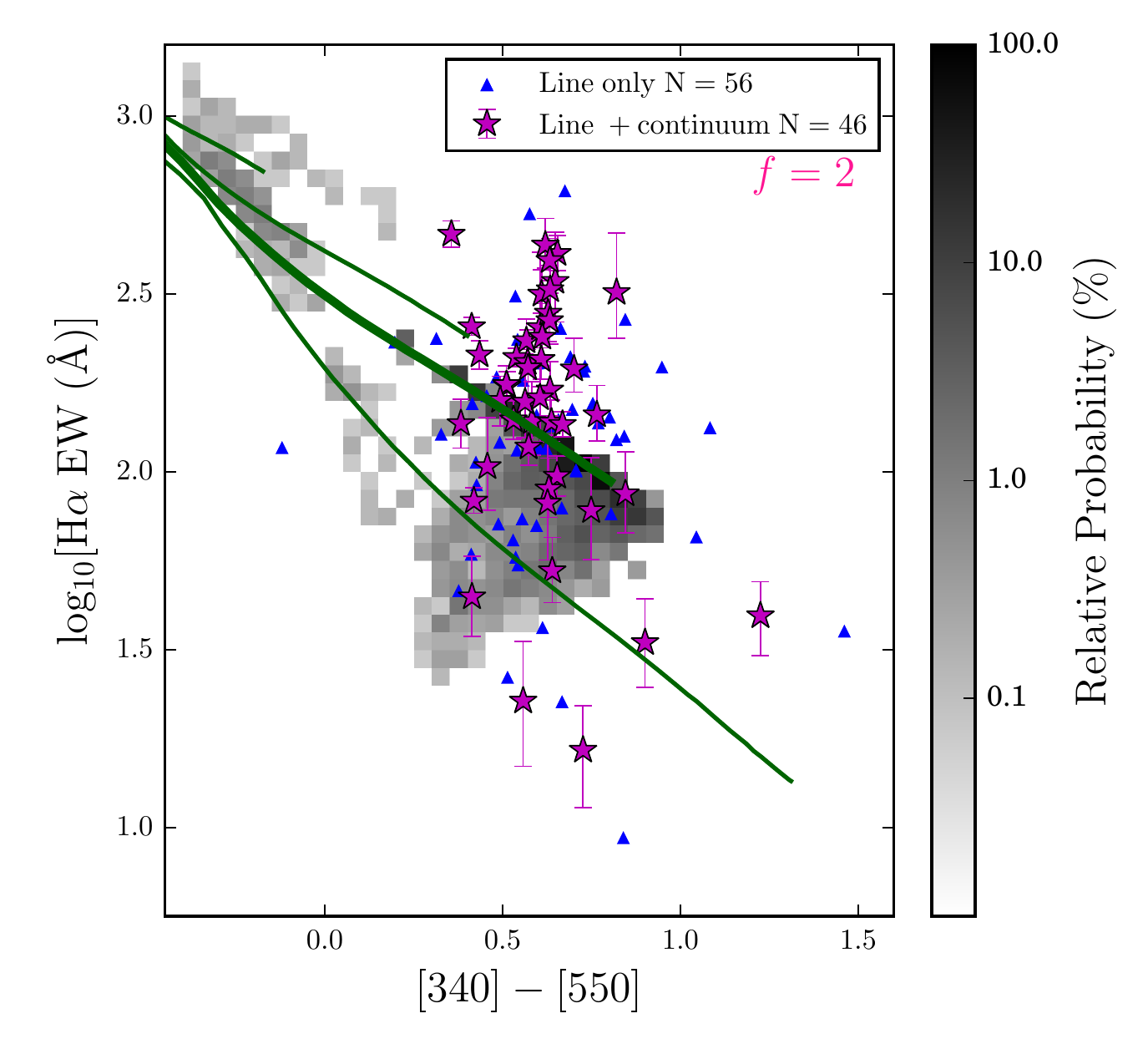}
\caption{{\bf Top left:} Relative probability distribution of galaxies occupying the \Halpha\ EW vs \boxfil\ colour space. The density distribution is made from 10,000 galaxies chosen randomly from the simulated galaxies with $t_b=  100-300$ Myr and $f_m= 0.1-0.3$. A darker shade suggests a higher probability. The \sample\ is overlaid on the figure with dust corrections applied with a $f=1$ following prescriptions outlined in Section \ref{sec:dust_corrections}. The PEGASE tracks shown are similar to Figure \ref{fig:EW_stacked}. 
{\bf Top right:} Similar to left but the \sample\ has been dust corrected using a $f=2$.
{\bf Bottom:} Similar to the top panels, but galaxies are selected from a simulation with smaller $t_b$ ($10-20$ Myr) but similar $f_m$ ($0.1-0.3$) values constrained to a redshift window between $2.0<z<2.5$. 
}
\label{fig:simulation_density}
\end{figure*}

To quantify the probability of starbursts dominating the scatter in the \Halpha\ EW vs \boxfil\ colour space, we select 10,000 galaxies randomly from the simulated sample to calculate the relative probability galaxies occupy in \Halpha\ EW vs \boxfil\ colour space. Figure \ref{fig:simulation_density} (top panels) shows the density distribution of the selected galaxy sample. 
The relative probability is calculated by normalizing the highest density bin to 100\%. To generate real values for the logarithmic densities we shift the distribution by 0.01 units. 
As evident from the figure, for both $f=1$ and $f=2$ dust corrections, there is a higher probability for galaxies to be sampled during the pre or post burst phase due to the very short time-scale the tracks take to reach a maximum \Halpha\ EW value during a starburst.
$\sim90\%$ of the galaxies in the \sample\ lie in regions with $\lesssim0.1\%$ probability. 
Therefore, we conclude that it is extremely unlikely that $\sim1/5$th ($f=1$) and $1/3$rd ($f=2$) of the galaxies in the \sample\ to be undergoing a starburst simultaneously and rule out the hypothesis that starbursts could explain the distribution of the \sample\ in the \Halpha\ EW vs \boxfil\ colour parameter space.

\begin{figure}
\includegraphics[trim = 10 10 10 10 , clip, scale=0.90]{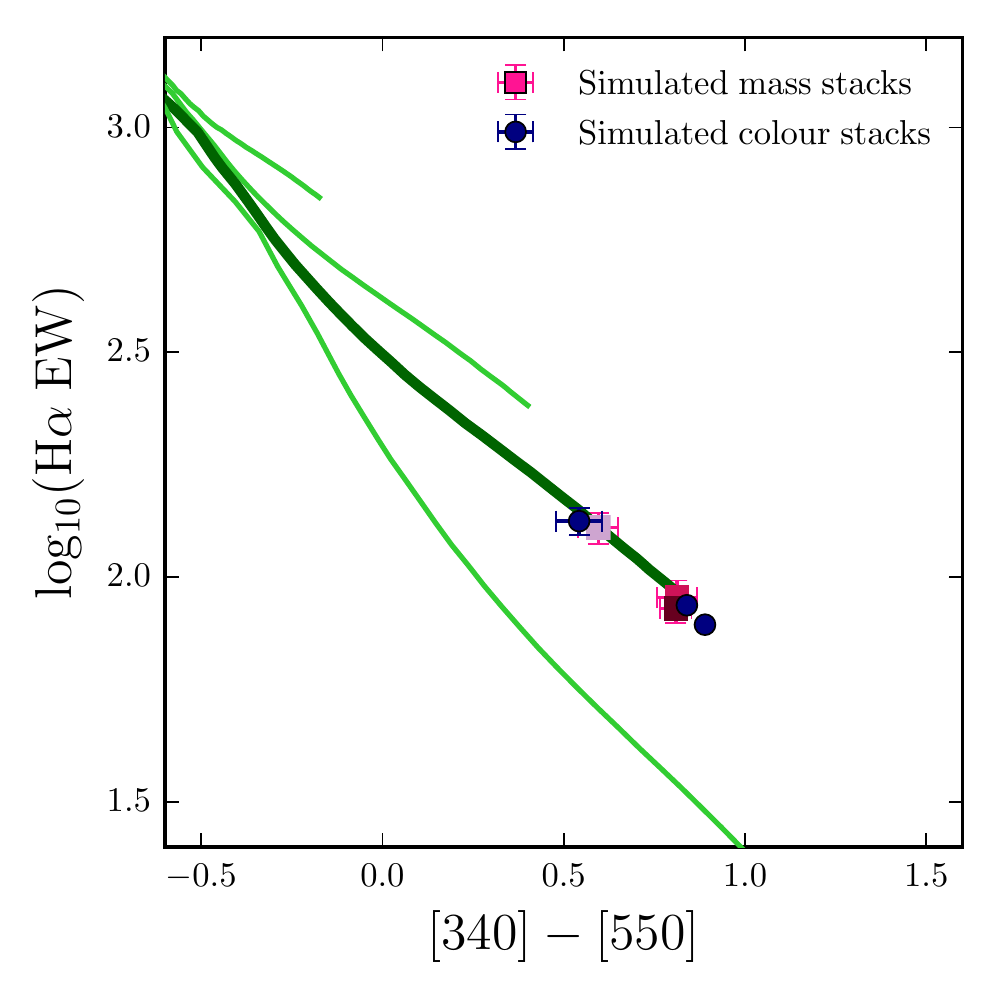}
\caption{\Halpha\ EW vs \boxfil\ colour distribution for stacked PEGASE galaxies used in the simulations in Figure \ref{fig:simulation_density}. Here we stack 100 randomly chosen galaxies from our simulated sample. The selected galaxies have constant SFHs with starbursts with varying strengths ($f_m=0.1-0.3$) and durations ($\tau_b=100-300$ Myr) overlaid at random times ($0-3250$ Myr) and have been stacked in stellar mass  and \boxfil\ colour bins. The errors of the stacks are computed using a bootstrap re-sampling. The PEGASE tracks shown have constant SFHs but varying IMFs.  From top to bottom the $\Gamma$ values of tracks are respectively $-0.5,-1.0, -1.35$, and -2.0. All tracks terminate at $t=3250$ Myr. 
}
\label{fig:simulation_stacks}
\end{figure}

In Figure \ref{fig:simulation_stacks}, we use our burst machinery to validate our stacking method.  
We select 100 galaxies randomly with replacement from the parent population of 100 galaxies. For each galaxy we select a random time to extract the galaxy spectra at the closest sampled time to retrieve stellar population parameters. 
We then stack the selected galaxies in stellar mass and \boxfil\ colour bins. The bins are generated in such a way that the selected galaxies are distributed evenly across the bins. 
We repeat the galaxy selection and stacking process 100 times to calculate bootstrap errors for the stacked data points. 
Galaxies containing SFHs with bursts stacked in mass and \boxfil\ colour show similar distribution to galaxy tracks with constant SFHs. Even with large $t_b$ values, the time-scale the tracks deviate significantly above the $\Gamma=-1.35$ is in the order of 1--5 Myr, and therefore it is extremely unlikely ($\lesssim5$ selected in the stacked sample of 100 galaxies) to preferentially select a large number of galaxies during this phase. 
Furthermore, stacked errors from bootstrap re-sampling do not deviate significantly from the $\Gamma=-1.35$ tracks. This further strengthens the point that repetitive sampling of galaxies does not yield stacks with higher \Halpha\ EW values for a given \boxfil\ colour.

\subsubsection{Smaller bursts}
\label{sec:small_bursts}

Galaxies at $z\sim2$ appear to be clumpy \citep[eg.,][]{Tadaki2013}. Therefore, it is possible for a single clump to have a high SFR that will add a significant contribution to the ionizing flux of a galaxy resulting in higher \Halpha\ EW values. To account for such scenarios, we perform the starburst simulations with smaller burst time-scales ($t_b\sim10-30$ Myr) and large burst strengths ($f_m\sim0.1-0.3$) and allow the galaxies to commence their constant SFH at a random time between 0 and 2500 Myr. We further constrain the bursts to redshifts between $2.0<z<2.5$ corresponding to a $\Delta t\sim650$ Myr, which is similar to the redshift distribution of our galaxies. 

As described previously, we randomly select 10,000 galaxies from our simulated sample, but constraining the selection to the redshift window of $2.0<z<2.5$. We show the density distribution of our randomly selected sample with the observed \sample\ in Figure \ref{fig:simulation_density} (bottom panels). A large fraction of galaxies are now selected during the post-burst phase, thus with lower \Halpha\ EWs compared to the reference IMF track, specially with $f=1$ dust corrections. 
Since the star-bursts are now short lived but have to generate the same fraction of stellar masses as the longer lived bursts, the fraction of mass generated by the burst per unit time is extremely high. Therefore, changes in \Halpha\ EW and optical colours are much more drastic as a function of time and makes it further unlikely to select galaxies with high \Halpha\ EWs. 
We further test scenarios including smaller bursts within short time-scales (see Table \ref{tab:simulation_param}) and find the distribution of selected galaxies to be similar to \ref{fig:simulation_density} (top panels). 
We conclude that, even limiting the starbursts to a narrow redshift window, does not yield a distribution of galaxies that would explain our high \Halpha\ EW sample.


\section{Considering Other exotica}
\label{sec:other_exotica}

In the previous sections we have shown that the distribution of the \sample\ galaxies in the \Halpha\ EW vs \boxfil\ colour cannot solely be described by dust or starbursts within a universal IMF framework. In this section, we investigate whether other exotic parameters related to SSP models such as stellar rotations, binary stars, metallicity, and the high mass cutoff could influence the distribution of the galaxies in our parameters to impersonate a varying IMF.

\subsection{Stellar rotation}
\label{sec:stellar_rotation}

First, we consider effects of implementing stellar rotation in SSP models.
Rotating stellar models produce harder ionizing spectra with higher amounts of photons that are capable of ionizing Hydrogen. This is driven by rotationally induced larger Helium surface abundances and high luminosity of stars \citep{Leitherer2012}, which results in $\sim\times5$ higher ionizing photon output by massive O stars at solar metallicity \citep{Leitherer2014} and can be $\gtrsim\times10$ towards the end of the main sequence evolution \citep{Szecsi2015}. 
The minimum initial mass necessary to form W-R stars is also lowered by stellar rotation resulting in longer lived W-R stars \citep{Georgy2012} increasing the number of ionizing photons. 
Therefore, stars with rotation shows higher \Halpha\ fluxes compared to systems with no rotation, resulting in higher \Halpha\ EW values. 
Additionally, stellar rotation also leads to higher mass loss in stars, which results in bluer stars in the red supergiant phase \citep{Salasnich1999}. 
Furthermore, stellar models with rotation results in longer lifetimes by $10\%-20\%$ \citep[eg.,][]{Levesque2012,Leitherer2014}. This allows a larger build up of short lives O and B stars compared with similar IMF and SFH models with no rotation resulting in higher \Halpha\ flux values and bluer stellar populations.

S99 supports stellar tracks from the Geneva group (explained in detail in \citet{Ekstrom2012} and \citet{Georgy2013} and references therein), which allows the user to compute models with and without invoking stellar rotation. Models with stellar rotation assumes an initial stellar rotation velocity (\vini) of 40\% of the break up velocity of the zero-age main-sequence (\vcrit).

\citet{Leitherer2014} notes that, \vini$=0.4$\vcrit\ for stellar systems is of extreme nature and should be considered as an upper boundary for initial stellar rotation values. 
\vini\ is defined as the rotational velocity the star possess when it enter the zero-age main-sequence. Depending on stellar properties and interactions with other stars \citep[eg.,][]{deMink2013} the initial rotational velocity of the star will be regulated with time (see Figure 12 of \citet{Szecsi2015}, where the evolution of stellar rotation has been investigated as a function of time for models with different stellar masses and initial velocities). 

A realistic stellar population will contain a distribution of \vini/\vcrit\ values.  
\citet{Levesque2012} investigated galaxy models with 70\% of stars with stellar rotation following \vini=0.4\vcrit\ and 30\% with no stellar rotation, thus allowing more realistic conditions. 
They found that such models show  $\sim0.5$ dex less Hydrogen ionizing photons compared to a stellar population with all stars with \vini=0.4\vcrit\ stellar rotation and $\sim1.4$ dex higher number of Hydrogen ionizing photons compared to a stellar population with no stellar rotation.

The extent of stellar rotation required to describe observed properties of stellar populations is not well understood. 
Gravitational torques have been shown to prevent stars from rotating $>50\%$ of its breakup velocity during formation \citep{Lin2011}. 
\citet{Martins2013} showed that Geneva models with stellar rotation does not reproduce the distribution of massive, evolved stars accurately and requires less amounts of convective overshooting thus lowering the required \vini.
However, recent studies demonstrate the requirement for populations of stars with extreme rotation in low-metallicity scenarios to explain the origin of narrow He emission in galaxies \citep{Grafener2015,Szecsi2015} and long Gamma-ray bursts \citep{Woosley2006,Yoon2006}. 
Stellar populations of the Large Magellanic Cloud have shown to be distributed following a two peak rotational velocity distribution with $\sim50\%$ of galaxies rotating at $\sim20\%$ of their critical velocities while $\sim20\%$ of the population having near-critical velocities \citep{Ramirez-Agudelo2013}. 
Furthermore, populations of Be stars \citep{Secchi1866,Rivinius2013}, which are near-critically rotating main-sequence B stars observed in local stellar populations \citep{Lin2015,Yu2016,Bastian2017}, have shown evidence for the existence of rapidly rotating stars in massive stellar clusters \citep[eg.,][]{Bastian2017}.

We show the evolution of galaxy properties in the \Halpha\ EW vs \boxfil\ colour in Figure \ref{fig:rotation_and_binary} (top panels). 
Due to limitations in rotational stellar tracks, the metallicity of the stars are kept at $Z=0.014$, but the stellar atmospheres are kept at $Z=0.020$. 
Invoking stellar rotation increases the \Halpha\ EW by \around0.1 dex for similar IMFs and shows slightly bluer colours for a given time \emph{t}. Further analysis of the sub-components shows us that these changes in the \Halpha\ EW vs \boxfil\ colours are driven by the increase in \Halpha\ flux and bluer optical colours.

Implementing stellar rotation results in similar effects of a shallower IMF ($\Gamma>-1.35$), but the deviations are not sufficient to explain the $f=2$ dust corrected \sample\ within a universal IMF scenario. 
However, with $f=1$ dust corrections, only $\sim5\%$ of the sample lie above the $\Gamma=-1.35$ track with stellar rotation models with a majority of galaxies showing steeper IMFs.

Having a large fraction of stars with extreme rotation will lead to a higher number of ionizing photons and bluer colours and could potentially explain the high-EW objects in our sample. 
Sustaining such high rotation requires extremely low metallicities, which we further discuss in Section \ref{sec:model_Z}.  
Even though we expect the actual variation of the \Halpha\ EW and \boxfil\ colours due to stellar rotation at near-solar metallicity to be much smaller than what is shown in Figure \ref{fig:rotation_and_binary}, we cannot rule out extreme stellar rotation dominant in at low metallicities (Z$\sim0.002$). 
Therefore, extreme stellar rotation may provide one explanation independent of the IMF to describe the distribution of our galaxies in the \Halpha\ EW and \boxfil\ colour space (see Section \ref{sec:model_Z}). 
Furthermore, stellar rotation can introduce fundamental degeneracies to IMF determination which we discuss further in Section \ref{sec:ssp_issues}.

\subsection{Binary system evolution}
\label{sec:binaries}

We consider the effect of implementing the evolution of binary stellar systems on our study. All SSP models described thus far only considered single stellar populations, i.e. there were no interactions between stars in a stellar population. 
However, recent observational studies in our Galaxy have shown that \around50\% of massive O stars are in binary systems and that the environment may have a strong influence on the dynamical and/or stellar evolution \citep{Langer2012,Sana2012,Sana2013}. Only a minority of O stars would have undisturbed evolution leading to supernovae \citep{Leitherer2014}, thus introducing additional complexities to SSP models and strong implications for studies using these models to infer observed stellar properties. Furthermore, \citet{Steidel2016} demonstrated the necessity of invoking models with massive star binaries to fit rest frame UV and optical features of star-forming galaxies at $z\sim2.5$.

\begin{figure*}
\includegraphics[scale=0.9]{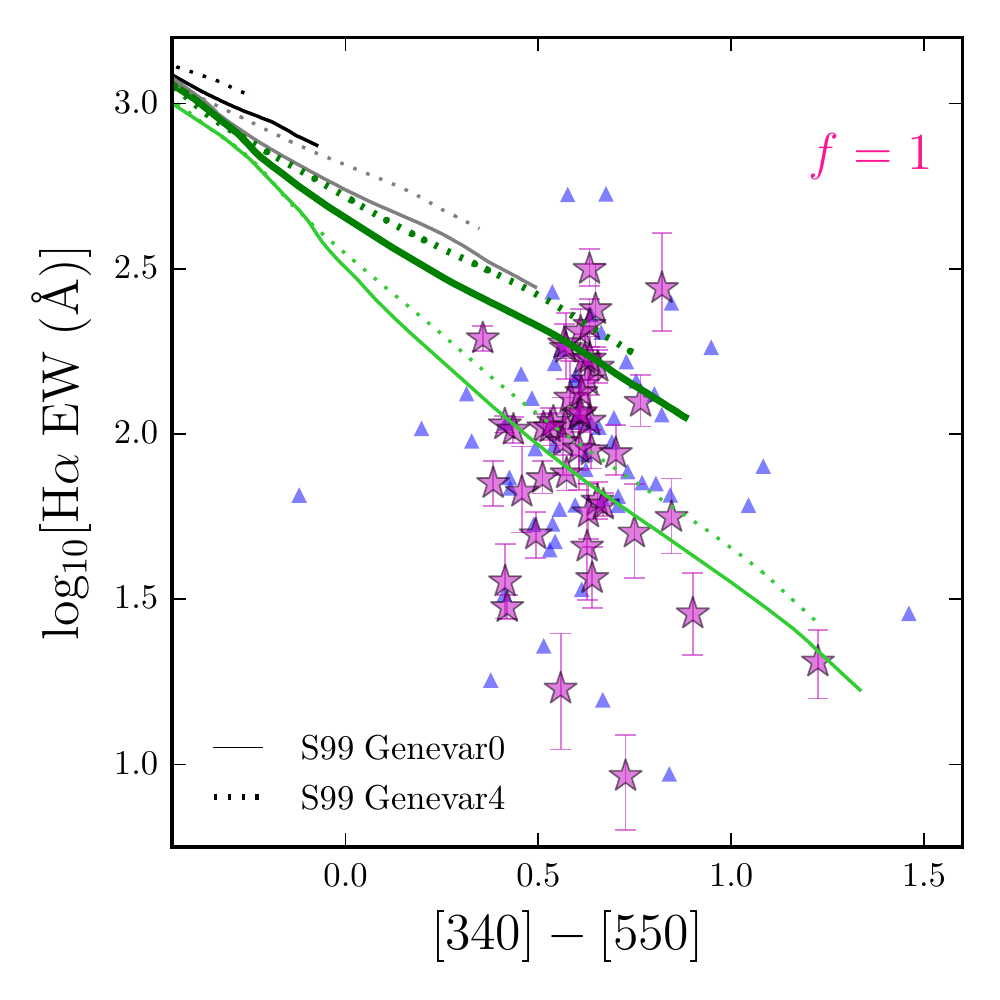}
\includegraphics[scale=0.9]{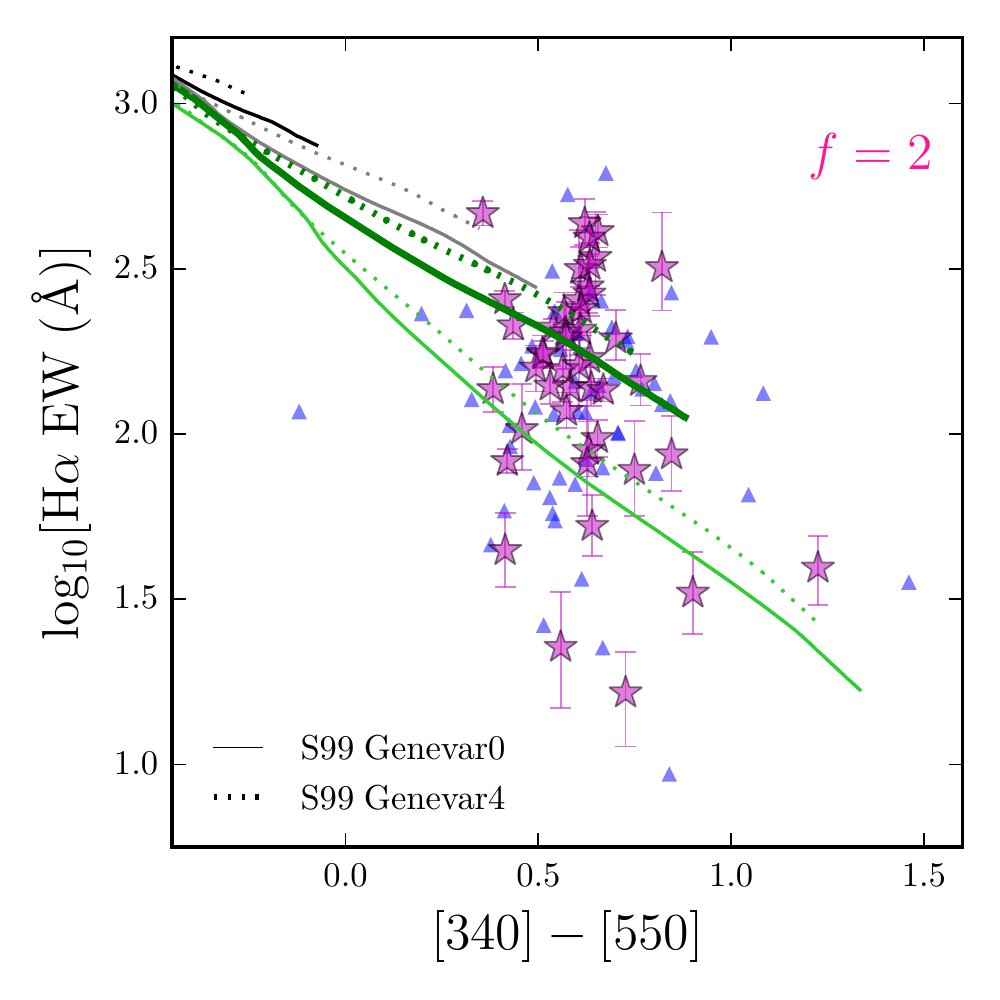}
\includegraphics[scale=0.9]{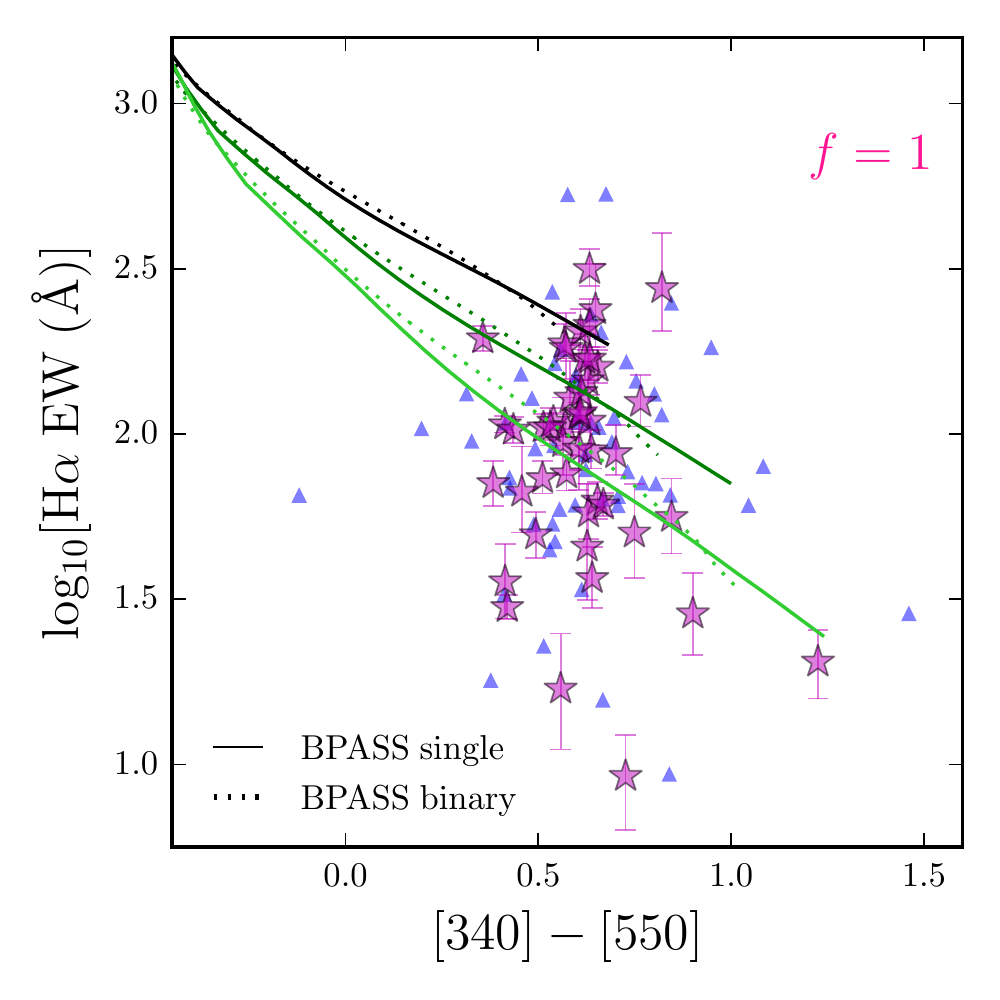}
\includegraphics[scale=0.9]{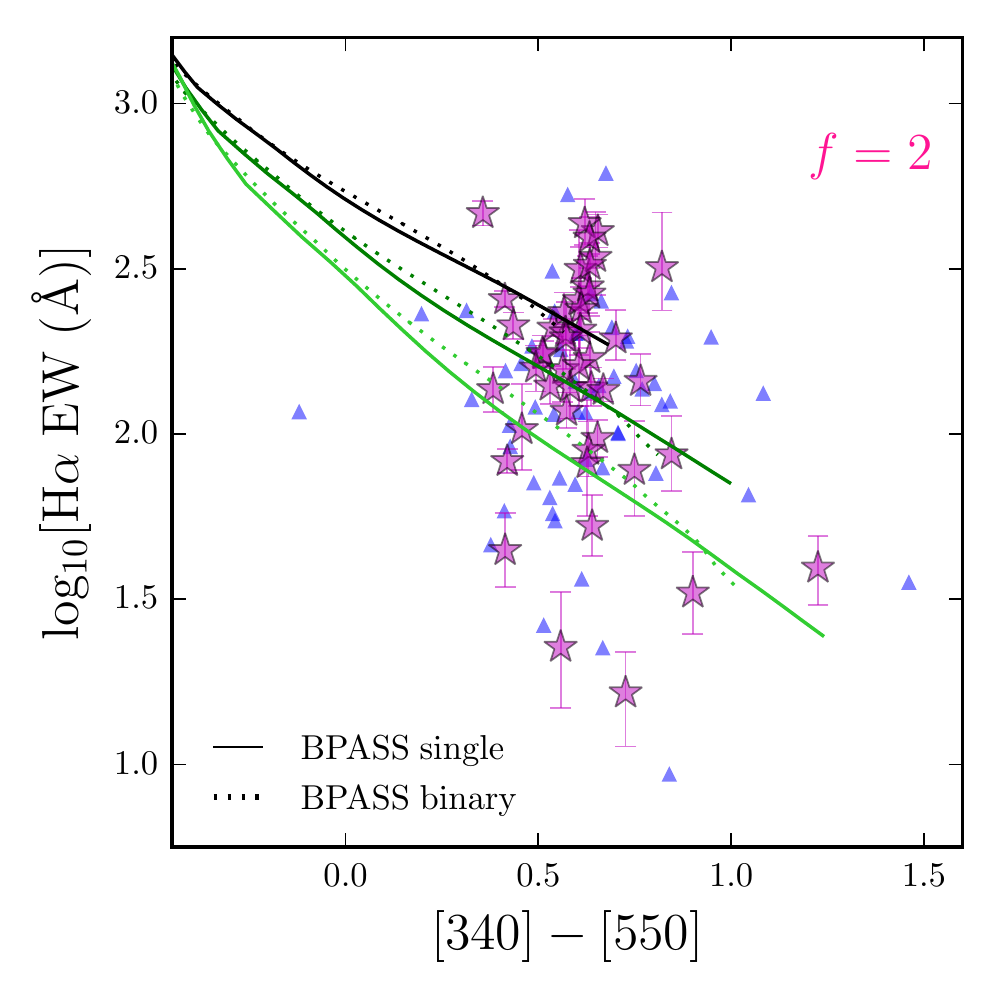}
\caption{ Effects of stellar rotation and evolution of binary stars in the \Halpha\ EW vs \boxfil\ colour space. 
{\bf Top left:} Here we show the evolution of  \Halpha\ EW vs \boxfil\ colours for Starburst99 SSP models using Geneva tracks with (dotted lines) and without (solid lines) stellar rotation. Each pair of tracks represent the same IMF and the IMFs plot are similar to Figure \ref{fig:EW_no_dust_corrections}. The overlaid \sample\ galaxies have been dust corrected using a $f=1$.
{\bf Top right:} Similar to the top left panel but the \sample  has been dust corrected using $f=2$.
{\bf Bottom left:} Here we show the effects of including the evolution of binary stellar systems in the \Halpha\ EW vs \boxfil\ colour space. The models shown here are from the \citet{Stanway2016} BPASSv2 models with Z=0.02. Each pair of tracks with same colour represent the same IMF with (dotted) and without (solid) considering the evolution of binary stellar systems. From top to bottom the IMFs plot are for $\Gamma$ values of$ -1.0, -1.35$ (Salpeter IMF), and $-1.70$. The overlaid \sample\ galaxies have been dust corrected using a $f=1$.
{\bf Bottom right:} Similar to the bottom left panel but the \sample\ has been dust corrected using $f=2$.}
\label{fig:rotation_and_binary}
\end{figure*}

We use the BPASS v2.0 models \citep{Stanway2016} to investigate the effects of invoking stellar binary evolution in the \Halpha\ EW vs \boxfil\ colour space. The computed models have been released by the BPASS team only for a limited set of IMF models. We use IMF models with Z=0.02 and $\Gamma=-1.00, -1.35,\ \mathrm{and}\ -1.70$ with a lower and upper mass cutoff at 0.5\msol\ and 100\msol,respectively. 
The IMF slope for stellar masses between 0.1\msol$-$0.5\msol\ is kept at $\Gamma=-0.30$ for all the models. We remind the reader that stars with $M_*\lesssim0.5M_\odot$ have negligible effect on the \Halpha\ EW vs optical colour parameter space.
Figure \ref{fig:rotation_and_binary} (bottom panels) compares the effect of considering stellar binary system evolution in this parameter space. 
Binary rotation with simple prescriptions for stellar rotation slightly increases the \Halpha\ EW (max increase for a given time is \around0.2dex) and make galaxies look bluer for a given IMF at a time \emph{t}. These changes are more prominent for galaxies with steeper IMFs and are driven by the \Halpha\ flux and optical colours of the galaxies. Furthermore, unlike effects of rotation, we see a trend on which the steeper IMFs show larger changes (up to $\sim\times2$) in \Halpha\ flux and \boxfil\ colours  compared to shallower IMFs.

Due to higher ionizing flux and longer lifetimes of massive O type stars in binary systems, galaxies look bluer at an older age compared to what is predicted by single-star models \citep{Eldridge2016}. The increase in ionizing flux is driven by transfer of mass between stars causing rejuvenation, generation of massive stars via stellar mergers, and stripping of Hydrogen envelope to form more hot Helium or W-R stars. Mass transfer and mergers between stars also result in larger, bluer stars at later times contributing to the stellar population to be bluer. 
The change of \Halpha\ EW and \boxfil\ colours due to binary system evolution at Z=0.02 is not sufficient to explain the distribution of the \sample\ galaxies and is significantly smaller than the contribution from stellar rotation.

Note that BPASS single stellar evolutionary models do not consider any form of stellar rotation. 
BPASS binary models do consider stellar rotation, but only if a secondary star accretes material from a companion. In such scenarios at Z $>0.004$ the secondary star is spun up, fully mixed, and is rejuvenated resulting it to be a zero-age main-sequence star. 
However, it is assumed that the star is spun down quickly and stellar rotation is not considered for the rest of it's evolution \citep[][J.J. Elridge., private communication]{Eldridge2012a,Stanway2016}.
Since the current version of BPASS binary models does not consider aspects of stellar rotation in the context of reduction in surface gravity and the driving of extra-mixing beyond the expectations from the standard mixing-length theory as discussed in Section \ref{sec:stellar_rotation} (also see papers in the series by \citet{Meynet2000} and \citet{Potter2012}), comparisons between S99 Geneva models and BPASS cannot be performed to constrain the net effect of introducing stellar binary stellar evolution to SSP models that consider stellar rotation.

\subsection{Stellar metallicity}
\label{sec:model_Z}

\citet{Hoversten2008} showed that the evolution of galaxies \Halpha\ EW vs optical colour parameter space was largely independent of the metallicity of the galaxies. However, these predictions were made using PEGASE models and did not account for the increase in mass loss via stellar winds and increase in ionizing flux predicted in low metallicity scenarios by models that consider stellar rotation and binary interactions.

The lack of elements such as Fe that dominate the opacity in radiation-driven stellar winds, stellar interiors, and atmospheres in low metallicity stars results in the generation of higher amount of ionizing photons \citep[eg.,][]{Pauldrach1986,Vink2005,Steidel2016}.
Furthermore, at lower metallicities due to weaker stellar winds the mass loss is rate is low, thus most massive stars retain their luminosity and continue to shine for an extended time. 

When stellar rotation is introduced to single stellar population models, due to the higher fraction of W-R stars in higher metallicity environments, rotational stellar models with higher metallicity show a larger increment ($\Delta$EW) in ionization flux compared to the increment seen in lower metallicity models  \citep{Leitherer2014}. 
However, at $t<3100$ Myr low metallicity rotating stellar models on average show higher amount of ionising flux compared to higher metallicities.

When binary interactions are considered, the mass transfer between the binaries result in the increase of angular momentum of the stars causing an increase in stellar rotation \citep{deMink2013}. 
Additionally, at Z $\leq0.004$ if stars with \mass\ $>20$\msol\ has accreted $>5\%$ of its original mass, BPASS assumes that the star maintains its rapid rotation throughout its main-sequence lifetime \citep{Stanway2016}. 
This is driven by weaker stellar winds that allow the stars to maintain their rapid rotations for a prolonged period. 
Furthermore, rotationally induced mixing of stellar layers causes Hydrogen burning to be efficient resulting in rejuvenation of the main sequence stars. 
As we show in Section \ref{sec:stellar_rotation}, stellar rotation increases the production of ionizing photons and therefore lower metallicity systems with binary interactions show higher \Halpha\ EW values. 
Lower cooling efficiencies prominent in lower metallicity environments, also result in the stars to be bluer and brighter. 
Comparisons between S99 Geneva models with Z=0.002 and Z=0.014 suggest metallicity to have a prominent effect in increasing the \Halpha-EWs compared to stellar rotation. 
BPASS models also show metallicity effects to be prominent compared to effects by stellar rotation and binary interactions. 

Therefore, we conclude that within the scope of current stellar models, metallicity to be the prominent driver in increasing the \Halpha-EWs with stellar rotation and binary interactions contributing to a lesser degree.

In Figure \ref{fig:Z_and_high_mass} (top panels), we show the evolution of a $\Gamma=-1.35$ IMF constant SFH stellar tracks from BPASS with varying metallicities. The variation in metallicity in the \Halpha\ EW vs \boxfil\ colour is degenerated with IMF variation. 
Models with lower metallicities favour higher EWs and bluer colours compared to their higher metallicity counterparts.

Next, we explore whether gas phase metallicities computed for our galaxies \citep{Kacprzak2015,Kacprzak2016} suggest sufficiently low stellar metallicities to  produce ionising flux to explain the high-EW galaxies within a $\Gamma=-1.35$ IMF scenario. 
Converting gas-phase oxygen abundance to stellar iron abundance in high-z galaxies is nontrivial.
First, there are considerable systematic uncertainties in the gas phase metallicities measured using \NII/\Halpha\ ratios.
There are uncalibrated interrelations between ionization parameter, electron density and radiation field hardness at $z\sim2$ \citep{Kewley2013a}. 
For example, at a fixed metallicity, the \NII/\Halpha\ ratio can be enhanced by a lower ionization parameter or the presence of shocks \citep[eg.,][]{Yuan2012,Morales-Luis2013} and it is unknown whether the N/O ratio evolves with redshifts \citep{Steidel2014}.  From \citet{Kacprzak2015}, the gas-phase oxygen  abundance of our sample at $\mathrm{log10(M_*/M_\odot)=9.5}$ is $\sim0.5$ \zsol, however, the  systematic uncertainty can be  a factor of $\times$2 because of the unknown calibrations.  Because of this, we emphasize that metallicity can be compared reliably in a relative sense, but not yet on an absolute scale \citep{Kewley2008}.

Second,  there is limited knowledge on how iron abundances relative to $\alpha-$element (e.g, O, Mg, Si, S and Ca) abundance change over cosmic time and in different galactic environments \citep[eg.,][]{Wolfe2005,Kobayashi2006,Yuan2015}. 
In addition, there is a lack of consistency in  abundance scale used in   stellar atmosphere modelling, stellar evolutionary tracks and nebular models \citep{Nicholls2016}.    There are considerable variations  in the [O/Fe] ratios that are not well-calibrated at the low metallicity end.  
For example, at [Fe/H] $< - 1.0$ the extrapolated [O/Fe] ratio based on Milky Way data is  0.5 \citep{Nicholls2016}, with a $\sim0.3$ dex uncertainty in conversions of individual values \citep[eg.,][]{Stoll2013}.   
\citet{Steidel2016} argued an average [O/Fe] ratio of 0.74 for $z\sim2$ UV selected galaxies at oxygen nebular metallicity of $\sim0.5$ \zsol, suggesting a substantially lower stellar metallicity of [Fe/H] $\sim-1.0$.    
If we adopt the [O/Fe] ratio of \citet{Steidel2016},  then we would reach the same conclusion as \citet{Steidel2016} that our stellar abundance is [Fe/H] $\sim -1.0$.  In this case, we cannot completely rule out extremely low metallicity scenarios to explain the distribution of galaxies in the \Halpha\ EW vs \boxfil\ colour space. 
With $f=2$ ($f=1$) dust corrections between BPASS binary models with stellar metallicity of Z=0.02 to Z=0.002, the amount of objects that lie $2\sigma$ above the reference $\Gamma=-1.35$ track changes from $\sim40\%$ ($\sim9\%$) to $\sim9\%$ ($\sim2\%$).

Given all the uncertainties mentioned above, we think it is premature to convert our gas-phase oxygen  abundance  to stellar iron abundance and draw meaningful conclusions.
We further note that there are significant uncertainties in massive star evolution in SSP codes and the treatment of stellar rotation and binary stars, which we discuss further in Section \ref{sec:ssp_issues}.

\subsection{High mass cutoff}
\label{sec:mass_cutoff}

\begin{figure*}
\includegraphics[scale=0.90]{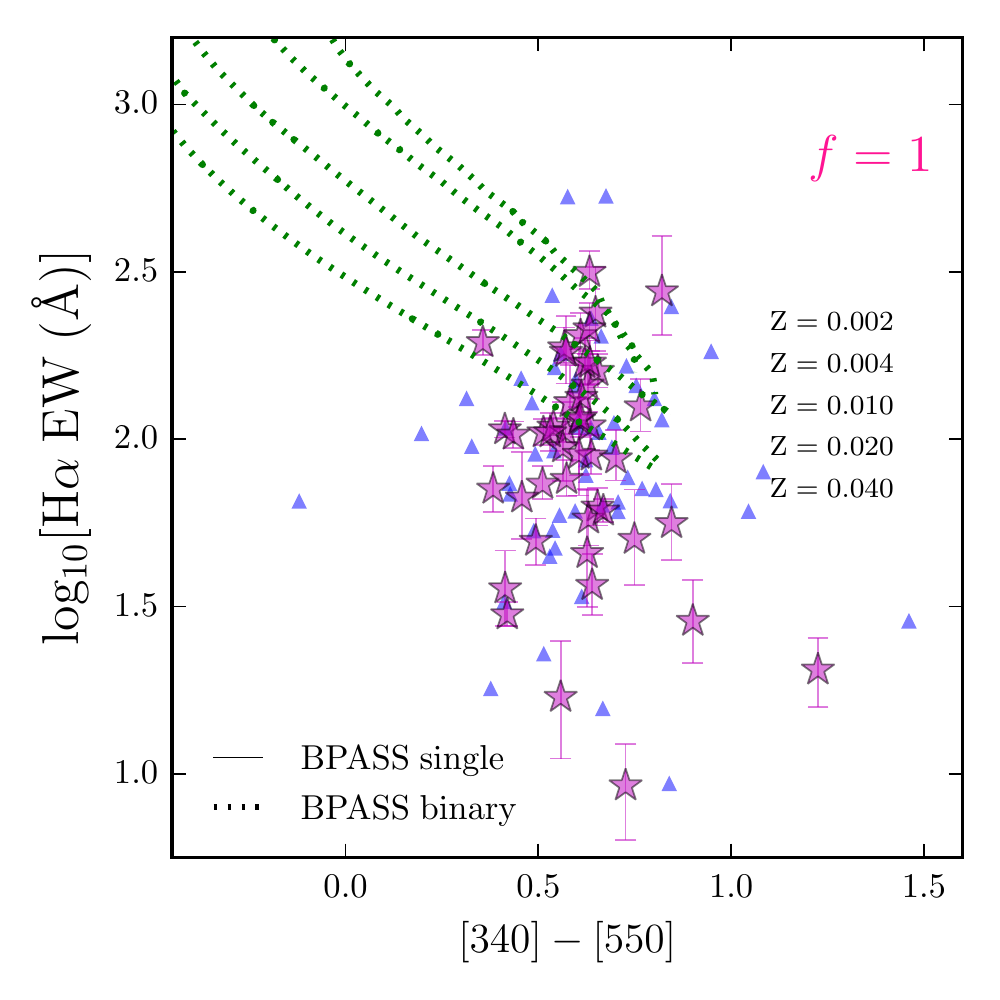}
\includegraphics[scale=0.90]{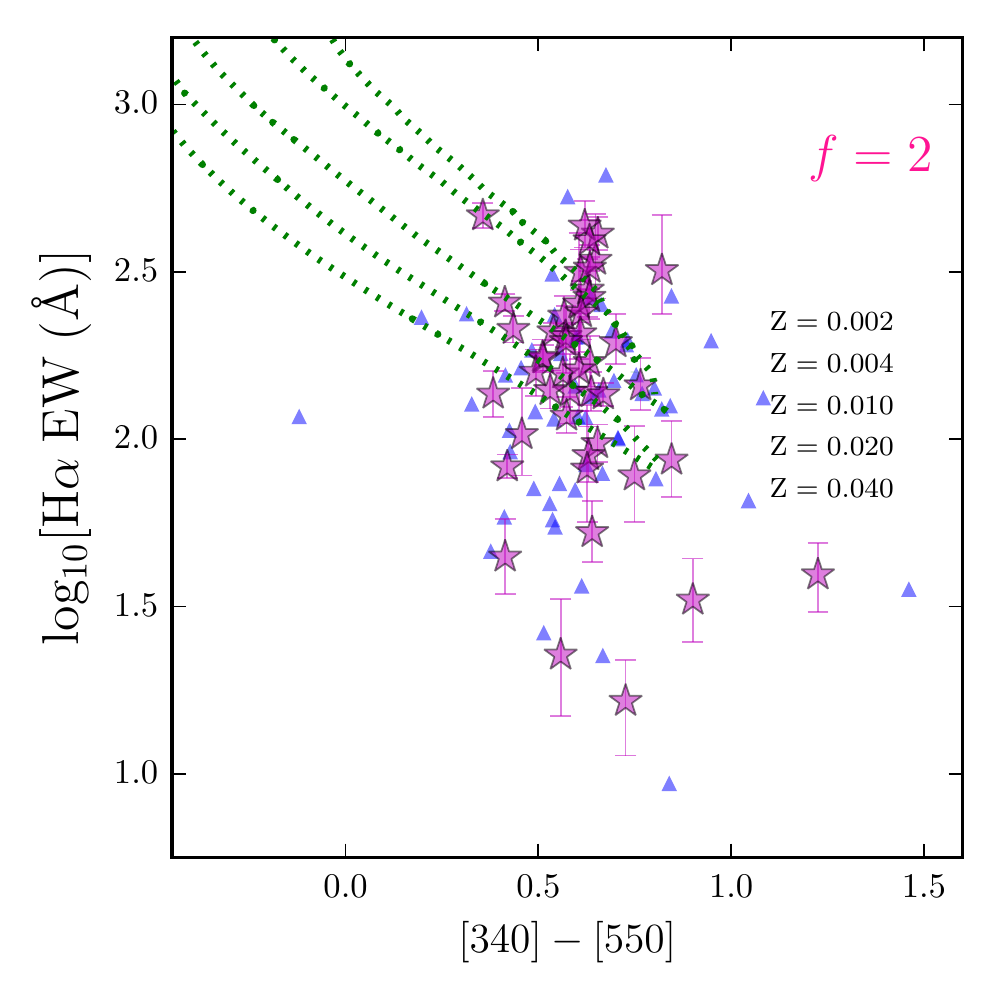}
\includegraphics[scale=0.90]{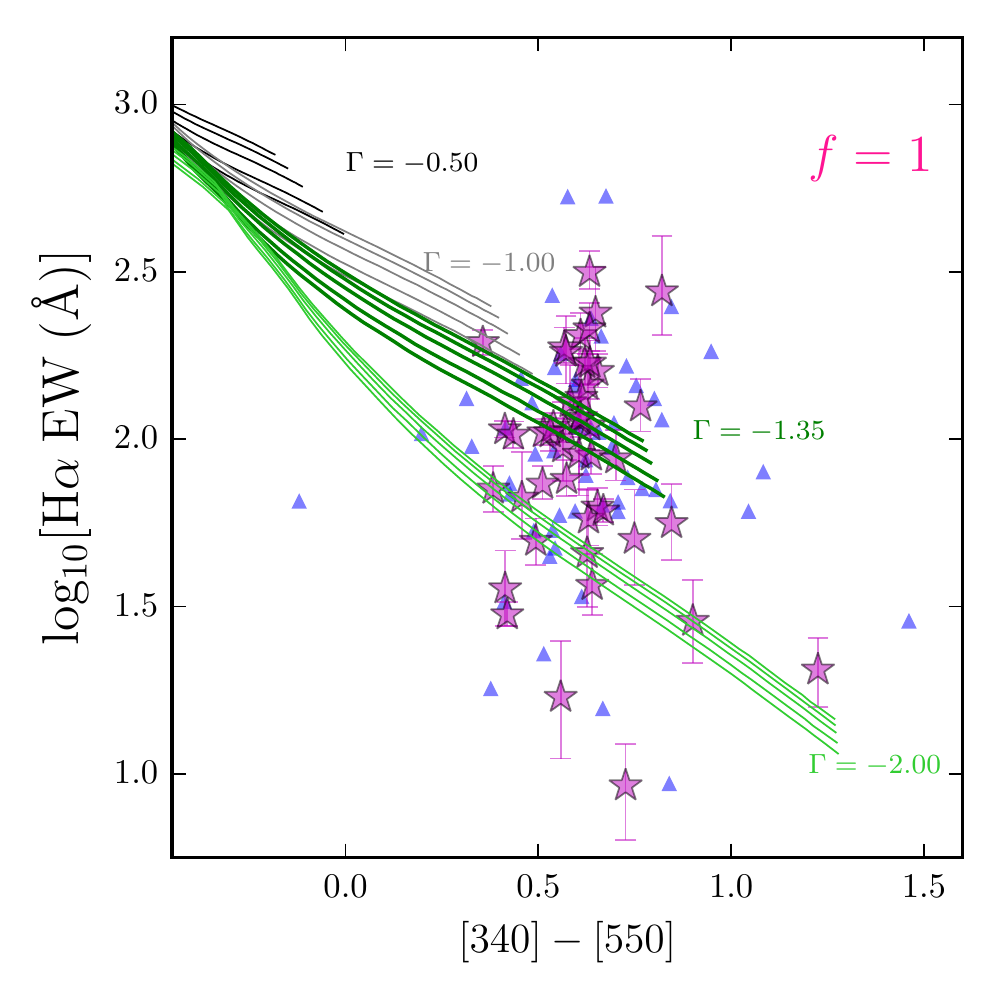}
\includegraphics[scale=0.90]{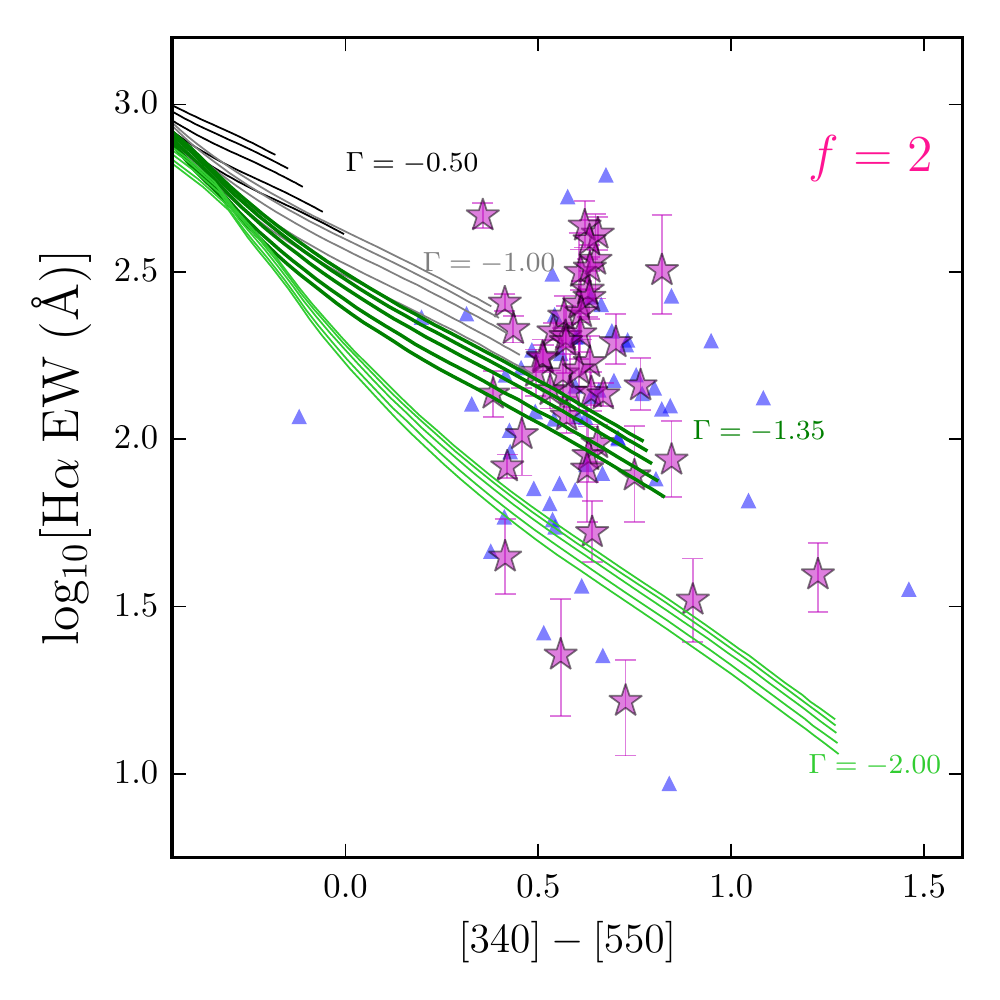}
\caption{ Effects of metallicity and  high mass cutoff of SSP models in the \Halpha\ EW vs \boxfil\ colour space. 
{\bf Top left:} Here we show the evolution of \Halpha\ EW vs \boxfil\ colours of the BPASS binary model galaxies with different metallicities. All BPASS models shown here have an IMF of slope $\Gamma=-1.35$ and a constant SFH. From top to bottom the metallicities of the tracks are Z=0.002, 0.004, 0.010, 0.020, and 0.040. The overlaid \sample\ has been dust corrected with a $f=1$.
{\bf Top right:} Similar to the top left panel but with the \sample\ dust corrected with a $f=2$.
{\bf Bottom left:}  Here we show the evolution of \Halpha\ EW vs \boxfil\ colours of PEGASE constant SFH models with varying IMFs and high mass cutoffs. From top to bottom each set of tracks with similar colour have IMf slopes $\Gamma=$ $-0.50, -1.00, -1.35$, and $-2.00$. Each set of IMFs are computed with varying high mass cutoffs. From top to bottom for each IMFs, the high mass cutoffs are respectively 120\msol, 110\msol, 100\msol, 90\msol, and 80\msol. The overlaid \sample\ has been dust corrected with a $f=1$.
{\bf Bottom right:} Similar to the bottom left panel but with the \sample\ dust corrected with a $f=2$.
}
\label{fig:Z_and_high_mass}
\end{figure*}

\citet{Hoversten2008} showed that the high mass cutoff is degenerated with IMF in the \Halpha\ EW vs \boxfil\ colour. In Figure \ref{fig:Z_and_high_mass} (bottom panels) we show various IMF slopes with constant SFHs computed for varying values of high mass cutoff. The deviation between tracks with 80\msol\ and 120\msol\ high mass cutoff varies as a function of IMF slope. Shallower IMFs will have a larger effect when the high mass cutoff is increased due to the high number of stars that will populate the high mass regions.

The maximum deviation for the high mass cutoffs for the $\Gamma=-1.35$ tracks is 0.17 dex, which cannot describe the scatter we notice in \Halpha\ EWs of our sample. Furthermore, at $z\sim2$ we expect the molecular clouds forming the stars to be of low metallicity \citep{Kacprzak2016}, which favours the formation of high mass stars. Therefore, we require the high mass cutoff to increase, but we are limited by the maximum individual stellar mass allowed by PEGASE. BPASSv2 does allow stars up to 300\msol, however, we do not employ such high mass limits due to our poor understanding of evolution of massive stars. We conclude that it is extremely unlikely that the high mass cutoff to have a strong influence on the distribution of the \sample\ galaxies in the \Halpha\ EW vs \boxfil\ colour parameter space.

\section{Dependencies with other observables}
\label{sec:other_observables}

In this section, we investigate if the \Halpha\ EW vs \boxfil\ colour distribution show any relationship with environment, stellar mass, SFR, and metallicity of the galaxies.

ZFIRE surveyed the \citet{Spitler2012} and \citet{Yuan2014} structure to great detail to probe the effects on environment on galaxy evolution. 
To date, there are 51 spectroscopically confirmed cluster candidates with ZFOURGE counterparts out of which 38 galaxies are included in our IMF analysis. The other 13 galaxies are removed from our analysis due to the following reasons: eight galaxies are flagged as AGN, two galaxies do not meet the $Q_z$ quality cut for our study, two galaxies give negative spectroscopic flux values and one object due to extreme sky line interference. 
We perform a 2-sample K-S test on the \Halpha\ EW values and \boxfil\ colours for the continuum detected cluster and field galaxies in our \sample\ and find the cluster and field samples to have similar parent properties. Therefore, we conclude that there are no strong environmental effects on the distribution of galaxies in our parameter space.


For the 22 continuum detected galaxies in common between \citet{Kacprzak2015} sample and \sample, we find no statistically significant differences between high and low metallicity samples for \Halpha\ EWs and \boxfil\ colours. 

\begin{figure*}
\includegraphics[scale=0.48]{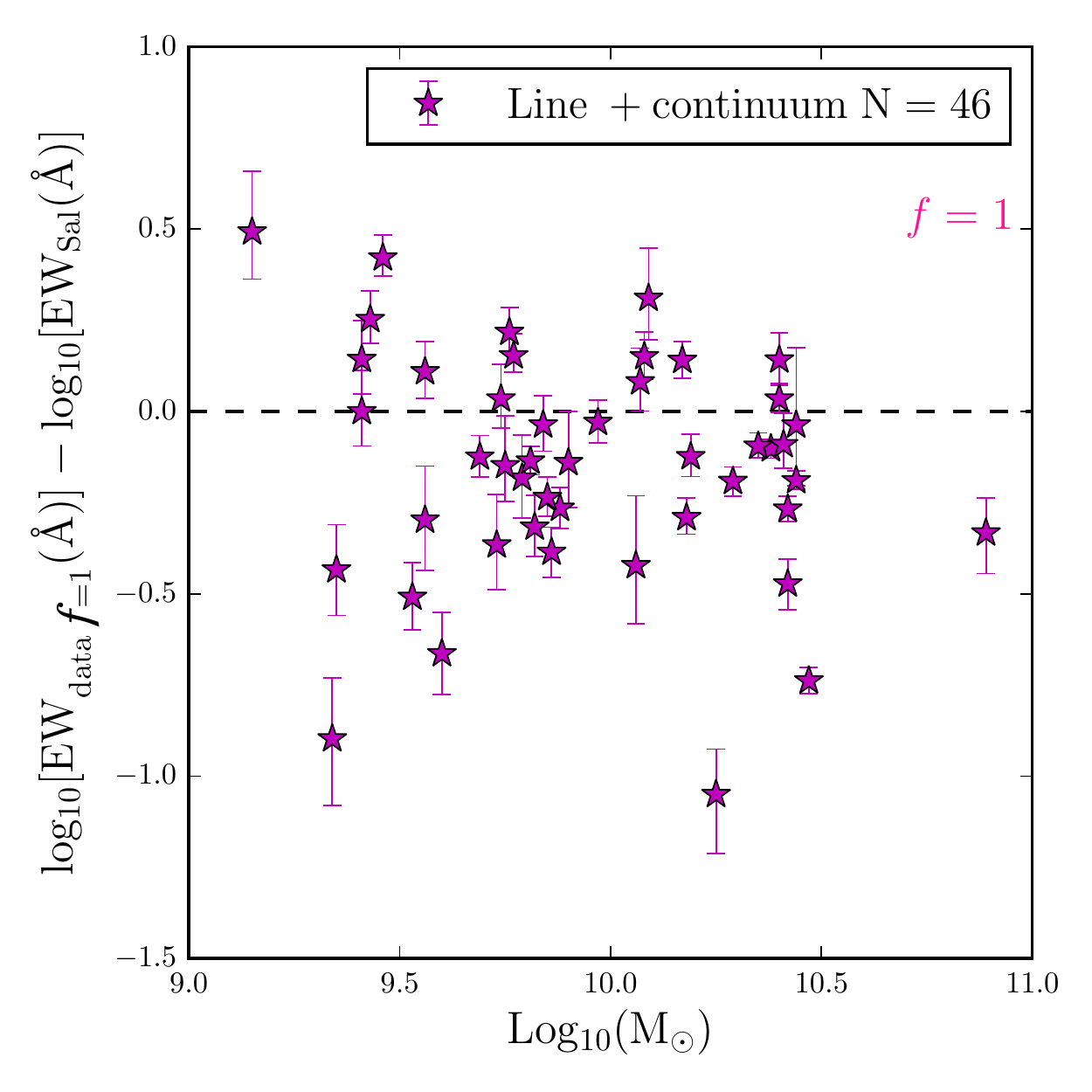}
\includegraphics[scale=0.48]{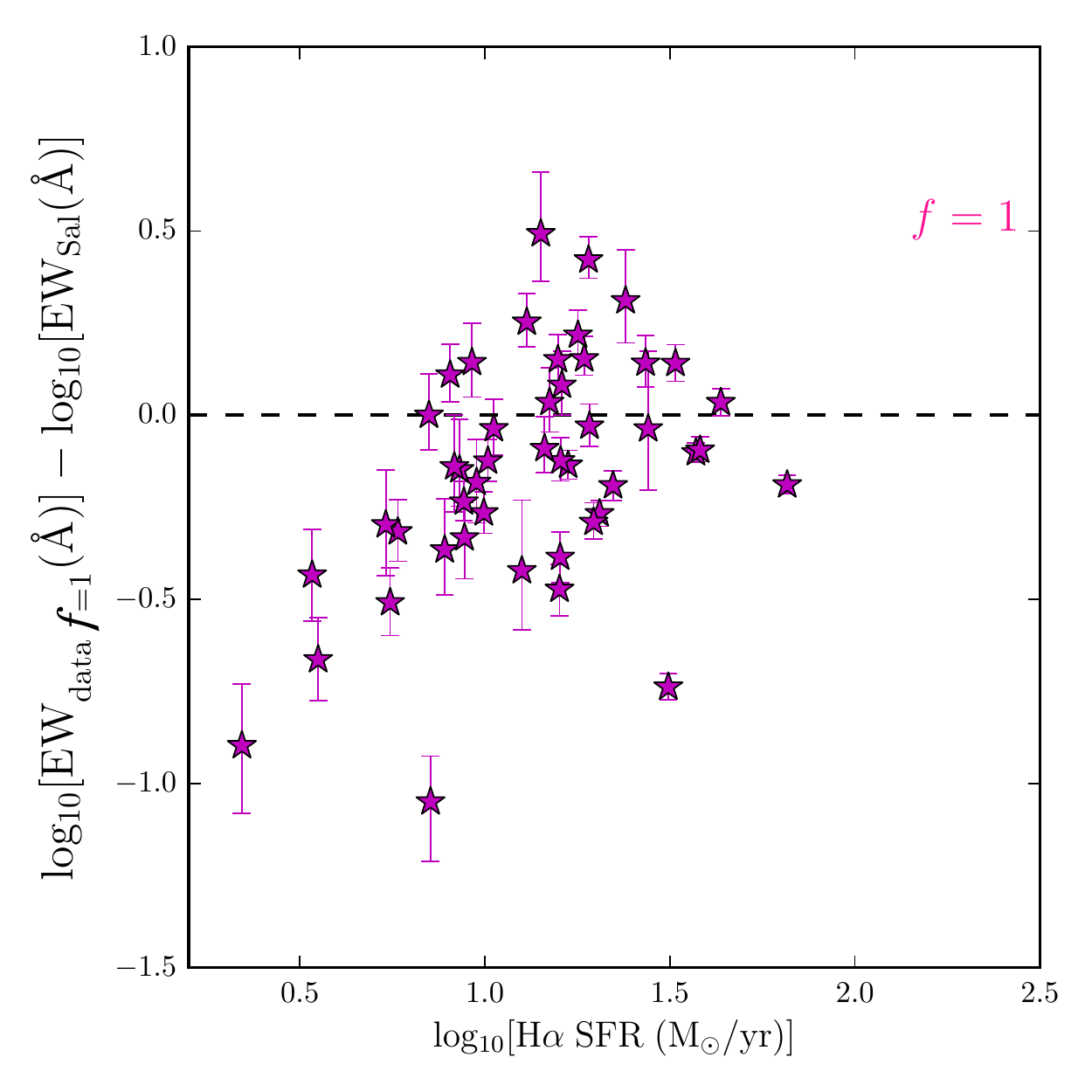}
\includegraphics[scale=0.48]{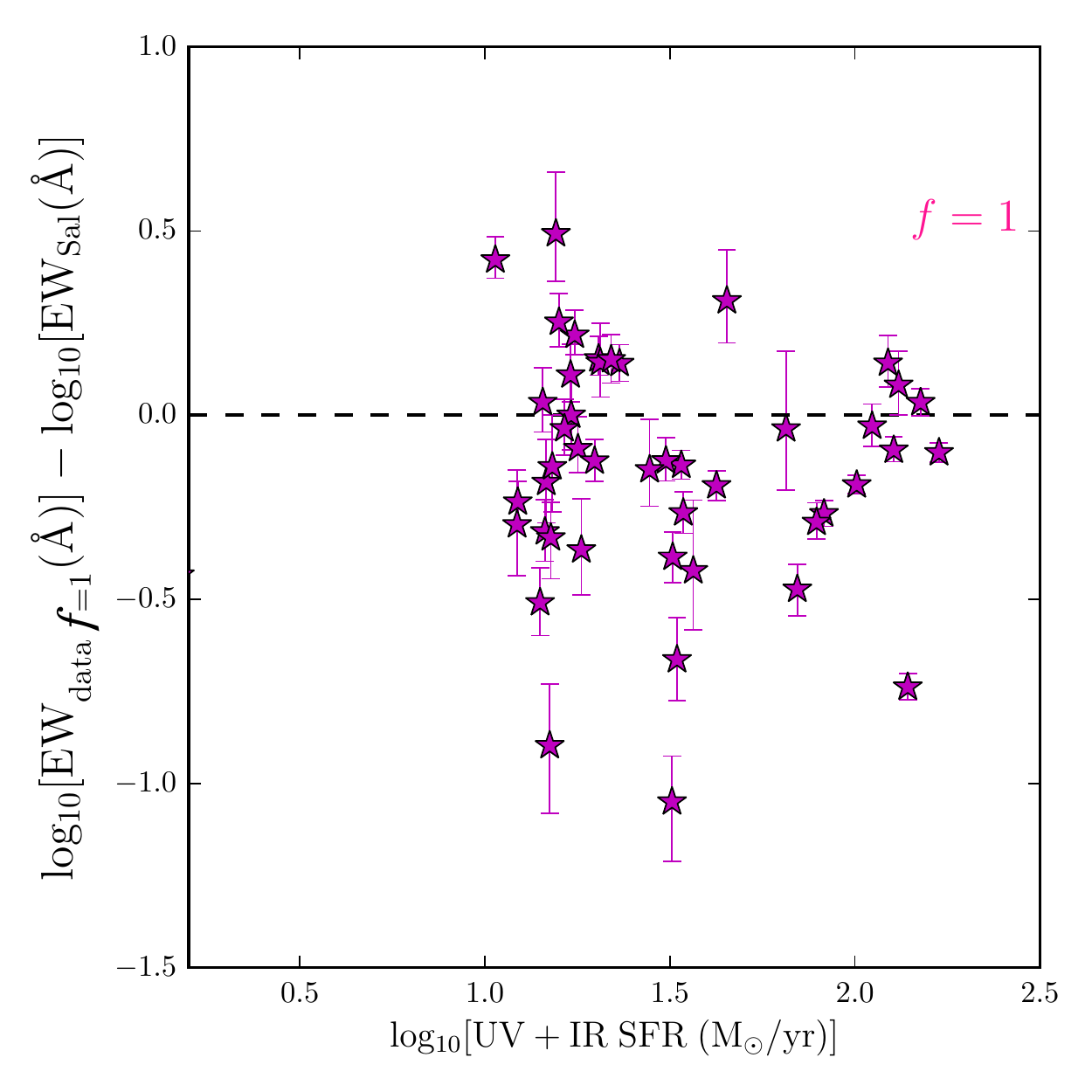}
\includegraphics[scale=0.48]{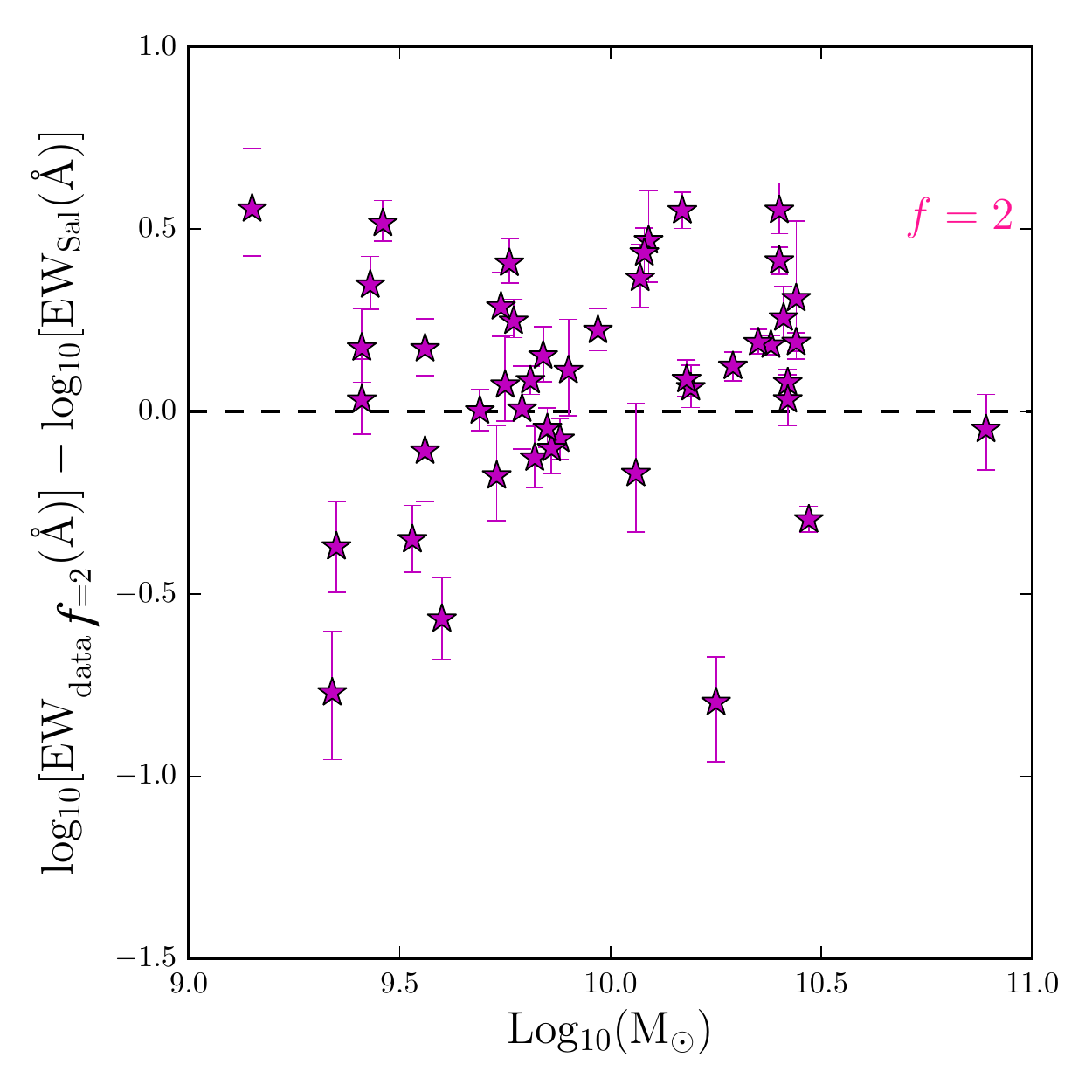}
\includegraphics[scale=0.48]{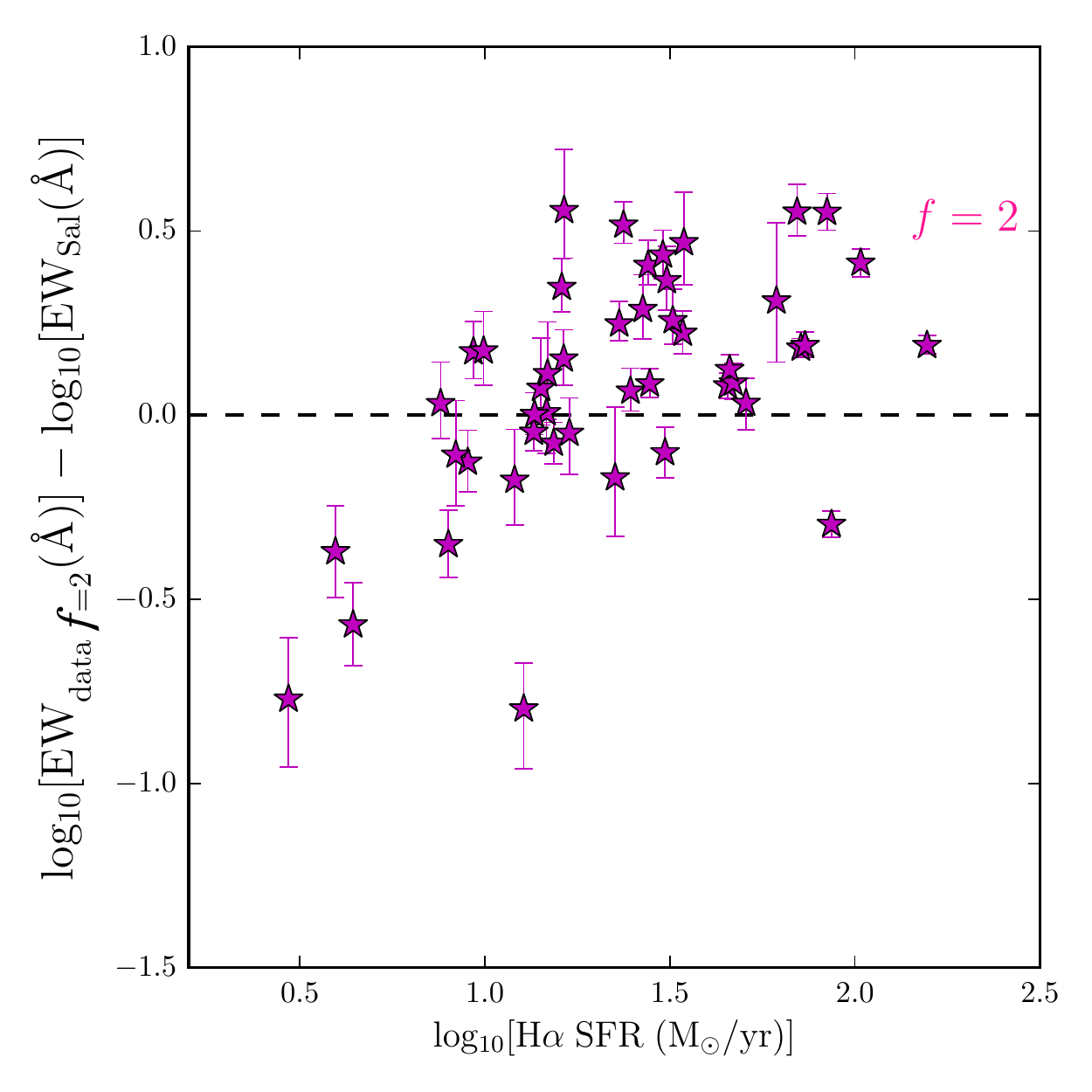}
\includegraphics[scale=0.48]{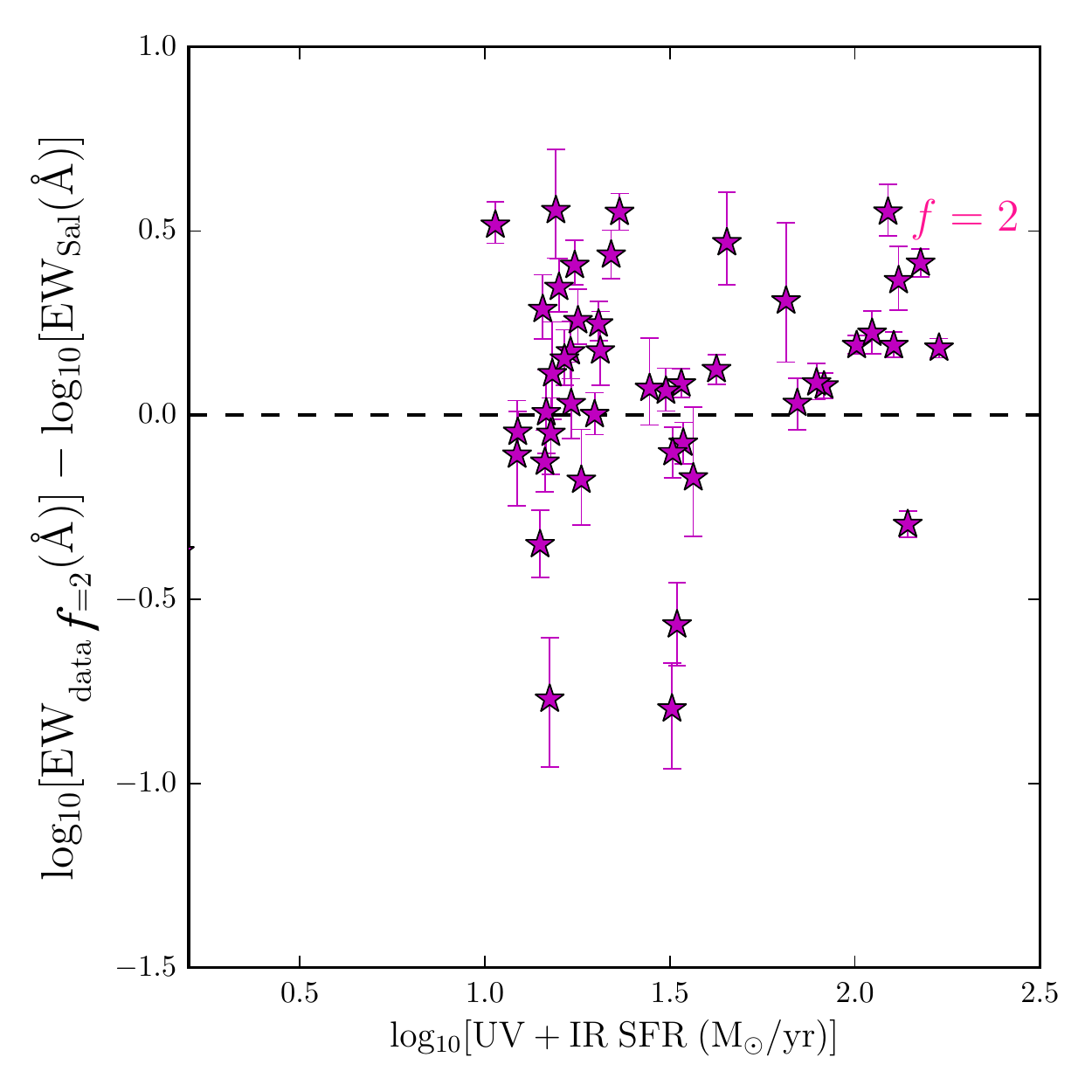}
\caption{ EW excess of the dust corrected (top panels $\rightarrow f=1$, bottom panels $\rightarrow f=2$) continuum detected sample  from a PEGASE track of $\Gamma=-1.35$ IMF slope with constant SFH.
{\bf Left panels:} EW excess as a function of stellar mass.
{\bf Centre panels:} EW excess as a function of \Halpha\ SFR. \Halpha\ SFR has been calculated using $f=1$ and $f=2$ in the top and bottom panels, respectively.
{\bf Right panels:} EW excess as a function of UV+IR SFR. 
In all panels the dashed line denotes y=0.
}
\label{fig:delta_EW_checks}
\end{figure*}

We further use the Salpeter IMF tracks with constant SFHs to compare the EW excess of our continuum detected sample with stellar mass and SFR. 
In Figure \ref{fig:delta_EW_checks} (left panels) we show the EW excess as a function of stellar mass. We divide the sample into low mass ($\mathrm{log_{10}(M_*/M_\odot)<10.0}$) and high mass ($\mathrm{log_{10}(M_*/M_\odot)\geq10.0}$) bins and compute the scatter in EW excess to find that there is a greater tendency for low mass galaxies to show larger scatter in EW offsets.

Looking for IMF change as a function of SFR is inherently problematic, esp. if SFR is itself computed from the \Halpha\ flux assuming a universal IMF. Nevertheless in order to compare with \citet{Gunawardhana2011} we show this in Figure \ref{fig:delta_EW_checks} (centre panels)  and confirm the trend they found of EW offset for higher ``SFR'' objects. However we refrain from interpreting this as a systematic trend for IMF variation. By using best fit SEDs from ZFOURGE, we compute the UV+IR SFRs \citep{Tomczak2014} and find that there is a greater tendency for low UV+IR SFR galaxies to show larger EW offsets, which is shown by Figure \ref{fig:delta_EW_checks} (right panels). We note that the difference in SFR between \Halpha\ and UV+IR is driven by the different time-scales of SFRs probed by the two methods.

\section{Discussion}
\label{sec:discussion}

\subsection{Comparison with local studies}
\label{sec:HG08_comp}

Our study follows a method first outlined by \citet{Kennicutt1983} and later implemented on large data sets by \citet{Hoversten2008} and \citet{Gunawardhana2011} to study the IMF of star-forming galaxies.  
We find that, the distribution of \Halpha\ EWs and optical colours at $z\sim2$ to be unlikely to be driven by a sample of galaxies with a universal Salpeter like IMF. 
\citet{Hoversten2008} found a trend with galaxy luminosity with low luminosity galaxies in SDSS  favouring a steeper IMF and the highest luminosity ones showing a Salpeter slope. \citet{Gunawardhana2011} found a systematic variance in IMF as a function of SFR in GAMA galaxies with the highest-SFR galaxies lying above the Salpeter track. However, we note that the use of \gr\ colour by the $z\sim0$ studies may have given rise to additional complexities in the analysis, by introducing significant emission line contamination, and the use of SFR as a variable in IMF change is problematic as its calculation 
 depends on IMF and \Halpha\ luminosity.

Comparing our results with the local galaxies of HG08 show distinctive differences in the \Halpha\ EW vs \gr\ colour distribution.
Since galaxies at $z\sim2$ had only $\sim3.1$ Gyr to evolve, we observe  younger, bluer stellar populations giving rise to tighter \gr\ colours (distributed around 0.082 mag with a standard deviation of 0.085 mag). However, HG08 galaxy sample comprise of much redder colours with a larger scatter in \gr.  In a smooth star-formation scenario, we interpret the large scatter of the HG08 sample in \gr\ colour space to be driven by the large variety of ages of the galaxies.

Galaxies at $z\sim2$ show a large range in \Halpha\ EWs compared to $z\sim0$ results. In our analysis, we investigated several key factors which may contribute to the large scatter of \Halpha\ EW at $z\sim2$. 
Compared to $z\sim0$ galaxy populations, we expect galaxies at $z\sim2$ to be young, actively star-forming in various environments and physical conditions that may be distinctively different from local conditions. 
Therefore, effects such as starbursts may be prominent and dust properties may have significant variation, which can influence the observed \Halpha\ EW values. Galaxy mergers and multiple starburst phases in the evolutionary history of $z\sim0$ galaxies add additional layers of complexity. Furthermore, the presence of old stellar populations requires \Halpha\ absorption to be corrected, which we expect to be negligible at $z\sim2$. 
Due to the limited evolutionary time-scale at $z\sim2$ (only 3 Gyr), we consider most of these effects to have no significant influence on our analysis. 
However, we cannot completely rule out effects of dust sight-lines to our analysis, which we discuss further in Section \ref{sec:dust_discussion}.

The development of much advanced stellar tracks and greater understanding of stellar properties allow us to explore uncertainties related to stellar modelling that may significantly influence the observed parameters of galaxies at $z\sim2$.

\subsection{What do we really find?}

In Section \ref{sec:observational_bias}, we showed that our ZFIRE selected sample was not preferentially biased towards extremely high star-forming galaxies and that our mass-complete $z\sim 2$ sample is sensitive to quite low EWs. Since observational bias appears not to be the explanation we
can investigate physical factors that drive the difference in the  \sample\ from universal Salpeter like IMF scenarios in the \Halpha\ EW vs \boxfil\ colour plane.

\Halpha\ flux is a direct probe of the SFR on time-scales of $\sim10$ Myr, the continuum level at 6563\AA\ provides an estimate of the mass of the old stellar populations, and therefore, \Halpha\ EW is a proxy for the sSFR. Similarly, for monotonic SFHs, the optical colours change smoothly with time, so the \boxfil\ colour 
is a second proxy for the sSFR, but with different IMF sensitivity. Of the two sSFR measures the \Halpha\ EW is the most sensitive to the highest mass stars,
so one way to express our result is to state that there is an excess of ionising photons (i.e. \Halpha) at a given SFR compared to a Salpeter-slope model.

In our sample, with $f=2$ dust corrections, \around50\%  galaxies have an excess of high mass stars for a given sSFR compared to the expectation by a Salpeter like IMF. 
By stacking galaxies in mass and \boxfil\ colour bins, we can average out stochastic variations in SFHs between galaxies. Our stacking results further confirmed that on average, for all masses and sSFR values, a universal IMF cannot produce the observed galaxy distribution in the \Halpha\ EW vs \boxfil\ colour space. 
We performed further analysis to understand other mechanisms which may drive this excess in \Halpha\ EW for a given sSFR.

\subsection{Dust and starbursts}
\label{sec:dust_discussion}

The dust extinction values in our analysis were derived using FAST, which uses underlying assumptions of IMF and SFH to produce best fit stellar parameters to galaxy observables. 
Our own analysis of dust showed SED derived extinction values from the UV slope to have a strong dependence on the assumed IMF. However for the purposes of testing consistency with a universal Salpeter-slope this suffices.

We further found that differential extinction in dust between the stellar continuum and nebular emission line regions can introduce significant scatter to galaxies in the \Halpha\ EW vs \boxfil\ colour parameter space. By analysing Balmer decrement values for a subset of galaxies in our sample, we found that, there was significant scatter in the relation between extinction of nebular and stellar continuum regions ($f$), which can be attributed to differences in dust sight-lines between galaxies. 
\citet{Reddy2015} showed that this scatter in extinction to be a function of \Halpha\ SFR, where galaxies with higher star-forming activity shows larger nebular extinction compared to galaxies with low SFRs. 
We test this by allowing $f$ values to vary as a free parameter in the \Halpha\ EW vs \boxfil\ colour space for each galaxy to force agreement with a universal Salpeter like IMF. Our results showed extreme values for the distribution of $f$, including unphysical negative values, suggesting that it is extremely unlikely that the scatter in the \Halpha\ EW vs \boxfil\ colour space is driven solely by the variation of $f$ values.

Starbursts in galaxies can introduce significant scatter in the \Halpha\ EW vs \boxfil\ colour space. We implemented a stacking procedure for the galaxies in mass and colour bins to remove stochastic SFHs of individual galaxies, treating them as an ensemble stellar population with a smooth SFH prior to $z\sim 2$. We found that, our stacks on average ($50\%$ of the $f=1$ stacks and $100\%$ of the $f=2$ stacks) favour shallower IMF slopes compared to the traditional $\Gamma =-1.35$ values from Salpeter. By performing Monte Carlo simulations of starbursts using PEGASE SSP models we found that time-scales of bursts makes it extremely unlikely for them to account for the galaxies which lie significantly above the $\Gamma=-1.35$ track.

\subsection{Dependencies on SSP models and stellar libraries}
\label{sec:ssp_issues}

We compared the evolution of model galaxies in the \Halpha\ EW vs \boxfil\ colours using PEGASE and Starburst99 SSP codes to conclude that the evolution of these parameters are largely independent of the SSP models used for a given stellar library. The \boxfil\ colours were designed in order to avoid strong emission lines regions in the rest frame optical spectra which averts the need of complicated photo-ionization codes to generate nebular emission lines. \Halpha\ flux is generated using a constant value to convert Lyman continuum photons to \Halpha\ photons, which is similar between PEGASE and S99.

We found that, stellar libraries play a vital role in determining the evolutionary tracks of galaxies in the \Halpha\ EW vs \boxfil\ colour parameter space. Stellar libraries with rotation show higher amounts of ionizing flux which results in higher \Halpha\ EW values for a given \boxfil\ colour. 
\cite{Leitherer2014} further showed that rotation  leads to larger convective cores in stars increasing the total bolometric luminosity, which can mimic a shallower IMF. 
At \boxfil= 0.61, introducing stellar rotation via Geneva stellar tracks with Z=0.014 results in $\Delta\mathrm{log_{10}[EW (log_{10}(\AA))]\sim0.09}$. Therefore, we found that rotation cannot itself account for the scatter of our sample in \Halpha\ EW vs \boxfil\ colour parameter space at near solar metallicities.

Consideration of binary stellar systems is imperative to understand the stellar properties of $z\sim2$ galaxies \citep{Steidel2016}. 
However, added complexity arises due to angular momentum transfer during binary star interactions. This may influence the rotation of the galaxies and therefore it is necessary to consider the evolution of binary stars with detailed prescriptions of stellar rotation. 
Metallicity of the stars become important in such scenarios, which is a strong factor that regulates the evolution of stellar rotation. 
However, adding additional degrees of freedom for SSP models makes it harder to constrain their values, thus resulting in extra uncertainties \citep{Leitherer2014}. 
At \boxfil= 0.61, introducing the effect of binaries via BPASS models resulted in $\Delta\mathrm{log_{10}[EW (log_{10}(\AA))]\sim0.01}$. 
Comparing results between S99 (single stellar population stellar tracks with and without rotation) and BPASS (single and binary stellar tracks with rotation), we found stellar rotation to have a larger contribution to the $\Delta$EW compared to binaries. 
Direct comparisons require further work to investigate differences in the evolution of stellar systems between S99 and BPASS SSP codes.

We found low stellar metallicities (Z$\sim$0.002) to have a strong influence in increasing the  \Halpha\ EWs for a given \boxfil\ colour. 
At \boxfil= 0.61, reducing the metallicity of BPASS binary models from Z=0.02 to Z=0.002 resulted in  $\Delta\mathrm{log_{10}[EW (log_{10}(\AA))]\sim0.36}$. 
This was largely driven by the increase in the number of ionization photons in the stellar populations due to lower opacities, lower mass loss via stellar winds, and sustained stellar rotation. 
Interactions between stars also contribute to an increase in ionising flux. 
When considering the ionization energy generated by a stellar population, effects of stellar rotation is degenerated with the abundance of high mass stars (see Figure 16 of \citet{Szecsi2015}). 
Therefore, we cannot completely rule out effects of stars with extremely low metallicities to describe the distribution of our galaxies in the \Halpha\ EW vs \boxfil\ colour parameter space.
In Section \ref{sec:model_Z}, we provided a thorough analysis of the gas phase metallicities derived for the ZFIRE sample by \citet{Kacprzak2015} and \citet{Kacprzak2016} and inferred the metal abundances of stellar systems, which is a primary regulator of ionising photons. 
However, uncertainties in deriving gas phase abundances of elements via nebular emission line ratios (driven by our limited understanding of the ionization parameter at low metallicities), uncertainties in computing relative abundances of $\alpha$ elements in stellar systems, and our limited understanding on linking gas phase metallicities to stellar metallicities in $z\sim2$ stellar populations constrains our ability to distinguish between effects of metallicity and IMF.

\subsection{Case for the IMF}
\label{sec:IMF_discussion}

So far we have investigated various scenarios (summarised in Table \ref{tab:summary_table}) that could explain the distribution of the \sample\ galaxies in the \Halpha\ EW vs \boxfil\ colour without invoking changes in the IMF. However, none of the scenarios by itself could best describe the distribution of our galaxies.

The galaxies in our sample have stellar masses between $\log_{10}(M_*/M_\odot)=9-10$ and we expect these galaxies to grow in stellar mass during cosmic time to be galaxies with stellar masses of $\sim\log_{10}(M_*/M_\odot)=10-11$ at $z\sim0$ \citep{DeLucia2007,vanDokkum2013b,Genel2014,Papovich2015}. 
Recent studies of ETGs with physically motivated models have shown the possibility for a two phase star-formation \citep[eg.,][]{Ferreras2015}. Furthermore recent semi-analytic models have shown that a varying IMF  best reproduces observed galaxy chemical abundances of ETGs \citep[eg.,][and references therein]{Lacey2016,Fontanot2017}. According to these models, ETGs, during their starburst phases at high-redshift are expected to produce higher fraction of high mass stars (shallower IMFs). \citet{Gunawardhana2011} showed that $z\sim0$ star-forming galaxies also show an IMF dependence, where highly star-forming galaxies prefer shallower IMFs.

If we consider a varying IMF hypothesis, our results are consistent with a scenario where star-forming galaxies form stars with a high fraction of high mass stars compared to their local ETG counterparts. With lower metallicities and higher SFRs prominent at $z\sim2$, we expect the fragmentation of molecular clouds to favour the formation of larger stars due to lower cooling efficiencies and higher heating efficiencies due to radiation from the young massive stars \citep{Larson2005}. \citet{Krumholz2010} showed that radiation trapping prominent in high star-forming regions of dense gas surface density can also favour the formation of massive stars. 
If we allow the IMF to vary in our analysis, the distribution of the \sample\ in \Halpha\ EW vs \boxfil\ colour space can be explained, however values as shallow as $\Gamma=0.5$ could be required. This could be problematic for chemical evolution models and have implications to how galaxies form and evolve \citep{Romano2005}. 
We note that invoking extremely shallow IMFs can have a significant influence on the inferred evolution of the universe. Therefore, it is imperative to fully understand these observations and test alternate explanations.

\subsection{Effect of IMF variation on fundamental quantities} 

If the IMF does vary, we need to consider the potential effect on the basic parameters in our input ZFIRE survey, which were calculated using a Chabrier IMF \citep{Chabrier2003}. 
First we consider possible effects on the calculation of our rest frame \boxfil\ colours. This should not have a significant effect for several reasons: first we are only
using the spectral models as an interpolator, and by design we are interpolating only across a small redshift range. At $z=2.1$ the interpolated and observed colours
agree well as discussed in Appendix B. Second we note that the main effect is an increased scatter in the EW axis (Figure \ref{fig:EW_HG08_comp}), once dust corrected the colours are
quite tight.  Finally we note that at these young ages everything is quite blue, hence quite flat spectra are being interpolated.

Next is the effect on SFR and stellar mass, which have been used in many of the previous ZFIRE papers \citep{Yuan2014,Tran2015,Kacprzak2015,Alcorn2016,Kacprzak2016,Kewley2016,Nanayakkara2016}, and here in our own mass selection. To quantify the change
in mass we run PEGASE  for $\Gamma$ and constant star-formation history models and estimate the change in $R$-band mass-to-light ratio ($\simeq$K-band at
$z\simeq 2$) for ages 1--3 Gyr. We find for $-1.35<\Gamma<-0.5$ the change in mass-to-light is $<0.7$ dex, with shallower IMFs resulting in a lower stellar mass.
Thus we conclude that our stellar mass selection is only slightly effected by the possible IMF variations we have identified.

The effect is much more severe for \Halpha\ derived star-formation rates \citep{Tran2015,Tran2017} as these directly count the number of the most massive stars, a sensitivity we have exploited
in this paper to measure IMF. For $-1.35<\Gamma<-0.5$ the change is $\sim1.3$ dex. 
UV and far-IR derived SFRs are more complicated. The rest frame UV is more sensitive to intermediate mass stars, at 1500\AA\ the change in flux is $\sim0.4$ dex
for $-1.35<\Gamma<-0.5$. The far-IR is from younger stars in deeper dust-enshrouded regions, at least in local galaxies \citep{Kennicutt1998}. It is common at high-redshift to use an indicator, which combines UV and far-IR \citep[eg.,][]{Tomczak2014}. These are often calibrated using stellar population models with idealized SFHs, and traditional IMFs  and for a fixed dust mass the balance between UV and IR luminosities will depend on dust geometry, IMF, and SFH \citep{Kennicutt1998,Calzetti2013}. Therefore, IMF change could lead to difficulties in predicting the true underlying SFR of stellar populations.

\begin{turnpage}
\begin{deluxetable*}{  l  l  l  l  l   l   l  l   l  l l}
\tabletypesize{\scriptsize}
\tablecolumns{3}
\tablewidth{0pt} 
\tablecaption{Summary of scenarios investigated to explain the distribution of the \sample\ in the \Halpha\ EW vs \boxfil\ colour parameter space within a universal IMF framework.
\label{tab:summary_table}}
\tablehead{\colhead{Effect}                                    &
           \colhead{Section}                             	   &
           \colhead{SSP model}                             	   &
           \colhead{SFH}                             	   	   &
           \colhead{Z}                             	           &
           \multicolumn{2}{c}{Median($\Delta$EW)}              &
           \multicolumn{2}{c}{\%\tablenotemark{a}}     	   	   &
           \colhead{Conclusion}                                \\
           \colhead{}                                          & 
           \colhead{}                                          & 
           \colhead{}                                          & 
           \colhead{}                                          & 
           \colhead{}                                          &        
           \colhead{$f=1$}                                     &
           \colhead{$f=2$} 									   &
           \colhead{$f=1$}                                     &
           \colhead{$f=2$} 									   &
           \colhead{}                                          & 
           }
\startdata
& \\
{\bf Dust} 		& \ref{sec:dust} & PEGASE & Exp declining ($\tau=1000$Myr)	& 0.020	& -0.13 & 0.10  & 20\% & 46\% &  unlikely\tablenotemark{b} \\
{\bf Observational bias} & \ref{sec:observational_bias} & PEGASE & Exp declining ($\tau=1000$Myr) & 0.020 & $--$ & $--$ & $--$ & $--$ &   excluded \\ 
{\bf Star bursts} & \ref{sec:star_bursts} & PEGASE & Constant & 0.020 & $--$ & $--$ & $--$ & $--$ &   excluded \\ 
{\bf Stellar rotation} & \ref{sec:stellar_rotation} & S99 & Constant & 0.020 & -0.39 & -0.15 & 2\% & 13\%  &  probable\tablenotemark{c} \\
{\bf Binaries} & \ref{sec:binaries} & BPASS & Constant & 0.020 & -0.18 & 0.05 & 9\%  & 39\% &  future work\tablenotemark{c} \\
\sidehead{\bf Metallicity}
Single stellar systems &  \ref{sec:model_Z} & BPASS & Constant & 0.020 & -0.17 & 0.06 & 13\% & 37\%  & unlikely \\
(with rotation)		   &   					&  		& Constant & 0.010 & -0.29 & -0.06 & 4\% &  22\% & probable \\ 
   					   &					&  		& Constant & 0.002 & -0.47 & -0.23 & 0\% &  6\%  & probable \\

Binary stellar systems & \ref{sec:model_Z} & BPASS & Constant & 0.020 & -0.18  & 0.05 & 9\% & 39\% &  unlikely\tablenotemark{c} \\
(with rotation)		   &  				   &  	   & Constant & 0.010 & -0.32 & -0.08 & 4\% & 22\% &  probable\tablenotemark{c} \\
					   & 				   &  	   & Constant & 0.002 & -0.51 & -0.28 & 2\% & 9\% &   probable\tablenotemark{c} \\
\sidehead{\bf High mass cutoff}
{80\msol} & \ref{sec:mass_cutoff} & PEGASE & Constant  & 0.020 & -0.01 & 0.22 & 28\% & 54\% &   excluded\tablenotemark{c} \\
{120\msol} & \ref{sec:mass_cutoff} & PEGASE & Constant & 0.020 & -0.14 & 0.09 & 17\% & 39\% &   excluded\tablenotemark{c} \\
\tablenotetext{a}{The fraction of \sample\ galaxies with $>2\sigma$ positive deviations from the $\Gamma=-1.35$ tracks. 
}
\tablenotetext{b}{Even though we cannot exclude effects from various dust sight-lines, we demonstrated that effects from dust cannot explain the excess of high \Halpha\ EW galaxies.}
 \tablenotetext{c}{Conclusions driven within the bounds of current SSP models, however, more sophisticated models are required on stellar rotation, binary evolution, and high mass evolution to fully constrain the effects.}
\end{deluxetable*}
\end{turnpage}


\section{Summary \& Future Work}
\label{sec:summary}

We have used data from the ZFIRE survey along with the multi-wavelength photometric data from ZFOURGE to study properties of a sample of star-forming galaxies in cluster and filed environments at $z\sim2$. By using the \Halpha\ EW and rest frame optical colours of the galaxies we performed a thorough analysis to understand what physical properties could drive the distribution of galaxies in this parameter space. We have improved on earlier analysis by deriving synthetic rest frame
filters that remove emission line contamination. We analysed effects from dust, starbursts, metallicity, stellar rotation, and binary stars in order to investigate whether the distribution of the \sample\ galaxies could be explained within a universal IMF framework.\\
 We found the following:\\
\begin{itemize}
\item  \sample\ galaxies show a large range of \Halpha\ EW values, with $\sim1/3$rd of the sample showing extremely high values compared to expectation from a $\Gamma=-1.35$ Salpeter like IMF. Compared to the HG08 SDSS sample, galaxies at $z\sim2$ show bluer colours with a larger scatter in \Halpha\ EW values.

\item  The difference in extinction between nebular and stellar emission line regions ($f$) in galaxies can have a strong influence in determining the distribution of galaxies in the \Halpha\ EW vs \boxfil\ colour space. Our Balmer decrement studies for a sub-sample of galaxies showed a large scatter in $f$ values. However, we showed that considering $f$ value as a free parameter cannot describe the distribution of galaxies in the \Halpha\ EW vs \boxfil\ colour space.

\item  Starbursts can increase the \Halpha\ EW to extreme values providing an alternative explanation to IMF for a subset of our galaxies with high \Halpha\ EW values. By implementing a stacking technique to remove stochastic SFHs of individual galaxies we concluded that on average our \sample\ still shows higher \Halpha\ EW values for a given \boxfil\ colour compared to a $\Gamma=-1.35$ Salpeter like IMF. We further used Monte Carlo simulations to study time-scales of starbursts to conclude that it was extremely unlikely that starbursts could explain the \Halpha\ EW vs \boxfil\ colour distribution of a large fraction of our galaxies. 

\item  Stellar rotation, binaries, and the high mass cutoff of SSP models could influence the distribution of galaxies in the \Halpha\ EW vs \boxfil\ colour parameter space. However, the individual effects of these were not sufficient to explain the distribution of the observed galaxies.

\item Considering multiple effects together can describe the galaxies in our parameter space. We showed that the fraction of galaxies above the $\Gamma=-1.35$ tracks reduces to $\sim5\%$ when considering stellar tracks with high initial rotations (\vini=0.4\vcrit) and equal dust extinction between nebular and stellar regions.

\item Including single or binary stars with stellar rotation in extreme low metallicity  scenarios can significantly increase the \Halpha\ EWs and is also one explanation to describe the distribution of our galaxies in \Halpha\ EW vs \boxfil\ colour parameter space. However, gas phase metallicity analysis of the ZFIRE sample by \citet{Kacprzak2015} and \citet{Kacprzak2016} rules our such low metallicities for our sample. 
We note that calibration of emission line ratios and differences between stellar and ionized gas metallicities at $z\sim2$ are uncertainties that may impact our inference about the stellar metallicity of our sample.  


\item  A non-universal high-mass IMF, varying between galaxies, could explain the distribution of galaxies in this parameter space. The \Halpha\ excess shows a broad trend with larger offsets for the less massive $z\sim 2$ galaxies.
We also confirm the same systematic trend in IMF slope with Chabrier-derived SFR as shown by \citet{Gunawardhana2011} but we refrain from interpreting this. 

\item Within the scope of our study, for $-1.35<\Gamma<-0.5$ the variation in high-mass IMF slope can lead to changes in mass-to-light ratios of up to $\sim0.7$ dex. Furthermore, ignoring calibration offsets we compute that \Halpha\ SFRs can show deviations up to $\sim1.3$ dex. 

 \end{itemize}

IMF change is an important topic as the IMF determines basic parameters such as stellar mass and star-formation rate, which are used to derive broad conclusions
about galaxy evolution. What we observe is a population of galaxies with high \Halpha\ equivalent widths, i.e. an excess of ionising photons for a given colour, 
and we have ruled out intermittent starbursts and alternate stellar population models as an explanation. Such high-EW objects appear to become more common at high-redshift, for example similar observations have been reported at $z\sim4$ by multiple studies \citep[eg.,][]{Malhotra2002,Finkelstein2011b,McLinden2011,Hashimoto2013,Stark2013} and have even been invoked at $z>5$ as an explanation for cosmological re-ionisation \citep{Labbe2013,Labbe2015,Schenker2015_thesis,Stark2017}. It seems reasonable to hypothesize that the abundance of high-EW objects is evolving towards high-redshift and we are seeing this at $z\sim 2$.

Is  IMF change responsible? This currently seems to be the only explanation that cannot be ruled out, but we do not yet understand what would drive it to vary between individual galaxies. Further study is required in order to fully comprehend the stellar population parameters of the $z\sim2$ galaxies to determine whether IMF is the main driver for the distribution of galaxies in the \Halpha\ EW vs rest frame optical colour parameter space. 

Future work should consider a more thorough statistical analysis using all the broad-band colour information and multiple spectral diagnostics including simultaneously modelling of possible effects of dust, starbursts, metallicity, stellar rotation, binary star evolution, and high mass cutoff of stellar systems together with systematic variances of the IMF.
A new generation of stellar models are allowing many of these parameters to be varied and tested.
The launch of the James Webb Space Telescope 
in 2018 will provide the opportunity to probe rest frame optical stellar and near-infrared populations via high signal/noise absorption lines
and will revolutionise our understanding of the processes of star-formation in the $z\sim 2$ universe.

 \acknowledgments
The data presented herein were obtained at the W.M. Keck Observatory, which is operated as a scientific partnership among the California Institute of Technology, the University of California and the National Aeronautics and Space Administration. The Observatory was made possible by the generous financial support of the W.M. Keck Foundation.
The authors wish to recognize and acknowledge the very significant cultural role and reverence that the summit of Mauna Kea has always had within the indigenous Hawaiian community.  We are most fortunate to have the opportunity to conduct observations from this mountain and without the generous hospitality of the indigenous Hawaiian community this study would have not been possible. T.N., K.G., and G.G.K. acknowledge Swinburne-Caltech collaborative Keck time, without which this study would have not been possible.
We thank the anonymous referee for the constructive comments on our analysis. 
We thank Madusha Gunawardhana, Elisabete da Cunha, Richard McDermid, Luca Cortese, John Eldridge, and Selma de Mink for insightful discussions. We thank Colin Jacobs for writing a Python wrapper around PEGASE.2, which was instrumental to perform the SSP simulations. We thank Erik Hoversten for providing us with the SDSS data used in our analysis. We thank Naveen Reddy for providing us with the MOSDEF data used in his \citet{Reddy2015} analysis. 
We thank the Lorentz Centre and the scientific organizers of the ``The Universal Problem of the Non-Universal IMF'' workshop held at the Loretnz Centre in December 2016, which promoted useful discussions among the wider community on a timely concept. 
K.G. acknowledges the support of the Australian Research Council through Discovery Proposal awards DP1094370, DP130101460, and DP130101667. G.G.K. acknowledges the support of the Australian Research Council through the award of a Future Fellowship (FT140100933).
\\

Facilities: \facility{Keck:I (MOSFIRE)}

\bibliographystyle{apj}
\bibliography{bibliography.bib}



\appendix


\section{Robustness of continuum fit and \Halpha\ flux}
\label{sec:cont_Halpha_test}

The study presented in this paper relies on accurate computation of \Halpha\ flux and the underlying continuum level. In this section, we compare our computed continuum level and \Halpha\ flux values using two independent methods to investigate any systematic biases to our \Halpha\ EW values.

We examine the robustness of our continuum fit to the galaxies by using ZFOURGE imaging data to estimate a continuum level from photometry. 
By using a slit-box of size 0.7$''\times 2.8''$ overlaid on the $0.7''$ point spread function convolved FourStar Ks image, we calculate the photometric flux expected from the galaxy within the finite slit aperture.   Justification of this slit-size comes from the spectrophotometric calibration of the ZFIRE data which is explained in detail in \citet{Nanayakkara2016}. Since we remove slits that contain multiple galaxies within the selected aperture, only 38 continuum detected galaxies and 39 galaxies with continuum limits are used in this comparison. 
\begin{subequations}
We then convert the magnitude to $f_\lambda$ as follows:
\begin{equation}
f_\lambda = 10^{-0.4(mag +48.60)} \times \frac{3\times10^{-2}}{\lambda_{c}^2}\ \mathrm{erg/s/cm^2/\AA}
\end{equation}
where $\lambda_c$ is the central wavelength of the MOSFIRE K band which we set to 21757.5\AA. 
Next we compute the \Halpha\ flux contribution to $f_\lambda$ by using the photometric bandwidth of FOURSTAR Ks band ($\Delta \lambda_{FS}$=3300\AA). 
\begin{equation}
F_{H\alpha_{cont}} = \frac{F_{H\alpha}}{\Delta \lambda_{FS}}
\end{equation}
We then remove the \Halpha\ flux contribution to the photometric flux to compute the inferred continuum level from photometry as shown below:
\begin{equation}
F_{cont_{photo}} = f_\lambda-F_{H\alpha_{cont}}
\end{equation}
Since the \Halpha\ flux is the dominant emission line for star-forming non-AGN galaxies, we ignore any contributions from other nebular emission lines to the photometric continuum level. 
\end{subequations}

\begin{figure}
\includegraphics[trim  = 10 10 0 10, clip, scale=0.9]{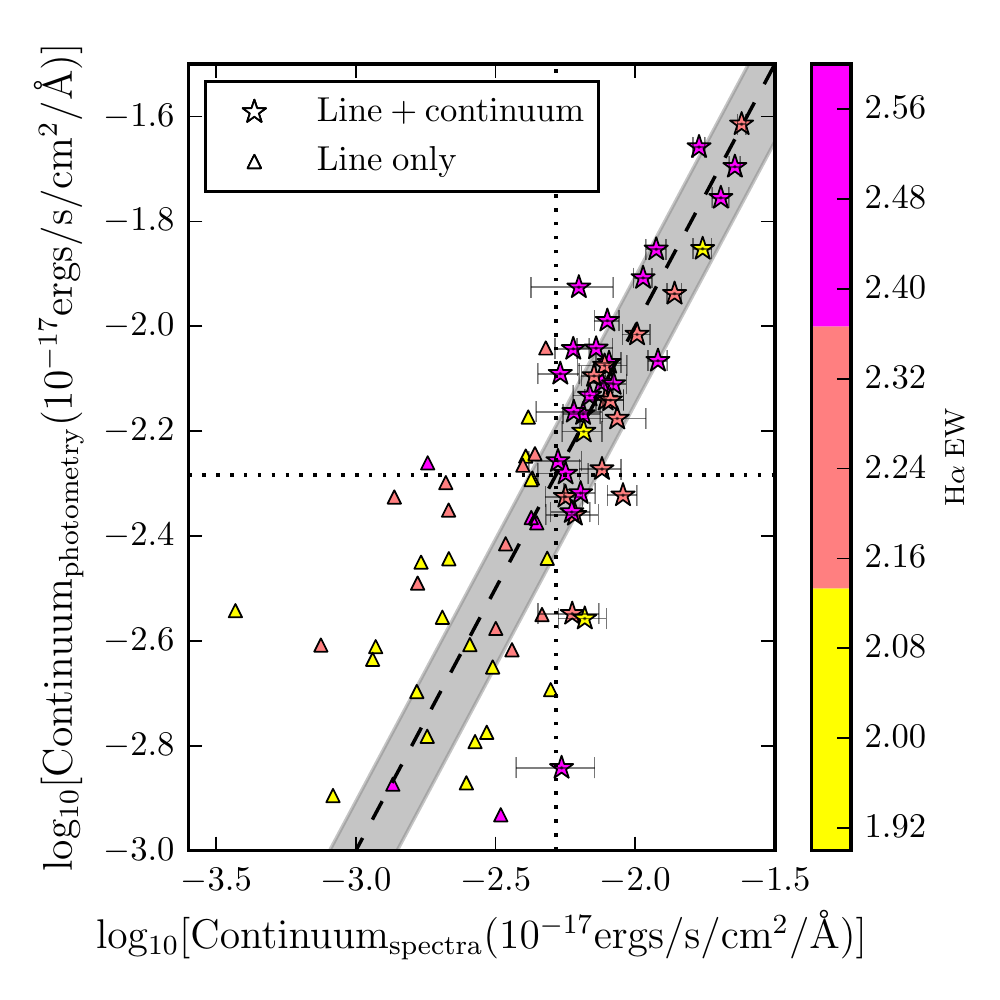}
\includegraphics[trim  = 0 10 10 10, clip, scale=0.9]{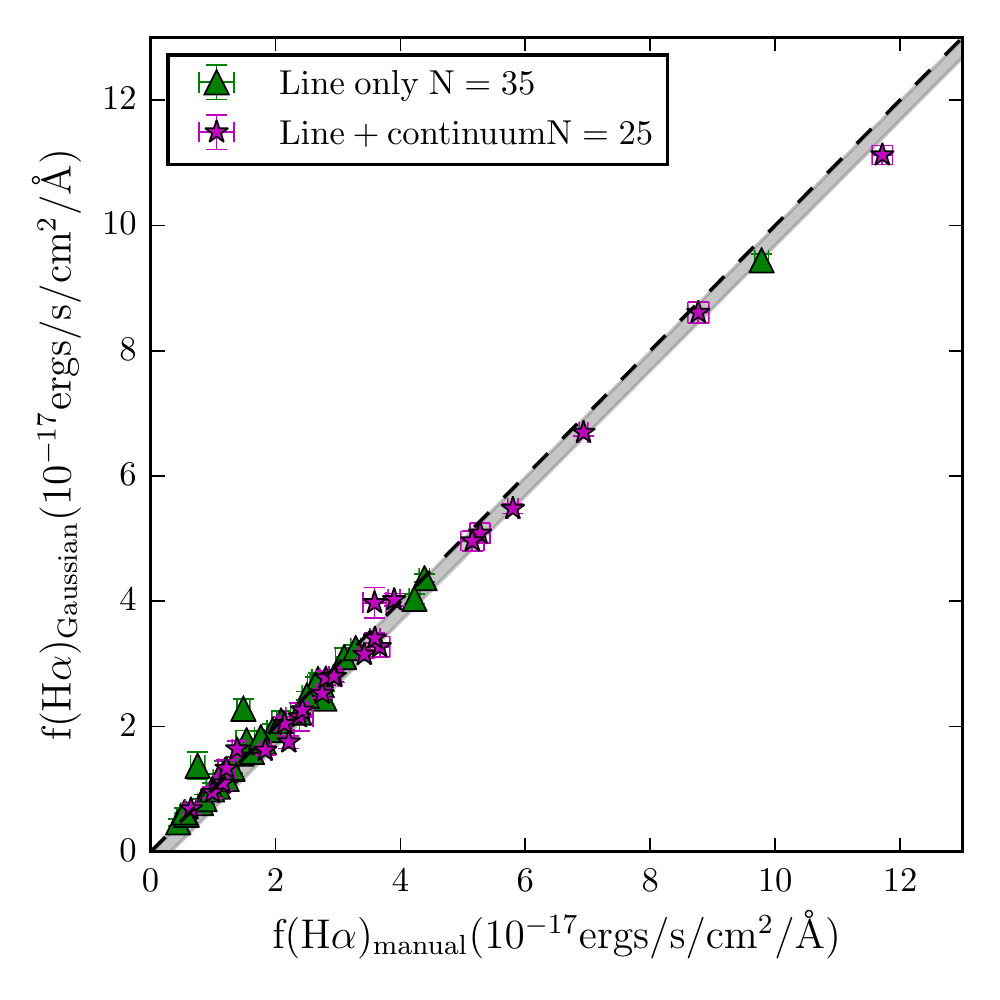}
\caption{{\bf Left:} We compare the continuum level derived from spectra with the expected continuum level from photometry. 
The stars represent objects with robust continuum detections. The remaining sample is shown as triangles. The error bars for the continuum detections come from bootstrap re-sampling. The one to one line is shown as a black dashed line. 
The median from one to one deviation of the continuum detected objects is 0.06 logarithmic flux units with  \NMAD = 0.14 (shown by the grey shaded region). 
The horizontal and vertical dotted lines are the continuum detection levels. 
{\bf Right:} We compare the \Halpha\ flux values computed using Gaussian fits to that of visually identified limits for galaxies which show no strong sky contamination. 
The magenta stars represent objects with robust continuum detections.  The remaining sample is shown as green triangles. The error values are from the integration of the error spectrum. The one to one line is shown as a black dashed line. 
The median one to one deviation of the continuum detected objects is 0.19 flux units with \NMAD= 0.20 (shown by the grey shaded region).
}
\label{fig:continuum_Halpha_checks}
\end{figure}

We compare this photometrically derived continuum level with the spectroscopic continuum level in Figure \ref{fig:continuum_Halpha_checks} (left panel). 
The median deviation of the detected continuum values are \around 0.026 in logarithmic flux values with a 1\NMAD scatter of 0.12, which leads us to conclude that the photometrically derived continuum values agree well with the spectroscopic continuum detections, thus confirming the robustness of our continuum calculations. 
We further note that the large scatter of galaxies below the continuum detection level is driven by the increased fraction of sky noise, which is expected and further confirms that we have robustly established the continuum detection limit. 
There is no strong dependence of the \Halpha\ EW on the continuum detection levels, which suggests that the \Halpha\ EW values are not purely driven by weak continuum levels. Further analysis of such detection biases are shown in Section \ref{sec:observational_bias}.

In order to test the robustness of the measured \Halpha\ flux, we compare the \Halpha\ fluxes 
between this study and the ZFIRE catalogue. Nebular line fluxes in the ZFIRE catalogue are measured by integrating a Gaussian fit to the emission lines \citep{Nanayakkara2016}. We follow a similar technique to calculate \Halpha\ fluxes for emission lines unless they show strong velocity structures.

It is vital to ascertain if Gaussian fits to emission lines would give drastically different \Halpha\ flux values compared to our visually integrated fluxes. 
In Figure  \ref{fig:continuum_Halpha_checks} (right panel), we show this comparison for \sample\ galaxies which do not have strong sky line residuals.  
For the continuum detections in the above subset, the median deviation between the manual limits and Gaussian fits is 0.19$\mathrm{\times 10^{-17} ergs/s/cm^2/A}$ with \NMAD= 0.20$\mathrm{\times 10^{-17} ergs/s/cm^2/A}$. 
Therefore,  \Halpha\ flux values agree with each other within error limits with minimal scatter. 
Single Gaussian fits would fail to describe \Halpha\ emission profiles of galaxies with strong rotations or galaxy that have undergone mergers. 
These features will require complicated multi-Gaussian fits to accurately provide the underlying \Halpha\ flux. All 3\NMAD\ outliers of continuum detections contain profiles that cannot be described using single Gaussian fits. 
Therefore, we expect the direct integration to be the most accurate method to calculate the \Halpha\ flux for galaxies with velocity structures because it is independent of the shape of the \Halpha\ emission. We note that all galaxies with sky line contamination shows profiles that are well described by single Gaussian fits. 

Due to above tests, we are confident that neither the \Halpha\ flux nor the continuum level calculations would give rise to systematic errors in our analysis. Therefore, we conclude that the \Halpha\ EW values derived for out continuum detected \sample\ are robust.

\newpage

\subsection{AGN contamination to \Halpha\ flux}
\label{sec:AGN}

As described in Section \ref{sec:sample_selection}, we flag AGN of the ZFIRE sample following \citet{Coil2015} selection criteria. 
However, it is possible that weak AGN that are not flagged by our selection may still contaminate the \sample\ and contribute to higher \Halpha\ emission. 
In order to investigate effects from such AGN, we use \citet{Coil2015} selection and the measured \NII\ fluxes to compute upper limits to \Halpha\ fluxes required for the galaxies to be flagged as AGN as follows:
\begin{equation}
\label{eq:AGN_Ha_limit}
f(H\alpha)_{inf} < \frac{f([NII])}{0.316}
\end{equation} 
where f(\NII) is the measured \NII\ flux for our galaxies. 
We find that our measured \Halpha\ fluxes are $\sim\times2$ higher than the inferred \Halpha\ fluxes ($f(H\alpha)_{inf}$) computed using the above equation. 
Using the ratio of the measured and inferred \Halpha\ fluxes, we find the upper limit to the fraction of \Halpha\ photons produced by the strongest possible AGN that would not be flagged by the \citet{Coil2015} selection to be  $\sim0.4$. 
Therefore, if our sample is contaminated by weak sub-dominant AGN, we expect the AGN to be responsible for $\sim50\%$ of the observed \Halpha\ flux.


\section{Derivation of box-car filters for IMF and dust analysis}
\label{sec:box car filters}

\subsection{The choice of 340 and 550 filters}
\label{sec:filter choice 340}

\begin{figure}
\includegraphics[trim=10 10 10 0, clip, scale=0.90]{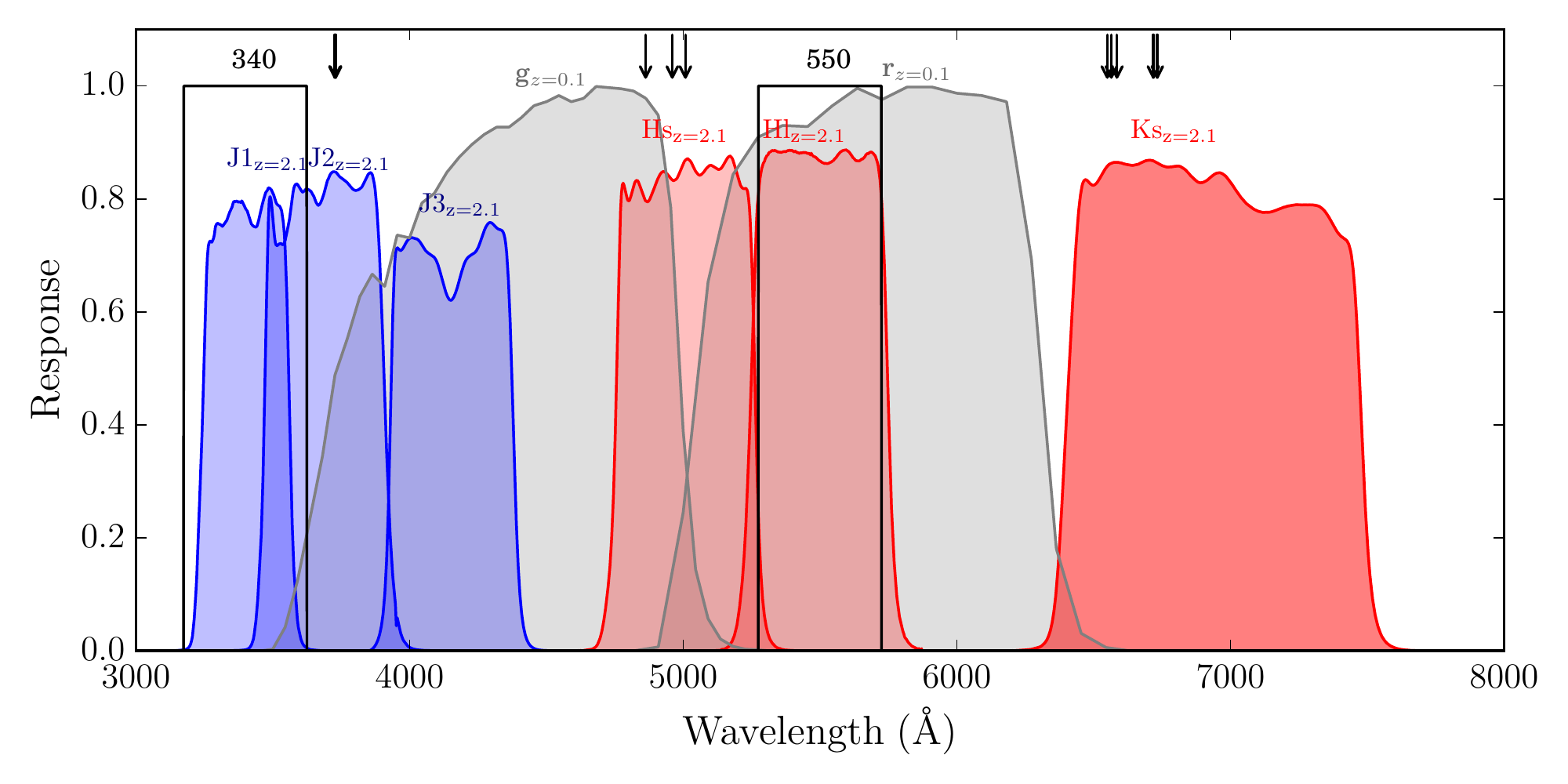}
\includegraphics[trim=10 10 10 0, clip, scale=0.88]{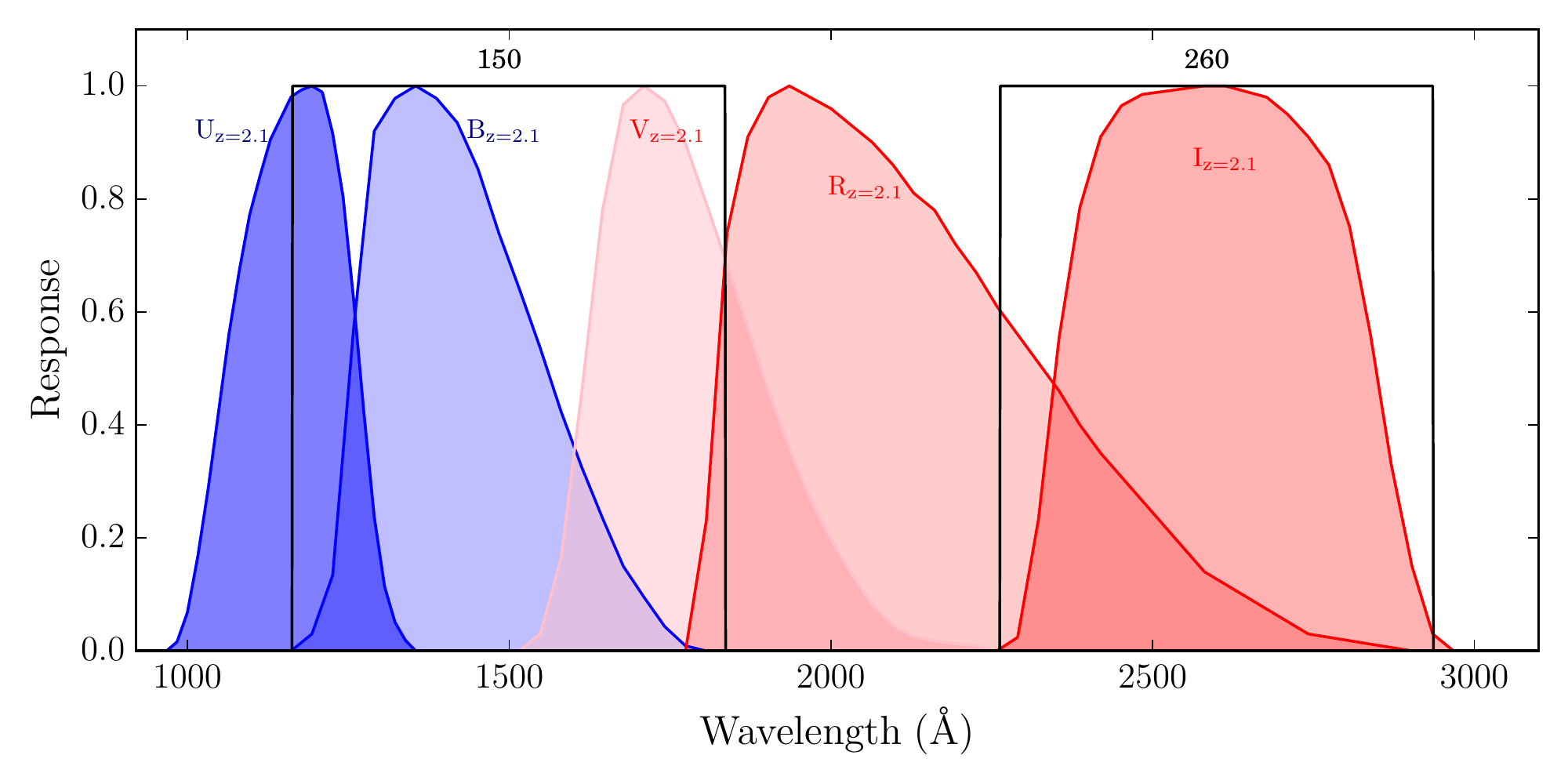}
\caption{ {\bf Top:} The wavelength coverage of the [340] and [550] filters compared with the wavelength coverage of rest frame FourStar NIR filters for a galaxy at $z=2.1$. 
We also show the wavelength coverage of the $g_{z=0.1}$ and $r_{z=0.1}$ filters used by the HG08 analysis. The arrows denote locations of strong emission lines. 
{\bf Bottom:} The wavelength coverage of the [150] and [340] filters along with  \citet{Bessell1990} filters de-redshifted from $z=2.1$. 
}
\label{fig:filter_coverage}
\end{figure}

Due to the strong dependence of nebular emission line properties in the \gr\ colour regime, we shift our analysis to synthetic box car optical filters that avoid regions of strong emission lines. Figure \ref{fig:filter_coverage} shows the wavelength coverage of our purpose built [340] and [550] box car filters along with the wavelength coverage of the FourStar filters in the rest frame of a galaxy at $z=2.1$. It is evident from the figure that $\mathrm{J1_{z=2.1}}$, $\mathrm{J3_{z=2.1}}$, and $\mathrm{Hl_{z=2.1}}$ filters avoid wavelengths with strong emission lines. 
We choose the median wavelength of $\mathrm{J1_{z=2.1}}$ (3400\AA) and $\mathrm{Hl_{z=2.1}}$ (5500\AA) filters to develop  box-car filters with a wavelength coverage of 4500\AA. 
These box-car filters are used to compute optical colours for the ZFIRE-IMF sample. 


\subsection{The choice of 150 and 260 filters}
\label{sec:filter choice 150}

To be consistent with our IMF analysis, which employs the \boxfil\ colours, we compute UV filters  employing a similar technique described in Appendix \ref{sec:filter choice 340}. 
\citet{Bessell1990} B and I filters that samples the optical wavelength regime are chosen for this purpose. For a typical galaxy at $z=2.1$, these filters sample the UV wavelength regime. 
Therefore, by dividing the wavelength coverage of the B and I filters by 3.1, we define a filter set that samples the UV region in the rest frame of a galaxy at $z=2.1$

We then define two box-car filters that has similar wavelength coverage of $\sim6700$\AA\ to the blue-shifted B and I filters. The bluer filter is centred at 1500\AA\ while the redder filter is centred at 2600\AA, both with a width of 673\AA. We name these filters [150] and [260] respectively, and are used in our analysis to investigate the IMF dependence of the dust parameters derived by FAST.

\subsection{Comparison between observed colours and EAZY derived rest frame colours}
\label{sec:EAZY colour comparision}

\begin{figure}
\includegraphics[trim=10 10 10 0, clip, scale=0.925]{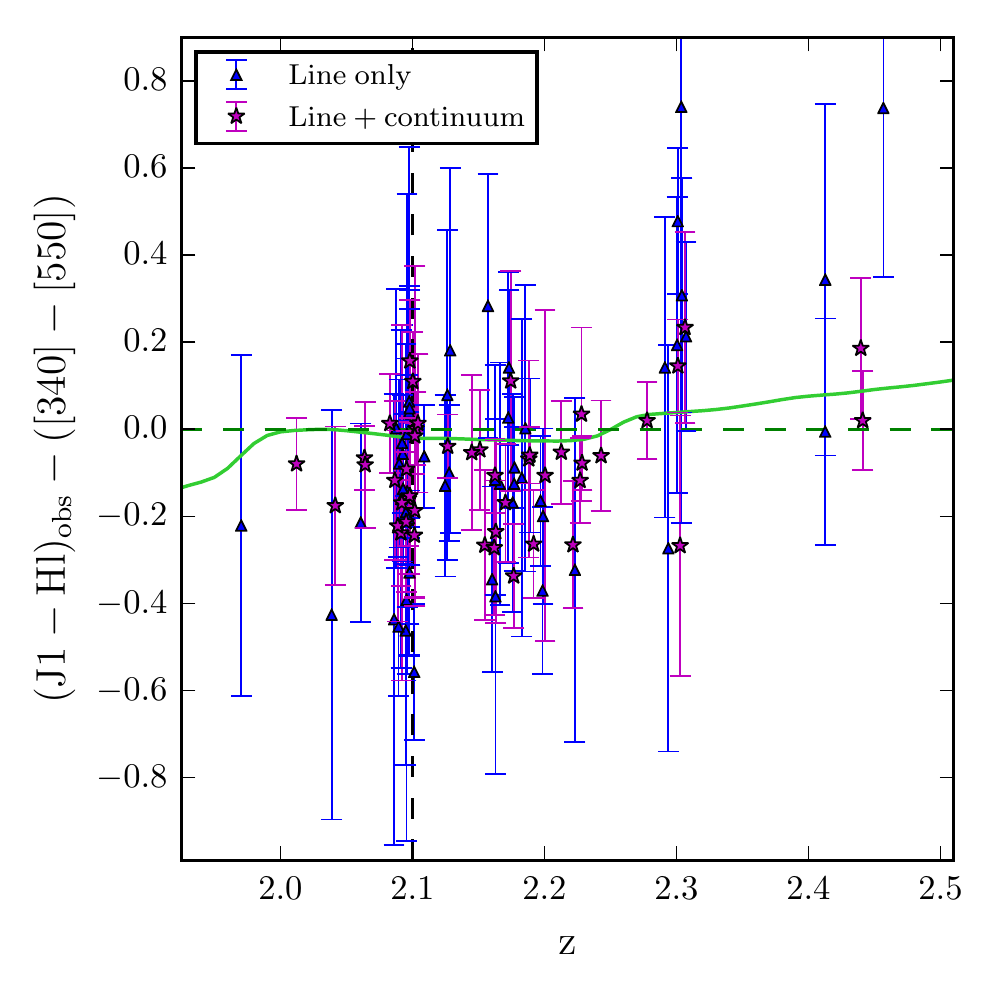}
\includegraphics[trim=0 10 10 0, clip, scale=0.925]{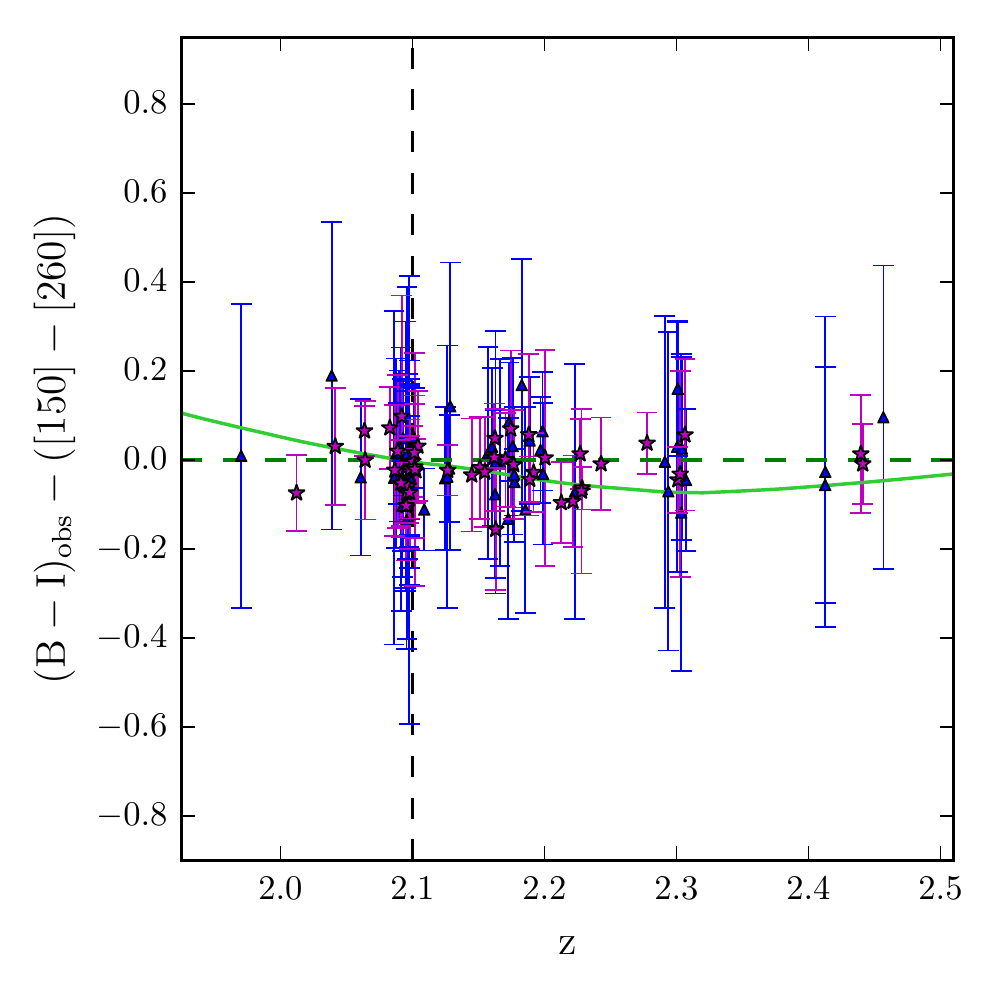}
\caption{ The colour difference between observed colours and rest frame colours derived from best-fitting SED templates from EAZY. rest frame colours are computed in such a way that the wavelength coverage in the observed frame at $z=2.1$ is approximately similar to the wavelength coverage of the rest frame filters at $z=0$.
{\bf Left:} The colour difference between the observed (J1$-$Hl) colours and the \boxfil\ colours of the galaxies in our IMF sample. Galaxies with continuum detections are shown as magenta stars while galaxies with only \Halpha\ emission are shown as blue triangles. The error bars are from the errors in J1 and Hl filters from the ZFOURGE survey photometry. The green line shows the evolution of the colour difference of a PEGASE model galaxy. The vertical dashed line denotes $z=2.1$, which is the redshift used to de-redshift the NIR filters in order to compute rest frame colours. The horizontal dashed line shown is the $\Delta(colour)=0$ line. Galaxies lying on this line shows no difference between the observed colours and the rest frame colours derived via EAZY. 
{\bf Right:}  Similar to left but for (B$-$I) and \dustfil\ colours. 
}
\label{fig:rest_frame_colour_comparision}
\end{figure}

We use the observed FourStar J1 and Hl fluxes and the best-fitting SED fits of our \sample\ to test the robustness of the observed colours with the EAZY derived rest frame colours. 
In Figure \ref{fig:rest_frame_colour_comparision} (left panel), we show the differences between the observed (J1$-$Hl) colours and the rest frame \boxfil\ colours  (as described in Appendix \ref{sec:filter choice 340}) computed from the best-fitting EAZY SED templates.
We compare J1$-$Hl with \boxfil\ colours and expect them to approximately agree by construction at $z=2.1$.

Using a PEGASE model spectrum, we compare the difference with $z$ of J1$-$Hl with \boxfil\ colours with what we expect from SED templates. Lines go through zero at $z=2.1$ as we expect. 
The model spectrum is extracted at $t = 3100$ Myr from a galaxy with an exponentially declining SFH with a p$_1$ = 1000 Myr, $\gamma=-1.35$ IMF, and no metallicity evolution. 
We use the model spectrum to compute the \boxfil\ colour.  We then make a grid of redshifts between $z=1.8$ to $z=2.7$ with $\Delta z = 0.01$ and redshift the wavelength of the model spectra for redshifts in this grid by multiplying the wavelength by $(1+z)$. For each redshift we compute the (J1$-$Hl) colours and since we only investigate the colour difference, there is no need to consider the redshift dimming or K corrections etc.

Since the rest frame filters assume that the galaxies are at $z=2.1$, we expect the observed colours and rest frame colours to agree at this redshift. Figure \ref{fig:rest_frame_colour_comparision} (left panel) shows that for the model galaxy this expectation holds with a maximum deviation of \around$\pm0.1$ mag in colour difference between $z=1.8$ to $z=2.7$. Galaxies in the \sample\ shows a much larger deviation of \around$\pm0.5$ mag, which we attribute to errors in photometry as evident from the large error bars. 
Furthermore, zero-point corrections in the SED fitting techniques can give rise to additional systematic variations.

A similar analysis is performed on the (B$-$I) colours using the observed fluxes from the ZFOURGE survey and \dustfil\ colours (as described in Appendix \ref{sec:filter choice 150}) on the best-fitting EAZY SED templates. The same PEGASE model galaxy used for the (J1$-$Hl) comparison is used to derive the (B$-$I) colours by red-shifting the spectra to redshifts between $z=1.8$ to $z=2.7$. Figure \ref{fig:rest_frame_colour_comparision} (right panel) shows the comparison between observed  and EAZY derived colours along with the $ideal$ expectation computed from PEGASE spectra. Due to the intrinsic shape of the SED, the redshift evolution of $\Delta$(B$-$I) is opposite to that of $\Delta$(J1$-$Hl).

\section{Does SSP models give identical results?}
\label{sec:SSP comparision}

In order to investigate whether there is a strong dependence of the \Halpha\ EW and/or \gr\ colour evolution of model galaxies on the SSP models used, we compare the galaxy properties from PEGASE with that of Starburst99 \citep{Leitherer1999}. 
S99 models support the use of multiple stellar libraries. For this analysis we consider the Padova AGB stellar library which is an updated version of the \citet{Guiderdoni1988} stellar tracks that includes cold star parameters and thermally pulsating asymptotic giant branch (AGB) and post-AGB stars.

We compute PEGASE models using a constant SFR of $1\times10^{-4}$\msol/Myr with various $\Gamma$ values. PEGASE models are scale free and are generated using a baryonic mass reservoir normalized to 1\msol. Having higher SFRs in PEGASE will result in the SFR exceeding the maximal amount of SFR possible for the amount of gas available in the galaxy reservoir before reaching $z\sim2$. 
The other parameters of the PEGASE models are kept as mentioned previously.

S99 models employ a different approach to compute synthetic galaxy spectra where the SFR is not normalized and therefore the SFR should be kept at a reasonable level which allows the HR diagram to be populated with sufficient number of stars during the time steps the models are executed. For S99, we use a SFR of 1\msol using Padova AGB stellar libraries with a Z of 0.02, and similar IMFs to the PEGASE models. We do not change any other parameters in the S99 models from its default values. 
We state the S99 parameters below. \\
$\bullet$ Supernova cutoff mass is kept at 8\msol.\\
$\bullet$ Black hole cutoff mass is kept at 120\msol.\\
$\bullet$ Initial time is set to 0.01 Myr and time 1000 time steps are computed with logarithmic spacing up to 3100 Myr. \\
$\bullet$ We consider the full isochrone for mass interpolation. \\
$\bullet$ We leave the indices of the evolutionary tracks at 0. \\
$\bullet$ PAULDRACH/HILLIER option is used for the atmosphere for the low resolution spectra. \\
$\bullet$ Metallicity of the high resolution models are kept at 0.02.\\
$\bullet$ Solar library is used for the UV line spectrum.\\
$\bullet$ In order to generate spectral features in the NIR, we use microturbulent velocities of 3kms$^{-1}$ and solar abundance ratios for alpha-element to Fe ratio.

To account for the difference in the SFRs between the SSP models, we scale the PEGASE \Halpha\ flux and the corresponding continuum level to the 1\msol/yr value by multiplying by $10^{10}$. The scaling process assumes that the \Halpha\ luminosity $\propto$ SFR as shown by \citet{Kennicutt1983}. 
By interpolating the S99 models to the PEGASE time grid, we calculate the difference in the parameters between the two models for a given time.

All models are computed using a constant SFR. Since, the number of O and B stars that contribute strongly to the \Halpha\ flux is regenerated at a constant speed, the \Halpha\ flux reaches a constant value within a very short time-scale and maintains this value. The lifetime of these O and B stars are in the order of $\sim10$ Myr and therefore, there is no effect from the accumulation of these stars to the \Halpha\ flux. \Halpha\ flux between the SSP models show good agreement with shallower IMFs showing larger \Halpha\ flux values.  
This is driven by the increase in the fraction of larger O and B stars, which contributes to the increase in ionizing photons to boost the \Halpha\ flux. The \Halpha\ flux generated by the two SSP models agree within \around0.03 dex. The discrepancy is slightly higher for steeper IMFs, perhaps driven by minor differences in the mass distribution of the stars in the SSP models.

The continuum level at 6563\AA\ also show good agreement between the SSP models and the differences are $\lesssim 0.1$ dex. Unlike \Halpha\ flux, the continuum levels do not reach a constant value within 3 Gyr. This is driven by the larger lifetime of the A and G stars, which largely contributes to the galaxy continuum. 
The rate at which the continuum level increase is dependant on the IMF, where galaxies with steeper IMFs take longer times to reach a constant continuum level.
However, having a higher fraction of smaller stars eventually leads to a higher continuum level compared to a scenario with a shallower IMF. Since the fraction of A and G stars are higher in a steeper IMF, the higher continuum level is expected. PEGASE and S99 models follow different time-scales for stellar evolution.  For a given IMF, PEGASE models evolves the continuum level faster to reach a higher value compared to the S99 models. The discrepancy between the models increases up to \around1500 Myr,  after which it decreases to reach a constant value.

\begin{figure*}
\includegraphics[scale=0.60]{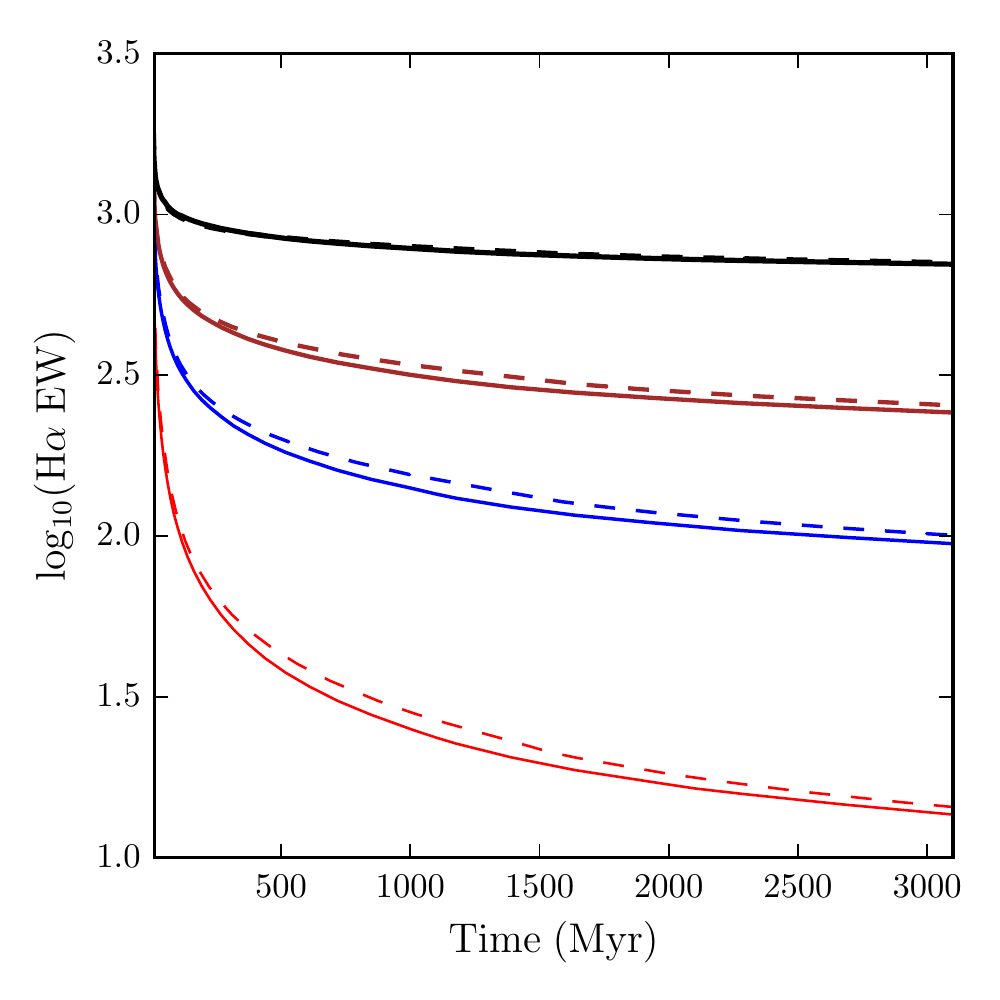}
\includegraphics[scale=0.60]{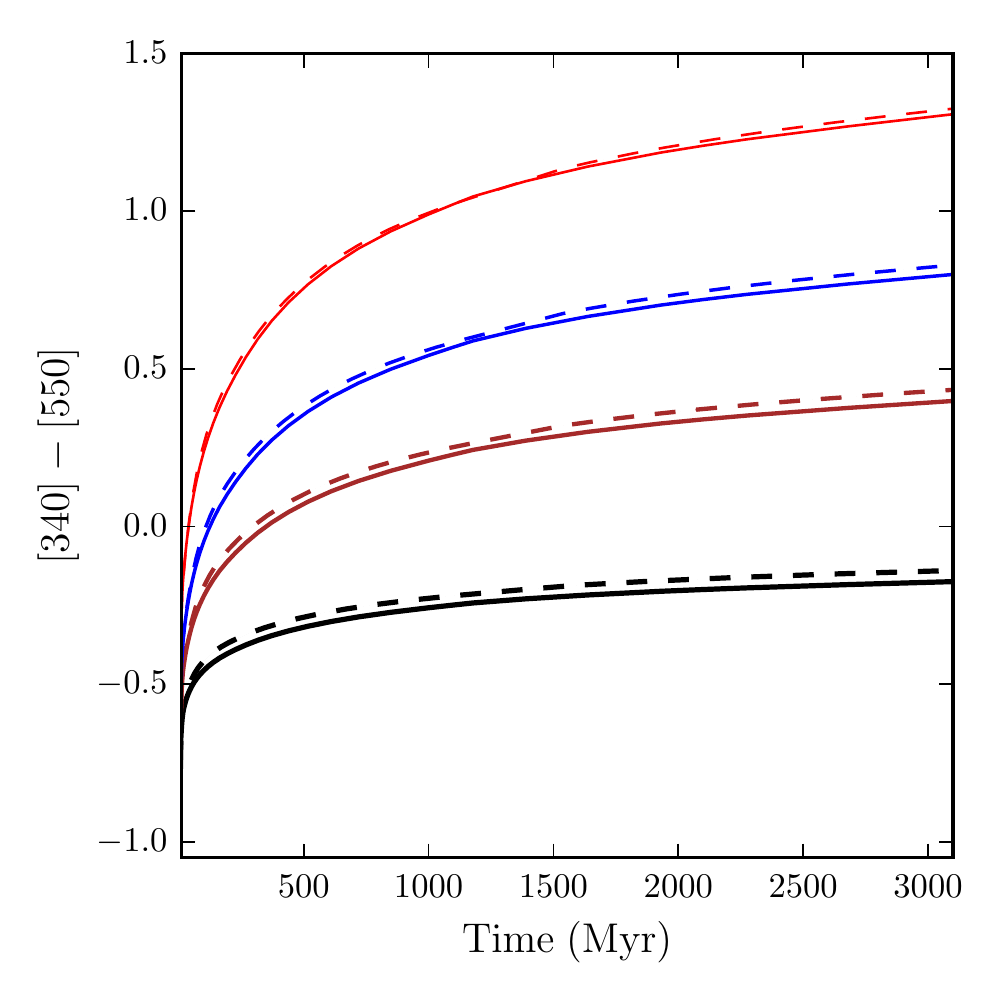}
\includegraphics[trim=10 0 0 0, clip, scale=0.6]{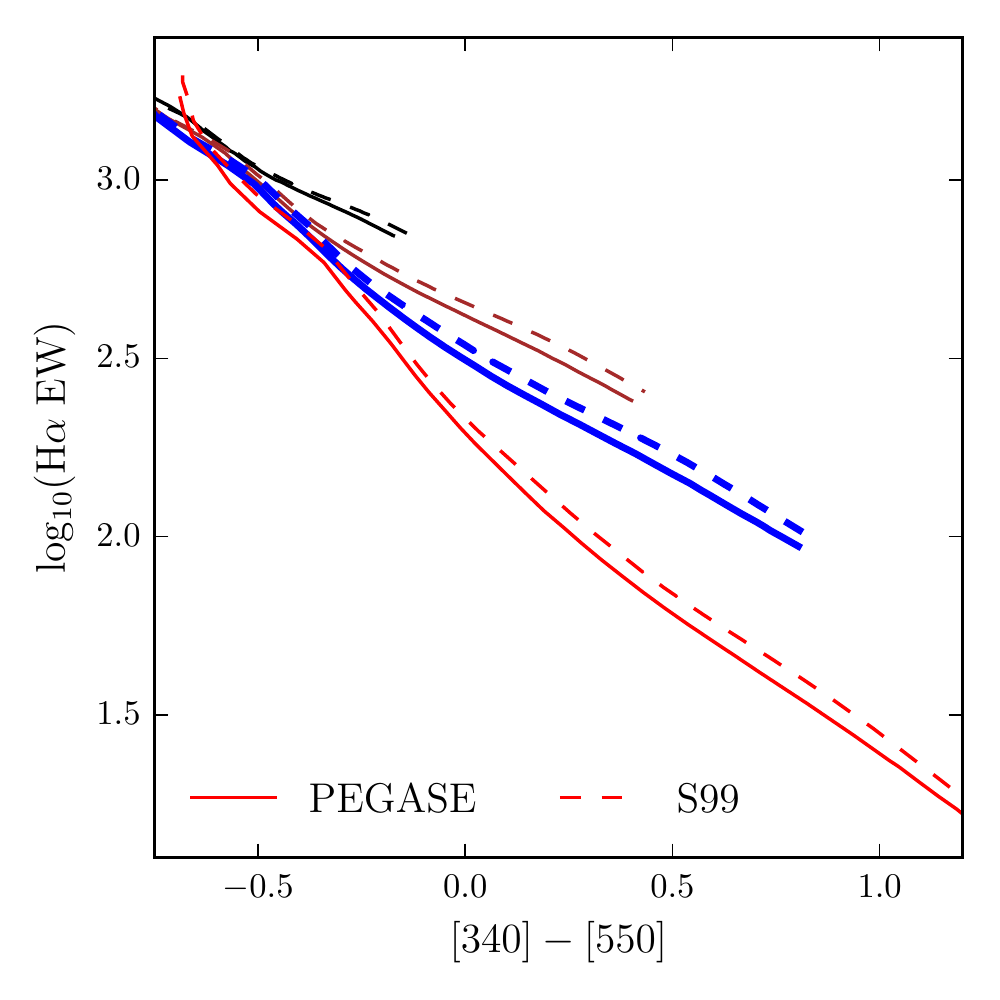}
\caption{Comparison of model parameters between PEGASE and Starburst99 (S99). We compare the evolution of \Halpha\ EW and \boxfil\ colours between PEGASE and S99  models. The S99 models are computed with similar to PEGASE using the updated Padova stellar tracks following prescriptions in Section \ref{sec:PEGASE_models}. Each SSP model has been computed using varying $\Gamma$ values of $-0.5$ (black), $-1.0$ (brown), $-1.35$ (blue), and $-2.0$ (red). The PEGASE models are shown as solid lines while S99 models are shown as dashed lines. 
{\bf Left:} The evolution of  \Halpha\ EW with time for PEGASE and S99 models. 
{\bf Centre:} The evolution of PEGASE and S99 \boxfil\ colours with time. 
{\bf Right:} The evolution of PEGASE and S99 \Halpha\ EW and \boxfil\ colours. The wavelength coverage used for \boxfil\ colours do not include any strong emission lines and therefore the colours are independent of photo-ionization properties of the galaxies. 
}
\label{fig:PEGASE_S99_comp}
\end{figure*}

The change of the \Halpha\ EW between the two SSP models (shown by \ref{fig:PEGASE_S99_comp} left panel) is driven by the differences in the \Halpha\ flux and the continuum level. Both the models behave similarly by decreasing \Halpha\ EW with time. 
Shallower IMFs show higher EW values driven by higher \Halpha\ flux and lower continuum values and the shape of the $\Delta$EW function is driven by the differences in the continuum evolution.

Furthermore, in Figure \ref{fig:PEGASE_S99_comp} (centre panel) we investigate the evolution of \boxfil\ colours derived between PEGASE and S99. 
Since the wavelngth regime covered by [340] and [550] colours do not include emission lines, a direct \boxfil\ colour comparison between S99 models (with no emission lines) with PEGASE models (with emission lines) is possible. 
Models with different IMFs show distinctive differences between the derived \boxfil\ colours. Driven by the excess of higher mass blue O and B stars, galaxies with shallower IMFs show bluer colours compared to galaxies with steeper IMFs for a given time.
Steeper IMFs show a better agreement between the two SSP models. Both, PEAGSE and S99 use the same stellar tracks from the Padova group and  therefore, we attribute the differences between the SSP models to differences in methods used by PEAGSE and S99 to produce the composite stellar populations.

In Figure \ref{fig:PEGASE_S99_comp} (right panel), we compare the evolution of \Halpha\ EW with \boxfil\ colours for PEGASE and S99 models. 
Following on close agreement between the evolution of \Halpha\ EW and \boxfil\ colours between the two SSP models, in the \Halpha\ EW vs \boxfil\ colour plane galaxies from both  PEGASE and S99  show similar evolution. Therefore, our conclusions in this study are not affected by the choice of SSP model (PEGASE or S99) but we note that stellar libraries do play a more prominent role, which we discuss in detail in Section \ref{sec:other_exotica}.

\section{Nebular extinction properties of ZFIRE $z\sim2$ sample}

\subsection{\Hbeta\ detection properties}
\label{sec:Balmer decrement extended}

\begin{figure*}
\includegraphics[trim = 10 0 10 0, clip,scale=0.9]{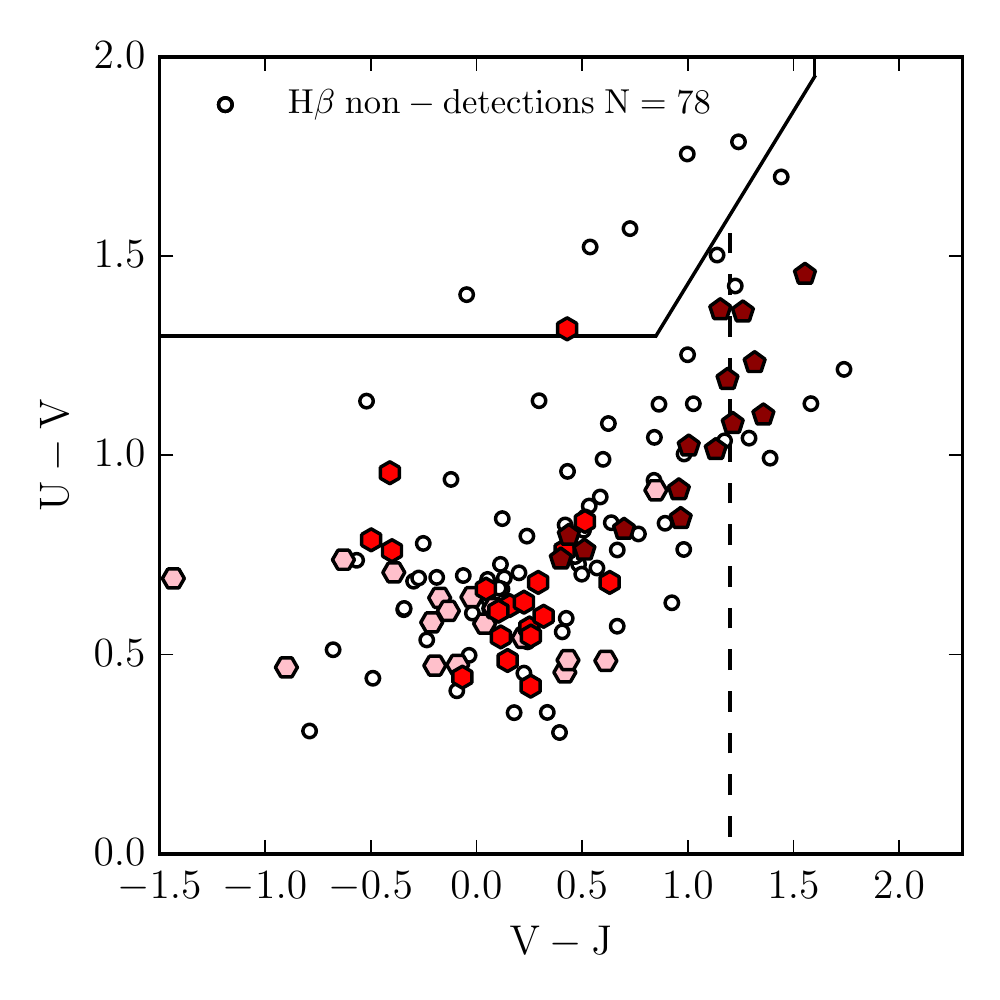}
\includegraphics[trim = 10 0 10 0, clip,scale=0.9]{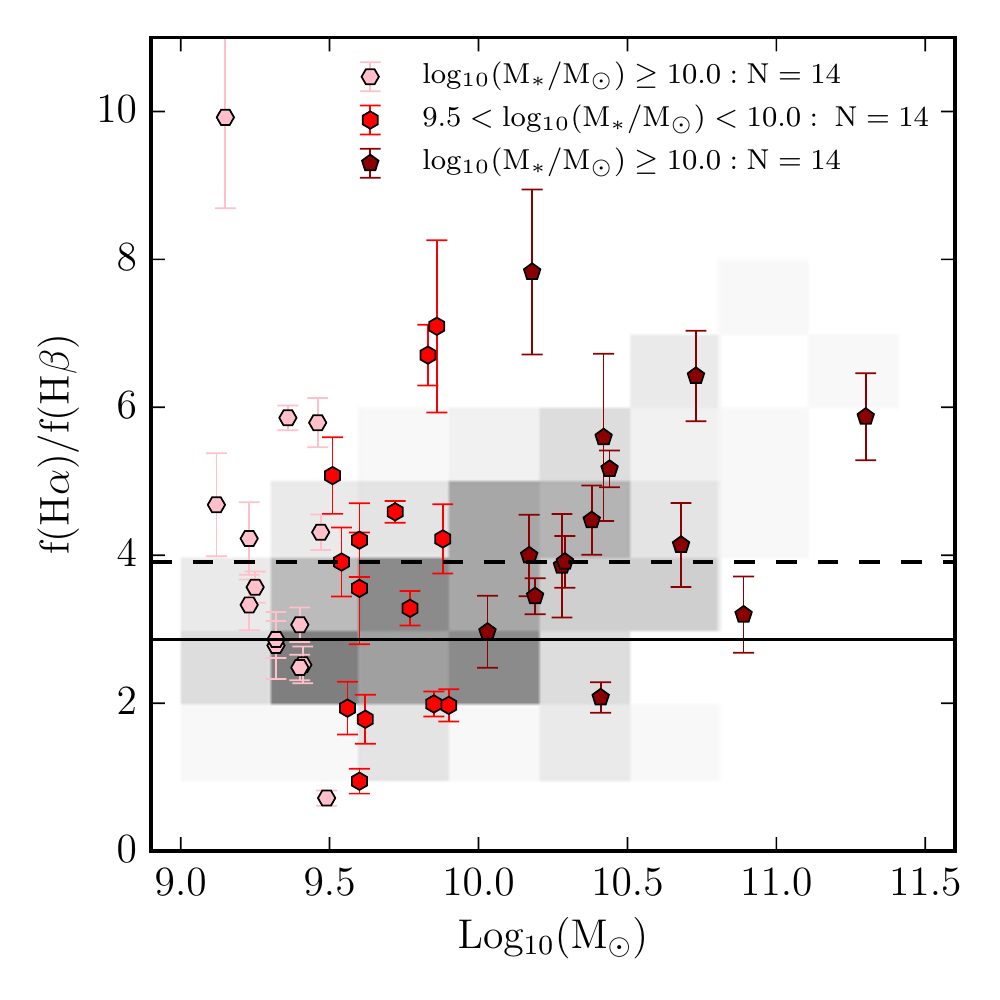}
\caption{
{\bf Left:} rest frame UVJ diagram of the ZFIRE \Hbeta\ targeted sample.  Galaxies shown by green filled symbols have been detected in \Hbeta\ with SNR$>5 $ while black open circles show \Hbeta\ non-detected galaxies. Note that not all galaxies with \Hbeta\ detections have been targeted in K band. The rest frame colours have been derived using spectroscopic redshifts where available. 
{\bf Right:} Balmer decrement vs mass of the ZFIRE sample. 
The black dashed horizontal line is the median Balmer decrement value (3.9) of the sample and the black solid line is the Balmer decrement = 2.86 limit from Case B recombination at $T=10,000$ K \citep{Osterbrock1989}. 
All masses are derived from SED fitting by FAST. The 2D density histogram shows the distribution of values from \citet{Reddy2015}. 
In both panels, the \Hbeta\ detected sample is colour coded according to their stellar mass. 
}
\label{fig:balmer_decrement_properties}
\end{figure*}

Figure \ref{fig:balmer_decrement_properties} (left panel) shows the UVJ diagram \citep{Spitler2014} of the ZFIRE  \Hbeta\ targeted and detected sample.
Rest frame UVJ analysis  shows that our \Hbeta\ detected sample is a reasonably representative subset of our star forming galaxies.


Figure \ref{fig:balmer_decrement_properties} (right panel) shows Balmer decrement values as a function of stellar mass. We calculate the median Balmer decrement for our sample to be 3.9. Using a least-squares polynomial fit to the data we find that galaxies with higher mass are biased towards high Balmer decrement values. There are 9 galaxies with Balmer decrements below 2.86, which are below the theoretical minimum value for case B recombination at $T=10,000$ K \citep{Osterbrock1989}.

\subsection{Derivation of the dust corrections to the \sample}
\label{sec:dust_derivation}

 In this section, we show how we used the \citet{Calzetti2000} and \citet{Cardelli1989} attenuation laws to derive extinction values for the \sample. 

We first calculate the starburst reddening curve at $0.6563\mu$m using the following equation:  
\begin{subequations}
\begin{equation}
\label{eq:starburst_curve_calzetti_IR}
k'(\lambda) = 2.659(-1.857+ \frac{1.040}{\lambda}) +R'_v
\end{equation}
where $\lambda$ is in $\mu m$. This equation is only valid for wavelengths between $0.63\mu m<\lambda<2.2\mu m$. 
Following \cite{Calzetti2000} the total attenuation ($R^{'}_{v}$) is set to 4.05. 
We use the derived value for the reddening curve to calculate the attenuation of the continuum at $0.6563\mu$m\ ($A_c(0.6563)$).
\begin{equation}
\label{eq:cont attenuation}
A_{c}(0.6563) = k'(0.6563) \times \frac{A_v}{R'_v} = 0.82A_{c}(V)
\end{equation}
\end{subequations}
Next we use the \cite{Cardelli1989} prescription to calculate the attenuation of the nebular emission lines. This law is valid for both diffuse and dense regions of the ISM and therefore we expect it to provide a reasonable approximation to the ISM of galaxies at $z\sim2$. 
We use the following equations to evaluate the extinction curve at 6563\AA. 
\begin{subequations}
\begin{equation}
x = 1/\lambda
\end{equation}
where $\lambda$ is in $\mu$m and is between $\mathrm{1.1\mu m^{-1} \leq x \leq 3.3 \mu m^{-1}}$. 
Wavelength dependent values $a(x)$ and $b(x) $are defined as follows:
\begin{equation}
\begin{split}
a(x) = 1 + (0.17699 \times y) - (0.50447\times(y^2)) \\
- (0.02427\times(y^3))+ (0.72085\times(y^4)) + \\
(0.01979\times(y^5))-(0.77530\times(y^6))+ \\
(0.32999\times(y^7))
\end{split}
\end{equation}
\begin{equation}
\begin{split}
b(x) = (1.41338*y)+(2.28305*(y^2))+(1.07233*(y^3))-\\
(5.38434*(y^4))-(0.62251*(y^5))+\\
(5.30260*(y^6))-(2.09002*(y^7))
\end{split}
\end{equation}  
where $y = x-1.82$. 
Using $a(x)$ and $b(x)$ values, the attenuation of the nebular emission line at 0.6563$\mu m$\ ($A_{n}(0.6563)$)  can be expressed as follows: 
\begin{equation}
\label{eq:A_n}
A_{n}(0.6563) = A_{n}[a(0.6563^{-1}) + \frac{b(0.6563^{-1})}{R''_v}] = 0.82A_{n}(V)
\end{equation}
$R^{''}_{v}$ is set to 3.1 following \cite{Cardelli1989}. 
\end{subequations}
Colour excess is defined as:
\begin{equation}
\label{eq:colour excess}
E(B-V) = A(V) /R_v
\end{equation}
\cite{Calzetti1994} shows that at $z\sim0$ newly formed hot ionizing stars reside in dustier regions of a galaxy compared to old stellar populations. Ionizing stars mainly contribute to the nebular emission lines while the old stellar populations contribute the stellar continuum of a galaxy. Therefore, they find that nebular emission lines of a galaxy to be $\sim2$ times more dust attenuated than the stellar continuum. Here, we denote this correction factor as $f$. Using $n$ and $c$ subscripts to denote the nebular and continuum parts respectively,
\begin{equation}
\label{eq:cal fac}
E_n(B-V) = f \times E_c(B-V)
\end{equation}
Substituting equation \ref{eq:colour excess} to Equation \ref{eq:A_n}:
\begin{subequations}
\begin{equation}
A_{n}(0.6563) = 0.82 \times\ R''_v \times E_n(B-V)
\end{equation}
Using Equation \ref{eq:cal fac}:
\begin{equation}
A_{n}(0.6563) = 0.82 \times R''_v \times f \times E_c(B-V)	
\end{equation}
$E_c(B-V)$ is computed using the \citet{Calzetti2000} dust law,
\begin{equation}
A_{n}(0.6563) = 0.82 \times f \times A_c(V) \frac{R''_v}{R'_v}	
\end{equation}
\end{subequations}
Therefore, we express the dust corrected nebular line ($f_i$(\Halpha)) and continuum flux ($f_i$(cont)) as follows:
\begin{subequations}
\begin{equation}
\label{eq:Halpha dust corrected}
f_i(H\alpha)= f_{obs}(H\alpha) \times 10^{0.4(0.62\times f\times A_{c}(V))}
\end{equation}
\begin{equation}
\label{eq:cont dust corrected}
f_i(cont)= f_{obs}(cont) \times 10^{0.4(0.82\times A_{c}(V))}
\end{equation}
where the subscript $obs$ refers to the observed quantity while $i$ refers to the intrinsic quantity. 
\end{subequations}
Since $EW_i =f_i(H\alpha)/f_i(cont))$, finally the dust corrected \Halpha\ EW can be expressed as follows:
\begin{equation}
\label{eg:EW dust corrected}
\log_{10}(EW_i) = \log_{10}(EW_{obs}) + 0.4A_c(V)(0.62f-0.82)
\end{equation}
\\

Next we consider the dust correction for the $z=0.1$ optical colours. Using \cite{Calzetti2000} attenuation law we calculate the starburst reddening curve for these wavelengths using the following equation: 
\begin{equation}
\label{eq:starburst_curve_calzetti_optical}
k'(\lambda) = 2.659(-2.156+ \frac{1.509}{\lambda} - \frac{0.198}{\lambda^2} +\frac{0.011}{\lambda^3}) +R'_v
\end{equation}
This equation is different from Equation \ref{eq:starburst_curve_calzetti_IR}, since this is valid for more bluer wavelengths between $0.12\mu m<\lambda<0.63\mu m$. 
Similar to Equation \ref{eq:cont attenuation}, we work out the attenuation for median wavelengths of the [340] and [550] filters (by definition the filter medians are respectively at $0.34\mu$m and $0.55\mu$m) as follows:. 
\begin{subequations}
\begin{equation}
\label{eq:BC340 dust corrected}
f([340]) = f([340]_{obs}) \times 10^{0.4 \times 1.56 A_c(V)}
\end{equation}
\begin{equation}
\label{eq:BC550 dust corrected}
f([550]) = f([550]_{obs}) \times 10^{0.4 \times 1.00 A_c(V)}
\end{equation}
\end{subequations}
Dust corrected fluxes are used to recalculate the \boxfil\ colours.

The median wavelengths of the g$_{0.1}$ and r$_{0.1}$ filters are respectively at 0.44$\mu m$ and 0.57$\mu m$. Similar to [340] and [550] filters, we use Equation \ref{eq:starburst_curve_calzetti_optical} to calculate the attenuation for g$_{0.1}$ and r$_{0.1}$ filters as follows:
\begin{subequations}
\begin{equation}
\label{eq:g_dust_corrected_derivation}
f(g_i)_{0.1} = f(g_{obs})_{0.1} \times 10^{0.4 \times 1.25 A_c(V)}
\end{equation}
\begin{equation}
\label{eq:r_dust_corrected_derivation}
f(r_i)_{0.1} = f(r_{obs})_{0.1} \times 10^{0.4 \times 0.96 A_c(V)}
\end{equation}
\end{subequations}


\section{PEGASE simulations of starburst galaxies}
\label{sec:simulation_properties}

Here we describe the PEGASE simulations discussed in Section \ref{sec:simulations} to model the effects of starbursts in \Halpha\ EW vs \boxfil\ colour parameter space. 
We tune burst empirical parameters to maximize the number of high \Halpha\ EW objects.
We consider 4 scenarios in our simulations as shown in Table \ref{tab:simulation_param}.  For each scenario we model 100 galaxies and superimpose a single starburst on PEGASE model tracks with an IMF slope of $\Gamma=-1.35$ and a constant SFH. 

PEGASE model galaxies are generated from an initial gas reservoir of 1\msol. Therefore, we normalize the total mass generated by the constant SFH and the star burst to 1\msol\ in order to calculate the SFR for each time step. These values are used to calculate SSPs with a IMF slope of $\Gamma=-1.35$ and upper and lower mass cutoffs set at 0.5\msol\ and 120\msol\ respectively. We use a constant metallicity of 0.02  for all our simulations. The other parameters are kept similar at their default values as described in Section \ref{sec:PEGASE_models}. Following this recipe we generate the simulated galaxies with finer sampling of the time steps around the time of burst to better resolve the effects of the bursts.

\begin{deluxetable}{  c  r  r  r  r  }
\tabletypesize{\scriptsize}
\tablecaption{Summary of scenarios investigated in our starburst simulations.
\label{tab:simulation_param}}
\tablecolumns{5}
\tablewidth{0pt} 
\tablewidth{0pt}
\startdata
\hline
& \\
Scenario & Start of SFH (Myr) & Time of burst (Myr) & $\tau_b$ (Myr) & $f_m$ \\[+1ex] \hline \hline 
 &   \\
1 & 0       & 0--3250    & 100--300 & 0.10--0.30 \\
2 & 0--2500 & 2585--3250 &  10--30  & 0.01--0.03 \\
3 & 0--2500 & 2585--3250 &  10--30  & 0.10--0.30 \\
4 & 0--2500 & 2585--3250 & 100--300 & 0.01--0.03 \\
\end{deluxetable}

\begin{figure*}
\includegraphics[scale=0.6]{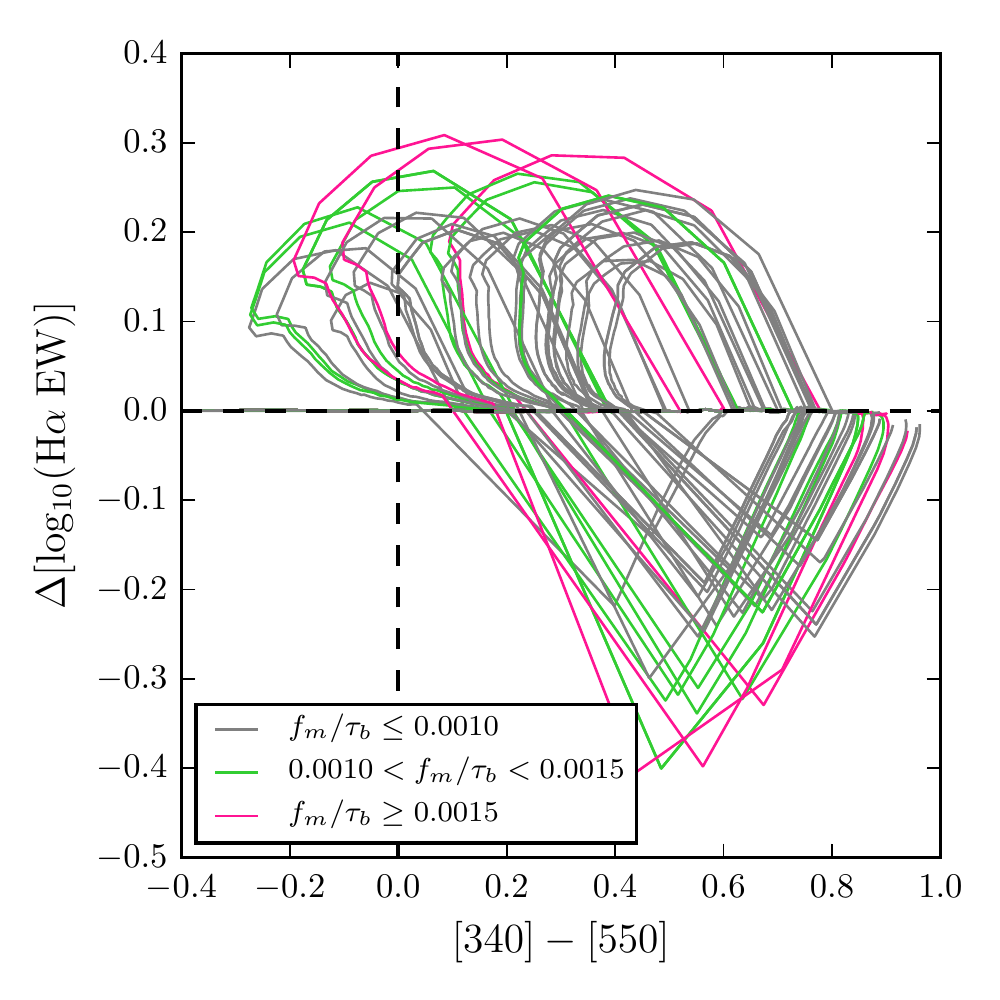}
\includegraphics[scale=0.6]{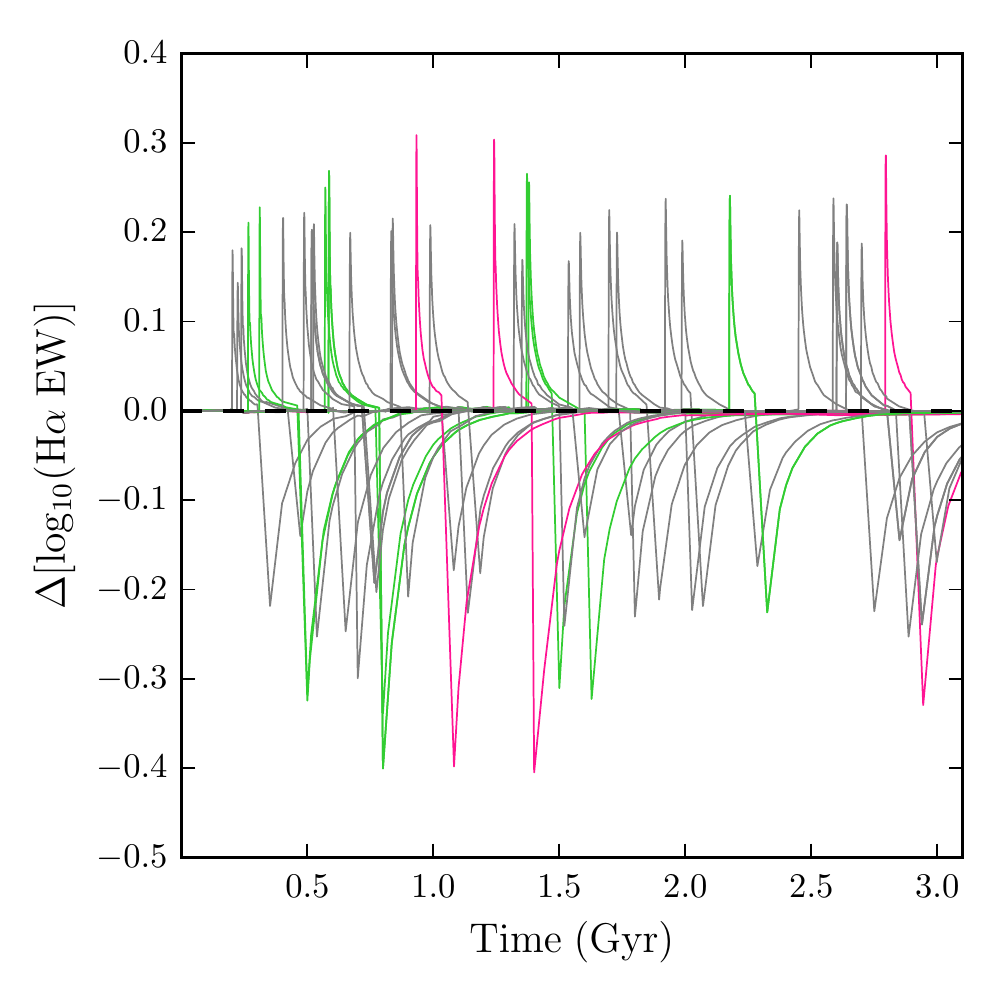}
\includegraphics[scale=0.6]{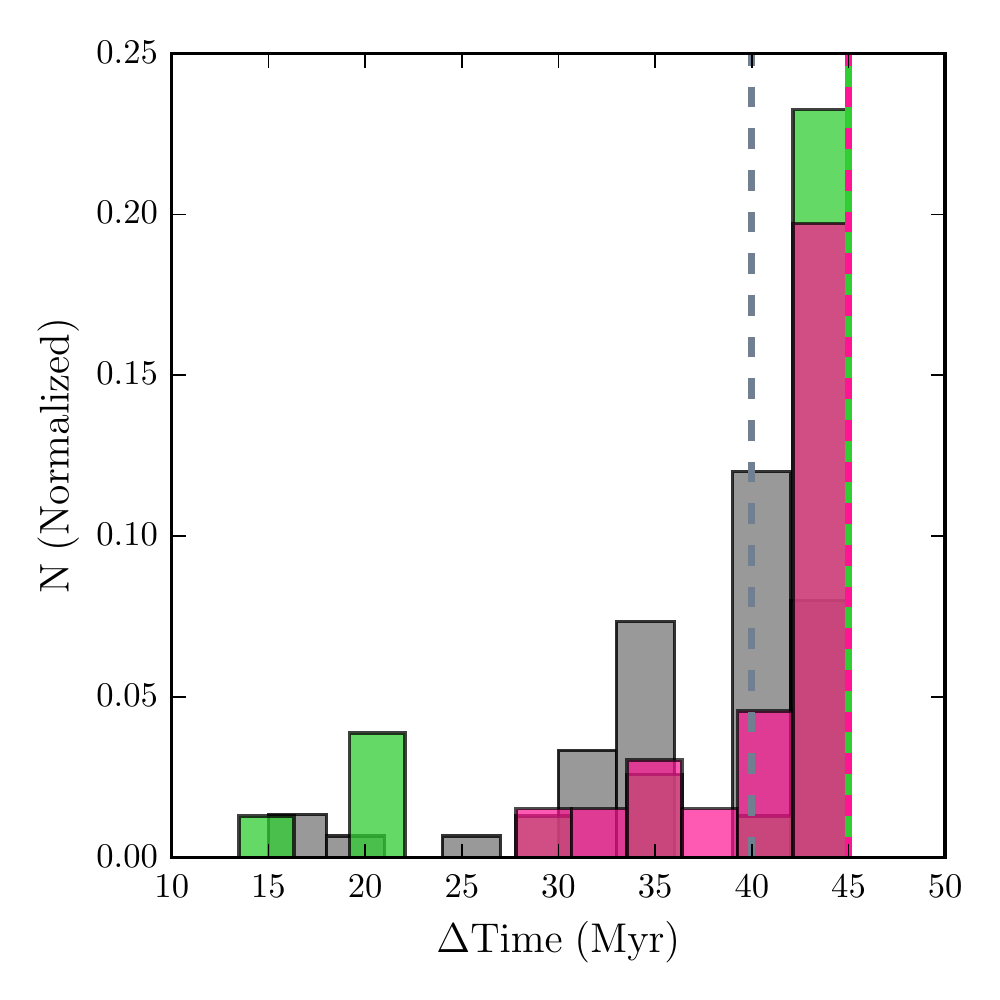}
\caption{ The \Halpha\ EW deviation of galaxies from our simulated sample (Table \ref{tab:simulation_param} Scenario 1). For each model galaxy, the deviations are calculated with respect to the $\Gamma=-1.35$ IMF constant SFH model at the same \boxfil\ colour. 
{\bf Left:} \Halpha\ EW deviations  as a function of \boxfil\ colour for 40 randomly selected galaxies. 
Galaxies are colour coded according to the fraction of stellar mass generated by the burst per unit time (measured in Myr$^{-1}$). The starbursts occur at random times, hence the galaxies deviate from the reference IMF track at random \boxfil\ colours. 
{\bf Centre:} \Halpha\ EW deviations as a function of time.  The galaxy sample is similar to the left panel and are similarly colour coded according to the fraction of stellar mass generated by the burst per unit time.
{\bf Right:} The normalized histogram of the amount of time the starburst tracks stay at least 0.05 dex above the $\Gamma=-1.35$ smooth SFH model. All 100 galaxies in the simulation are shown here and have been colour coded according to the fraction of stellar mass generated by the burst per unit time. The vertical dashed lines show the median for each bin and are as follows: $\mu(f_m/\tau_b\leq0.0010)=40$ Myr, $\mu(0.0010<f_m/\tau_b<0.0015)=45$ Myr, and  $\mu(f_m/\tau_b\geq0.0015)=45$ Myr.
}
\label{fig:simulations_delta_EW_vs_col_time}
\end{figure*}

We use the simulated galaxies from Scenario 1 (Table \ref{tab:simulation_param}) to further investigate the evolution of \Halpha\ EW and \boxfil\ colour of galaxies during a starburst phase. 
40 galaxies are chosen at random from the simulations in Figure \ref{fig:simulations_delta_EW_vs_col_time} to show the deviation of the \Halpha\ EW values from the $\Gamma=-1.35$ IMF track. In a smooth SFH, the \boxfil\ colours are correlated with the age of the galaxies due to stellar populations moving away from the main sequence making the galaxy redder with time. 
However, galaxies undergo bursts at random times resulting them to deviate from the smooth SFH track at random times. Galaxies with higher burst fractions per unit time ($f_m/\tau_b$), show larger deviations due to the extreme SFRs required to generate higher amount of mass within a shorter time period. 
The time-scale galaxies populate above the reference IMF track is small (in the order of $\lesssim50$ Myr) compared to the total observable time window of the galaxies at $z\sim2 (\sim3$ Gyr). The \Halpha\ EW increase significantly soon after a burst within a short time-scale (in order of few Myr) and decreases rapidly to be deficient in \Halpha\ EW compared to the reference constant SFH model. Afterwards, the \Halpha\ EW increases at a slower phase until the tracks join the smooth SFH model after the burst has past. 

In Figure  \ref{fig:simulations_delta_EW_vs_col_time} (right panel), we select all simulated galaxies to calculate the amount of time galaxies spend with higher \Halpha\ EW values ($0.05<\mathrm{\Delta[\log_{10}(H\alpha\ EW)]}$) compared to models with smooth SFHs. We bin the galaxies according to the burst fraction per unit time to find that there is no strong dependence of it on $\Delta$Time.

\end{document}